\documentclass[12pt]{article}
	\usepackage{graphicx} 
	\usepackage{epstopdf}
	\usepackage{amsthm,amsmath,amssymb, amsfonts, amscd, xspace, pifont}
	\usepackage{epsfig, psfrag,pstricks}
	\usepackage{times}
	\usepackage[authoryear,round]{natbib} 
	\newtheorem{theorem}{Theorem}
	\newtheorem{lemma}{Lemma}
	\newtheorem{proposition}{Proposition}
	\newtheorem{corollary}{Corollary}

	\def\bbeta{\mbox{\boldmath $\theta$}}
	\def\bbeta{\mbox{\boldmath $\Theta$}}
	\def\bSigma{\mbox{\boldmath $\Sigma$}}
	\def\bsigma{\mbox{\boldmath $\sigma$}}
	\def\veta{\mbox{\boldmath $\eta$}}
	\def\bbeta{\mbox{\boldmath $\beta$}}
	\def\bw{\mbox{\boldmath $\beta$}}
	
	\def\bJ{{\bf J}}
	\def\bR{\mathbb{R}}
	\def\bN{\mathbb{N}}
	\def\bP{\mathbb{P}}
	\def\vI{\mathbb{I}}
	\def\vE{\mathbb{E}}

	\def\ba{{\bf a}}
	\def\bb{{\bf b}}
	\def\be{{\bf e}}
	\def\bI{{\bf I}}

	\def\bS{{\bf S}}

	\def\bX{{\bf X}}

	\def\bw{{\bf w}}
	
	\def\vf{{\bf f}}

	\def\bv{{\bf v}}
	\def\br{{\bf r}}

	\def\bI{{\bf I}}
	\def\bw{{\bf w}}
	\def\bx{{\bf x}}
	\def\by{{\bf y}}
	\def\bz{{\bf z}}
	
	\def\nn{\nonumber}
	
	\def\v2{\vspace{0.2in}}

\begin{document}

	\title{LASSO risk and phase transition under dependence}

\author{Hanwen Huang \\\\
	{\it Department of Epidemiology and Biostatistics}\\
	{\it University of Georgia}\\
	{ Athens, GA 30602, USA}\\
	huanghw@uga.edu}
\date{}

\maketitle
	
	\begin{abstract}
We consider the problem of recovering a $k$-sparse signal $\bbeta_0\in\bR^p$ from noisy observations $\by=\bX\bbeta_0+\bw\in\bR^n$. One of the most popular approaches is the $l_1$-regularized least squares, also known as LASSO. We analyze the mean square error of LASSO in the case of random designs in which each row of $\bX$ is drawn from distribution $N(0,\bSigma)$ with general $\bSigma$. We first derive the asymptotic risk of LASSO for $\bw\ne 0$ in the limit of $n,p\rightarrow\infty$ with $n/p\rightarrow\delta\in[0,\infty)$. We then examine conditions on $n$, $p$, and $k$ for LASSO to exactly reconstruct $\bbeta_0$ in the noiseless case $\bw=0$. A phase boundary $\delta_c=\delta(\epsilon)$ is precisely established in the phase space defined by $0\le\delta,\epsilon\le 1$, where $\epsilon=k/p$. Above this boundary, LASSO perfectly recovers $\bbeta_0$ with high probability. Below this boundary, LASSO fails to recover $\bbeta_0$ with high probability. While the values of the non-zero elements of $\bbeta_0$ do not have any effect on the phase transition curve, our analysis shows that $\delta_c$ does depend on the signed pattern of the nonzero values of $\bbeta_0$ for general $\bSigma\ne\bI_{p\times p}$. This is in sharp contrast to the previous phase transition results derived in i.i.d. case with $\bSigma=\bI_{p\times p}$ where $\delta_c$ is completely determined by $\epsilon$ regardless of the distribution of $\bbeta_0$. Underlying our formalism is a recently developed efficient algorithm called approximate message passing (AMP) algorithm. We generalize the state evolution of AMP from i.i.d. case to general case with $\bSigma\ne\bI_{p\times p}$. Extensive computational experiments confirm that our theoretical predictions are consistent with simulation results on moderate size system. 
	\end{abstract}
	
	\section{Introduction}
	\subsection{LASSO phase transition}\label{sec11}
	
	Consider the problem of recovering a sparse signal $\bbeta_0\in\bR^p$ from a under-sampled collection of noisy measurements $\by=\bX\bbeta_0+\bw$, where the matrix $\bX$ is $n\times p$, the $p$-vector $\bbeta_0$ is $k$-sparse (i.e. it has at most $k$ non-zero entries), and $\bw\in\bR^n$ is random noise. One of the most popular approaches for this problem is called LASSO which estimates $\bbeta_0$ by solving the following convex optimization problem
	\begin{eqnarray}\label{lasso0}
	\hat{\bbeta}(\lambda)&=&\text{argmin}_{\bbeta\in\bR^p}\left\{\frac{1}{2}\|\by-\bX\bbeta\|^2+\lambda\|\bbeta\|_1\right\}.
	\end{eqnarray}
	In the noiseless case $\bw=0$, exact reconstruction of $\bbeta_0$ through (\ref{lasso0}) is possible when $n\ge p$ or $\bbeta_0$ is sufficiently sparse for the case of $n\textless p$. Knowing the precise limits to such sparsity for the case of $n\textless p$ is important both for theory and practice. 
	
\begin{figure}[hbtp]
		\vspace{0cm}
		\begin{center}
	\includegraphics[angle=-90,width=0.5\textwidth]{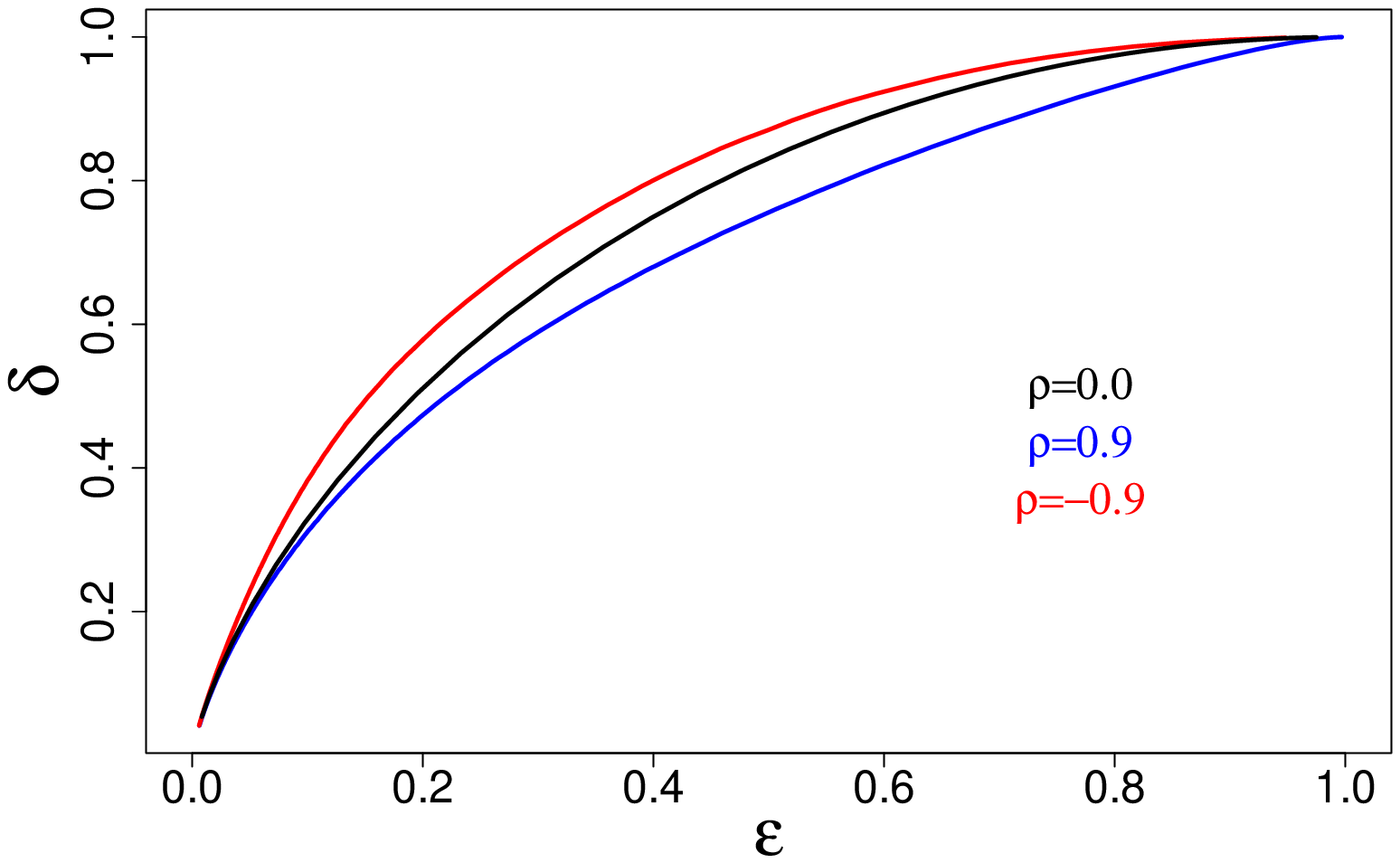}
		\end{center} 
		\vspace{-0.5cm}
		\caption{Phase transition boundary in the plane $(\delta,\epsilon)$ when the matrix $\bX$ consisting of i.i.d. Gaussian rows $\bx_i\sim N(0,\bSigma)$. Black curve: $\Sigma=\bI$. Red curve: $\bSigma$ is block-diagonal with AR(1) block structure $\bSigma_s$, i.e. $\Sigma_{s,ij}=\rho^{|i-j|}$ with block length $s=2$ and $\rho=-0.9$. Blue curve: $\bSigma$ is block-diagonal with AR(1) block structure, i.e. $\Sigma_{s,ij}=\rho^{|i-j|}$ with block length $s=2$ and $\rho=0.9$.}
		\label{figure1}
	\end{figure}

	In the noiseless case, the $\lambda=0$ limit of (\ref{lasso0}) is identical to the solution of the following $l_1$ minimization problem
	\begin{eqnarray}\label{mini}
	&&\min\|\bbeta\|_1,\\\nn
	&&\text{subject to }\by=\bX\bbeta.
	\end{eqnarray}
	The precise condition under which $\hat{\bbeta}(\lambda=0)$ can successfully recover $\bbeta_0$ has been obtained through large system analysis by letting $n,p,k$ tend to infinity with fixed rates $n/p$ and $k/p$. Let $\epsilon=k/p$ and $\delta=n/p$ denote the sparsity and under-sampling fractions for sampling $\bbeta_0$ and $\by$ according to $\by=\bX\bbeta_0$. Then $(\delta,\epsilon)\in[0,1]$ defines a phase space which expresses different combinations of under-sampling $\delta$ and sparsity $\epsilon$. When the elements of the matrix $\bX$ are generated from i.i.d. Gaussian, the phase space can be divided into two phases: "success" and "failure" by a phase transition curve $\delta=\delta_c(\epsilon)$ which has been explicitly derived in the literature (see e.g. \cite{Donoho9446,Donoho4273,Kabashima2009TypicalRL,dmm2009}) as shown by the black curve in Figure \ref{figure1}. Above this curve, LASSO perfectly recovers the sparse signal $\bbeta_0$ with high probability, i.e. $\hat{\bbeta}(\lambda=0)=\bbeta_0$. Below this curve, the reconstruction fails, i.e. $\hat{\bbeta}(\lambda=0)\ne\bbeta_0$ also with high probability.
	
	Our aim in this paper is to study the LASSO phase transition under arbitrary covariance dependence, i.e. $\bX$ consists of i.i.d. Gaussian rows $\bx_i\sim N(0,\bSigma)$ with general covariance matrix $\bSigma\succ 0$ and $\bSigma\ne\bI_{p\times p}$. We present formulas that precisely characterize the LASSO sparsity/undersampling trade-off for arbitrary $\bSigma$. Our numerical results show that LASSO phase transition depends on the form of $\bSigma$. For example, the red and blue curves in Figure \ref{figure1} correspond to the phase transition boundaries for block-diagonal covariance matrix $\bSigma$ with AR(1) block structure $\bSigma_s$, i.e. $\Sigma_{s,ij}=\rho^{|i-j|}$ with block length $s=2$ and $\rho=-0.9$ and $\rho=0.9$ respectively . These results indicate that for a given sparsity fraction $\epsilon$, the limits of allowable undersampling $\delta_c(\epsilon)$ of LASSO in the case when $\bX$ has non-independent entries can be either higher or lower than the corresponding value in the case when $\bX$ has i.i.d. entries. To the best of our knowledge, this is the first result to illustrate the LASSO phase transition for matrices $\bX$ that have non-independent entries. 
	
	%Our proof technique is able to deal with the output non-linearity and is hence of independent interest, opening ways to establish similar results for models of neural networks where nonlinearities are essential but in general difficult to account for.
	
	\subsection{Approximate Message Passing}\label{sec12}
	
	Our analysis is based on the asymptotic study of mean squared error (MSE) of the LASSO estimator, i.e. the quantity $\|\hat{\bbeta}(\lambda)-\bbeta_0\|^2/p$, in the large system limit $n,p\rightarrow\infty$ with $n/p=\delta\in [0,\infty)$ fixed. We derive the asymptotic MSE through the analysis of an efficient iterative algorithm first proposed by \cite{dmm2009} called approximate message passing (AMP) algorithm. The AMP algorithms can be considered as quadratic approximations of loopy belief propagation algorithms on the dense factor graph corresponding to the LASSO model. A striking property of AMP algorithms is that their high-dimensional per-iteration behavior can be characterized by a one-dimensional recursion termed $state~evolution$. 
	%State evolution is a formalism that allows to characterize the asymptotic behavior of AMP as the number of dimension tend to infinity. 
	The AMP's state evolution was first conjectured in \cite{dmm2009} and subsequently proved rigorously in \cite{bm2011} for i.i.d. Gaussian matrices. This result was extended to i.i.d. non-Gaussian matrices in \cite{bayati2015} under certain regularity conditions. \cite{10.1093/imaiai/iat004} further extended the AMP's state evolution to independent but non-identical Gaussian matrices. But there remains the important question of how AMP behaves with non-independent matrices. 
	
	In this paper, we establish the AMP's state evolution for non-independent Gaussian matrices whose fixed points are consistent with the replica prediction derived in \cite{Montanari}. 
	%Non-rigorous predictions for the optimal errors existed for special cases of GLMs, e.g., for the
	%perceptron, in the field of statistical physics based on the so-called replica method. 
	On the basis of this result, we first derive the MSE for AMP estimators using the fixed points of state evolution, then we obtain the MSE for LASSO by proving that, in the large system limits, the AMP algorithm converges to the LASSO optimum after enough iterations. Our analysis strategy is similar to the one used in \cite{BayatiM12} for i.i.d. Gaussian matrices. However, our main result cannot be seen as a straightforward extension of the ones in \cite{BayatiM12}. In particular, the proofs of some results for non-independent case are much more complicated than for i.i.d. case,  and our proof techniques are hence of independent interest, see e.g. the proof of Lemma 1 for the concavity and strict increasing of $\psi$ function defined in (\ref{taut}), the proof of Theorem {\ref{thm2}} for deriving the phase transition curve, and the proof of Lemmas \ref{csparsity} and \ref{lm1.2} for the structural property of LASSO under dependent designs.
	
	Note that although this study is motivated by the phase transition problem shown in Figure \ref{figure1} which is restricted to the case when $(\delta, \epsilon)\in [0,1]$, the AMP and main results derived in Theorem \ref{thm1} work fine for the entire range $\delta\in[0,\infty)$. The LASSO risk formulas derived in Theorem \ref{thm1} apply to both noiseless and noisy cases with quite general i.i.d. random error. The phase transition results derived in Theorem \ref{thm2} are only for the noiseless case. This result can also be generalized to the noisy case and we have some discussion about this in Section \ref{discussion}.
	
	\subsection{Related work}\label{sec13}
	
	\cite{NIPS2009_3635} derived expressions for the asymptotic mean square error of LASSO. Similar results were presented in \cite{5394838,Montanari}. Unfortunately, these results were non-rigorous and were obtained through the famous replica method from statistical physics \citep{Mezard}. Some rigorous proofs were given in \cite{Barbier2019,Reeves,BayatiM12} to show that the replica symmetric prediction for LASSO is exact. However, all these rigorous proofs are limited to settings with i.i.d. Gaussian measurement matrices.
	
	By now a large amount of empirical and theoretical studies have been conducted to understand the phase transitions of regularized reconstruction exhibited by different algorithms. In the noiseless case, the phase transition curve based on (\ref{mini}) was explored in \cite{Donoho9446} utilizing techniques of combinatorial geometry for entries of $\bX$ being i.i.d. Gaussians. The AMP algorithm was proposed in \cite{dmm2009} which produces the same phase transition curve. It has been proved in \cite{BayatiM12} that the limit of AMP estimate corresponds to the solution of LASSO in the asymptotic settings. Statistical physics methods were used to study $l_q$ ($0\le q\le 1$) based reconstruction methods in \cite{Kabashima2009TypicalRL}. \cite{7987040} and \cite{weng2018} studied the phase transition for $l_q$ penalized least square in the case of $0\le q\textless 1$ and $1\le q\le 2$ respectively. \cite{Krzakala} replaced the $l_1$ regularization with a probabilistic approach and studied its phase transition. \cite{AccuratePrediction} derived phase transition of AMP for a wide class of denoisers. In noisy case, \cite{dmm} studied the noise sensitivity phase transition of LASSO through deriving the minimax formulation of the asymptotic MSE. \cite{7987040,weng2018} studied the phase transition of $l_q$-regularized least squares using higher order analysis of regularization techniques. The phase transition in generalized linear models for i.i.d. matrices was characterized in \cite{glm1}. \cite{complex} generalized AMP to complex approximate message passing methods and used it to study phase transitions for compressed sensing with complex vectors.
	
	Most of the above results are for i.i.d. Gaussian matrices and some of them are for independent but non-identical Gaussian matrices. This paper performs the phase transition analysis of LASSO under dependent Gaussian matrices. We derive the basic relation between minimax MSE and the phase-transition boundary in the sparsity-undersampling plane. We adopt the message passing analysis whose state evolution allows to determine whether AMP recovers the signal correctly, by simply checking whether the MSE vanishes asymptotically or not. Most closely related to the current paper are results by \cite{Wainwright} that derives the sharp thresholds for LASSO sparsity recovery in the case of random designs in which each row of $\bX$ is drawn from a broad class of Gaussian ensembles $N(0,\bSigma)$. However, the major difference is that \cite{Wainwright} only provides the necessary and sufficient conditions for the recovery of sparsity pattern, while we focus on the recovery of complete signal including both signed support and magnitude. Recently, based on Gordon's inequality, \cite{celentano2020lasso} derived the LASSO risk under non-standard Gaussian design for i.i.d. Gaussian random error, i.e. $w_i\overset{i.i.d.}{\sim}N(0,\sigma_w^2)$. But they didn't study the phase transition problem and also we don't have Gaussian restriction here for random error $\bw$.	
	
	% Our proof assumes that the state evolution formalism developed in [DMM09, DMM10, DMM11] holds, in the precise terms stated below. This formalism was established rigorously for separable denoisers (under additional regularity assumptions) in [BM11a]. The latter problem does in turn reduce to a problem in real analysis. A crucial observation for state evolution is that the mean squared error of the AMP reconstruction at iteration $t$ is practically non-random for large system sizes $p$,  and has a well-defined limit as $p\rightarrow\infty$.  In particular, the limit exists almost surely. 	
	
	\section{LASSO risk}\label{method}
	The Gaussian random design model for linear regression is defined as follows. We are given $n$ i.i.d. pairs $(y_1,\bx_1),\cdots,(y_n,\bx_n)$ with $y_i\in\bR$, $\bx_i\in\bR^p$, and $\bx_i\sim N(0,\bSigma)$ for some positive definite $p\times p$ covariance matrix $\bSigma\succ 0$. Further, $y_i$ is a linear function of $\bx_i$, plus noise
	\begin{eqnarray}\nn
	y_i&=&\bx_i^T\bbeta_0+w_i,
	\end{eqnarray}
	where $w_i\stackrel{i.i.d.}{\sim} p_w$ with mean 0 and variance $\sigma_w^2$, and $\bbeta_0\in\bR^p$ is a vector of parameters to be estimated. The special case $\bSigma=\bI_{p\times p}$ is usually referred to as standard Gaussian design model. In matrix form, letting $\by=(y_1,\cdots,y_n)^T$, $\bw=(w_1,\cdots,w_n)^T$, and denoting by $\bX$ the matrix with rows $\bx_1^T,\cdots,\bx_n^T$, we have
	\begin{eqnarray}\nn
	\by&=&\bX\bbeta_0+\bw.
	\end{eqnarray}
%	We are interested in high-dimensional settings where the number of parameters exceeds the sample size, i.e. $p\textgreater n$, but the number of non-zero entries of $\bbeta_0$ is smaller than $n$. 
In this paper, our approach is based on the LASSO estimator 
	\begin{eqnarray}\label{lasso}
	\hat{\bbeta}&=&\text{argmin}_{\bbeta}{\cal C}(\bbeta),
	\end{eqnarray}
	where
	\begin{eqnarray}\nn
	{\cal C}(\bbeta)&=&\frac{1}{2}\|\by-\bX\bbeta\|^2+\lambda\|\bbeta\|_1.
	\end{eqnarray}
	
	We will consider sequences of instances of increasing sizes. The sequence of instances $\{\bbeta_0(p),\bw(p)$, $\bSigma(p),\bX(p)\}$ parameterized by $p$ is said to be a converging sequence if $\bbeta_0(p)\in\bR^p,\bw(p)\in\bR^n,\bSigma(p)\in\bR^{p\times p},\bX(p)\in\bR^{n\times p}$ with $n=n(p)$ is such that $n/p\rightarrow\delta\in(0,\infty)$, and in addition the following conditions hold:
	\begin{enumerate}
		\item The empirical distribution of the entries of $\bbeta_0(p)$ converges weakly to a probability measure $p_{\beta_0}$ on $\bR$ with bounded second moment. Further \\
		$\sum_{i=1}^p\beta_{0,i}(p)^2/p\rightarrow E_{p_{\beta_0}}\{\beta_0^2\}$.
		\item The empirical distribution of the entries of $\bw(p)$ converges weakly to a probability measure $p_{w}$ on $\bR$ with $\sum_{i=1}^nw_{i}(p)^2/n\rightarrow\sigma_w^2\textless\infty$.
		\item For any $\bv\in\bR^p$, $\|\bv\|^2_{\bSigma(p)}=O(\|\bv\|^2)$ and  $\|\bv\|^2_{{\bSigma(p)}^{-1}}=O(\|\bv\|^2)$, where $\|\bv\|^2_{\bSigma}=\bv^T\bSigma\bv$.
		\item The rows of $\bX(p)$ are drawn independently from distribution $N(0,\frac{1}{n}\bSigma(p))$.  
		\item The sequence of functions
		\begin{eqnarray}\label{c5}
		{\cal E}^{(p)}(a,b)\equiv\frac{1}{p}E\min_{\bbeta\in\bR^p}\left\{\frac{1}{2}\|\bbeta-\bbeta_0(p)-\sqrt{a}\bSigma(p)^{-1/2}\bz\|^2_{\bSigma(p)}+b\|\bbeta\|_1\right\}
		\end{eqnarray}
		admits a differentiable limit ${\cal E}(a,b)$ on $\bR_+\times \bR_+$ with $\frac{\partial{\cal E}^{(p)}(a,b)}{\partial a}\rightarrow\frac{\partial{\cal E}(a,b)}{\partial a}$ and $\frac{\partial{\cal E}^{(p)}(a,b)}{\partial b}\rightarrow\frac{\partial{\cal E}(a,b)}{\partial b}$, where $\bz\sim N(0,\bI_{p\times p})$ is independent of $\bbeta_0(p)$.
		\item For any $a_1,b_1,a_2,b_2\in\bR_+$ and any $2\times 2$ positive definite matrix $\bS$, the following limit exists and is finite  
		\begin{eqnarray}\nn
		\lim_{p\rightarrow\infty}\frac{1}{p}\left\langle\hat{\bbeta}^{(p)}_1,\hat{\bbeta}^{(p)}_2\right\rangle,
		\end{eqnarray}
		where $\langle\cdot,\cdot\rangle$ is the standard scalar product and 
		\begin{eqnarray}\nn
		\hat{\bbeta}^{(p)}_1&=&\text{argmin}_{\bbeta\in\bR^p}\left\{\frac{1}{2}\|\bbeta-\bbeta_0(p)-\sqrt{a_1}\bSigma(p)^{-1/2}\bz_1\|^2_{\bSigma(p)}+b_1\|\bbeta\|_1\right\},\\\nn
		\hat{\bbeta}^{(p)}_2&=&\text{argmin}_{\bbeta\in\bR^p}\left\{\frac{1}{2}\|\bbeta-\bbeta_0(p)-\sqrt{a_2}\bSigma(p)^{-1/2}\bz_2\|^2_{\bSigma(p)}+b_2\|\bbeta\|_1\right\},
		\end{eqnarray} 
		where $(\bz_1,\bz_2)\sim N(0,\bS\otimes\bI_{p\times p})$ and is independent of $\bbeta_0(p)$.		
	\end{enumerate}
Conditions 1 and 2 have appeared in \cite{BayatiM12} which indicate that the entries of $\bbeta_0$ and $\bw$ are drawn i.i.d. from certain distributions with bounded second order moment. Note that the entries of $\bw$ are not necessarily normal. Denote $\lambda_{min}(\bSigma(p))$ and $\lambda_{max}(\bSigma(p))$ the smallest and largest eigenvalues of $\bSigma(p)$ respectively, then Condition 3 is equivalent to that $1/\lambda_{min}(\bSigma(p))=O(1)$ and $\lambda_{max}(\bSigma(p))=O(1)$. Condition 5 indicates that the covariance matrix should satisfy such conditions that the $l_1$ penalized quadratic loss function specified in (\ref{c5}) has a differentiable limit, i.e. the derivative over $a,b$ and the limit of $p$ are exchangeable. It is worth stressing that Conditions 5 and 6 are satisfied by a larger family of covariance matrices. For instance, based on law of large number, it can be proved that it holds for block-diagonal matrices $\bSigma$ as long as the blocks have bounded length and the block's empirical distribution converges. This condition has also appeared in \cite{Montanari} and it ensures the existence of large dimensional limits of some functions such as (\ref{psi}), (\ref{falpha}), and (\ref{lambdaalpha}) that will be used in describing the main results of Theorems \ref{thm1} and \ref{thm2}. It also allows us to exchange the order of operations such as taking limit and derivative over these functions. In Section \ref{sec43}, we will discuss the specific choice of covariance structure such that this condition can be satisfied. We insist on the fact that $\bbeta_0(p)$, $\bw(p)$, $\bSigma(p)$, $\bX(p)$ depend on $p$. However, we will drop this dependence most of the time to ease the reading. 
	
	In order to present our main result, for any $\theta\textgreater 0$ and $\bSigma\succ 0$, we need to introduce the soft-thresholding operation $\veta_\theta:\bR^p\rightarrow \bR^p$ which is defined as
	\begin{eqnarray}\label{veta}
	\veta_\theta(\bv)=\text{argmin}_{\bbeta\in\bR^p}\left\{\frac{1}{2}\|\bbeta-\bv\|^2_{\bSigma}+\theta\|\bbeta\|_1\right\}.
	\end{eqnarray} 
	Then for a converging sequence of instances, we can define the function
	\begin{eqnarray}\label{psi}
	\psi(\tau^2,\theta)&=&\sigma^2_w+\lim_{p\rightarrow\infty}\frac{1}{p\delta}E\left(\|\veta_\theta(\bbeta_0+\tau\bSigma^{-1/2}\bz)-\bbeta_0\|_{\bSigma}^2\right),
	\end{eqnarray}
	where $\bz\sim N(0,\bI_{p\times p})$ is independent of $\bbeta_0$. Notice that the function $\psi$ depends implicitly on the law $p_{\beta_0}$. 
	
	Condition 5 allows us to verify the existence of the limit in (\ref{psi}). Toward this end, we start from (\ref{c5}) and have 
		\begin{eqnarray}\label{efun0}
		{\cal E}^{(p)}(\tau^2,\theta)&=&\frac{1}{p}E\left\{\frac{1}{2}\|\hat{\bbeta}-\bbeta_0-\tau\bSigma^{-1/2}\bz\|^2_{\bSigma}+\theta\|\hat{\bbeta}\|_1\right\},
\end{eqnarray}
where $\hat{\bbeta}=\veta_\theta(\bbeta_0+\tau\bSigma^{-1/2}\bz)$. In order to take derivative over $\tau^2$ and $\theta$, we need to conduct integrals over $\bz\in\bR^p$. We first divide the $p$-dimensional space into regions such that $\hat{\bbeta}$ is differentiable in each region and continuous across the entire space (see Figure \ref{figure2d} for a simple 2-dimensional illustration). Then the derivative of ${\cal E}^{(p)}(\tau^2,\theta)$ involves the explicit derivative inside each region and integrals over the boundaries among different regions over $p-1$-dimensional measure. According to Stokes's theorem, as in Theorem 1 of \cite{baddeley_1977}, we conclude that the boundary effects are canceled and have no contribution due to the continuity of $\hat{\bbeta}$ (see detailed discussion in \ref{a13}). Further note that, according to the definition of $\hat{\bbeta}$, the derivative of the integrand in (\ref{efun0}) over $\hat{\bbeta}$ is 0, therefore we only need to consider the explicit dependence of the integrand on $\tau^2$ and $\theta$ in deriving the corresponding derivatives. We obtain
		\begin{eqnarray}\label{efun1}
		\frac{\partial{\cal E}^{(p)}(\tau^2,\theta)}{\partial\tau^2}&=&-\frac{1}{2p\tau}E\left\langle\hat{\bbeta}-\bbeta_0,\bSigma^{1/2}\bz\right\rangle+\frac{1}{2},\\\label{efun2}
		\frac{\partial{\cal E}^{(p)}(\tau^2,\theta)}{\partial\theta}&=&\frac{1}{p}E\|\hat{\bbeta}\|_1.
\end{eqnarray}
From Condition 5, all the limits of ${\cal E}^{(p)}(\tau^2,\theta)$, $\frac{\partial{\cal E}^{(p)}(\tau^2,\theta)}{\partial\tau^2}$, and $\frac{\partial{\cal E}^{(p)}(\tau^2,\theta)}{\partial\tau^2}$ exist, therefore, $\lim_{p\rightarrow\infty}\frac{1}{p\delta}E\left(\|\hat{\bbeta}-\bbeta_0\|_{\bSigma}^2\right)$ also exists, which is just the right hand side of (\ref{psi}). Taking $\bbeta_0=0$, we immediately obtain that the limit of the following equation (\ref{falpha}) also exists.
 
	We choose $\theta=\alpha\tau$, then we have the following result in order to establish a calibration mapping between $\alpha$ and $\lambda$. 	
	
	\begin{proposition}\label{prop1}
	Define function
\begin{eqnarray}\label{falpha}
f(\alpha)\equiv\lim_{p\rightarrow\infty}\frac{1}{p\delta}E\left(\|\veta_\alpha(\bSigma^{-1/2}\bz)\|_{\bSigma}^2\right).
\end{eqnarray} 
Then the equation $f(\alpha)=1$ has a unique solution denoted by $\alpha_{min}(\delta)$ when $\delta\textless 1$.  Then for any $\delta\ge 1$ or $\delta\textless 1$ and $\alpha\textgreater\alpha_{min}(\delta)$, the fixed point equation 
		\begin{eqnarray}\label{fixedpoint}
\tau^2=\psi(\tau^2,\alpha\tau)
		\end{eqnarray}
admits a unique solution. 
%Denoting by $\tau_\star=\tau_\star(\alpha)$ this solution, we have  $|\frac{d\psi(\tau^2,\alpha\tau)}{d\tau^2}|\textless 1$ for $\tau=\tau_\star$.
	\end{proposition}
We then define a function $\alpha\rightarrow\lambda(\alpha)$ on $(\alpha_{min}(\delta),\infty)$ by
	\begin{eqnarray}\nn
	&&\lambda(\alpha)\\\label{lambdaalpha}
	&=&\alpha\tau_\star(\alpha)\left\{1-\lim_{p\rightarrow\infty}\frac{1}{p\delta}E\left[\text{div}\veta_{\alpha\tau_\star(\alpha)}(\bbeta_0+\tau_\star(\alpha)\bSigma^{-1/2}\bz)\right]\right\},
	\end{eqnarray}
where the divergence of the vector field is defined as $\text{div}\veta_\theta(\bv)=\sum_{j=1}^p\frac{\partial\eta_{\theta,j}(\bv)}{\partial v_j}$. This function defines a correspondence between $\alpha$ and $\lambda$. The existence of the limit of (\ref{lambdaalpha}) can be obtained from the existence of the limit of $\frac{\partial{\cal E}^{(p)}(\tau^2,\theta)}{\partial\tau^2}$ in (\ref{efun1}) following by integration by parts. In the following we will need to invert this function and define $\lambda\rightarrow\alpha(\lambda)$ on $(0,\infty)$ in such a way that
%\begin{eqnarray}\label{lalpha}
%	l(\alpha)\equiv\lim_{p\rightarrow\infty}\frac{1}{p\delta}E\left(\|\veta_\alpha(\bSigma^{-1/2}\bz)\|_0\right),
%	\end{eqnarray} 
%	Then in the following we invert this function and define $\lambda\rightarrow\alpha(\lambda)$ on $(\lambda_{min},\infty)$ in such a way that
	\begin{eqnarray}\label{alphalambda}
	\alpha(\lambda)\in\{a\in(\alpha_{min},\infty):\lambda(a)=\lambda\}.
	\end{eqnarray}
	The next result implies that the function $\lambda\rightarrow\alpha(\lambda)$ is well defined.
	\begin{proposition}\label{prop3}
		The function $\alpha\rightarrow\lambda(\alpha)$ is continuous on the interval $(\alpha_{min},\infty)$ and for any given $\lambda$ there exist a unique $\alpha$ such that $\lambda(\alpha)=\lambda$.
	\end{proposition}
	
	For two sequences (in $n$) of random variables $\bx_n$ and $\by_n$, write $\bx_n\overset{P}{\approx}\by_n$ when their difference convergences in probability to 0, i.e. $\bx_n-\by_n\xrightarrow{P}0$. For any $m\in\mathbb{N}_{\textgreater 0}$, we say a function $\varphi:\bR^m\times\bR^m\rightarrow\bR$ is pseudo-Lipschitz if there exist a constant $L\textgreater 0$ such that for all $\bx,\by\in\bR^m: |\varphi(\bx,\by)|\le L(1+\|\bx\|+\|\by\|)\|\bx-\by\|$. A sequence (in $m$) of pseudo-Lipschitz functions $\{\varphi_m\}_{m\in\mathbb{N}_{\textgreater 0}}$ is called uniformly pseudo-Lipschitz if, denoting by $L_m$ is the pseudo-Lipschitz constant, we have $L_m\textless\infty$ for each $m$ and $\sup_{m\rightarrow\infty}L_m\textless\infty$. Note that the input and output dimensions of each $\varphi_m$ can depend on $m$. We call any $L\textgreater\sup_{m\rightarrow\infty}L_m$ a pseudo-Lipschitz constant of the sequence. We can now state our main result. 
	\begin{theorem}\label{thm1}
		Let $\{\bbeta_0(p),\bw(p),\bSigma(p),\bX(p)\}_{p\in\bN}$ be a converging sequence of instances. Denote $\hat{\bbeta}(\lambda)$ the LASSO estimator for instance $\{\bbeta_0(p),\bw(p)$, $\bSigma(p),\bX(p)\}$ with $\lambda\textgreater 0$ and $P\{\bbeta_0(p)\ne 0\}\textgreater 0$. For any sequence $\varphi_p:\bR^p\times\bR^p\rightarrow\bR,~p\ge 1$, of uniformly pseudo-Lipschitz functions, we have
		\begin{eqnarray}\nn
		\varphi_p(\hat{\bbeta}(\lambda),\bbeta_0)\overset{P}{\approx}E\varphi_p(\veta_{\theta_\star}(\bbeta_0+\tau_\star\bSigma^{-1/2}\bz),\bbeta_0)
		\end{eqnarray}
		where $\bz\sim N(0,\bI_{p\times p})$ is independent of $\bbeta_0\sim p_{\beta_0}$, $\tau_\star=\tau_\star(\alpha(\lambda))$, and $\theta_\star=\alpha(\lambda)\tau_\star(\alpha(\lambda))$.
	\end{theorem}
	Using function $\varphi_p(\ba,\bb)=\frac{1}{p}\|\ba-\bb\|^2$, we obtain LASSO MSE $\frac{1}{p}\|\hat{\bbeta}(\lambda)-\bbeta_0\|^2$ which can be used to evaluate competing optimization methods on large scale applications. Using Theorem \ref{thm1}, we get
	\begin{eqnarray}\label{lrisk}
	\frac{1}{p}\|\hat{\bbeta}(\lambda)-\bbeta_0\|^2&\overset{P}{\approx}&\frac{1}{p}E\|\veta_{\theta_\star}(\bbeta_0+\tau_\star\bSigma^{-1/2}\bz)-\bbeta_0\|^2,
	\end{eqnarray}
	where $\bz\sim N(0,\bI_{p\times p})$ is independent of $\bbeta_0\sim p_{\beta_0}$, $\tau_\star=\tau_\star(\alpha(\lambda))$, and $\theta_\star=\alpha(\lambda)\tau_\star(\alpha(\lambda))$.
	
	Therefore, for fixed $\lambda$, LASSO MSE explicitly depends on $\tau^2_\star$ which can be obtained by solving the fixed point equation $\tau^2_\star=\psi(\tau^2_\star,\alpha\tau_\star)$ together with (\ref{lambdaalpha}). Closer to the spirit of this paper, \cite{Montanari} non-rigorously derived the LASSO MSE under the same setting considered here using the replica method from statistical physics. The present paper is rigorous and putting on a firmer basis this line of research.
	
	\section{Phase transition of LASSO under dependence}\label{sec3}
	
	Note that the LASSO risk results based on Theorem \ref{thm1} work fine for entire $\sigma_w^2,\delta\in[0,\infty)$. To study phase transition, we only need to consider $\delta\in[0,1]$ and evaluate the results in the noiseless setting $\sigma_w^2=0$ and understand the extend to which (\ref{lasso}) accurately recovers $\bbeta_0$ under this setting. Consider a class of distributions ${\cal F}_\epsilon$ whose mass at zero is equal to $1-\epsilon$, i.e.
	\begin{eqnarray}\nn
	{\cal F}_\epsilon\equiv\{p_{\beta_0}:p_{\beta_0}(\{0\})=1-\epsilon\}.
	\end{eqnarray}	 
	When the matrix $\bX$ has i.i.d. Gaussian elements, i.e. $\bSigma=\bI_{p\times p}$, phase space $0\le\delta,\epsilon\le 1$ can be divided into two components, or phases, separated by a curve $\delta_c=\delta(\epsilon)$, which does not depend on the actual distribution of $p_{\beta_0}$ and can be explicitly computed. Above this curve, LASSO perfectly recovers the sparse signal $\bbeta_0$ with high probability. Below this curve, we have $\hat{\bbeta}\ne\bbeta_0$ with high probability. 
	
	For non-standard Gaussian design, i.e. $\bSigma\ne\bI_{p\times p}$, we need to consider a more general class of distributions ${\cal F}_{\epsilon,\Delta}$ defined as
	\begin{eqnarray}\nn
	{\cal F}_{\epsilon,\Delta}\equiv\left\{p_{\beta_0}:p_{\beta_0}(\{0\})=1-\epsilon\text{ and }\frac{|p_{\beta_0}(\{\textgreater 0\})-p_{\beta_0}(\{\textless 0\})|}{|p_{\beta_0}(\{\textgreater 0\})+p_{\beta_0}(\{\textless 0\})|}=\Delta\right\}.
	\end{eqnarray}	 
	Here we introduce an extra parameter $\Delta=\frac{|P(\beta_0\textgreater 0)-P(\beta_0\textless 0)|}{|P(\beta_0\textgreater 0)+P(\beta_0\textless 0)|}$ which represents the positive-negative asymmetry for the nonzero components of $\bbeta_0$. Clearly, $0\le \Delta\le 1$, and if $\Delta=0$, we have $P(\beta_0\textgreater 0)=P(\beta_0\textless 0)$, i.e. $\bbeta_0$ has positive and negative nonzero components with equal probability. 
	
	We denote by $[p]=\{1,\cdots,p\}$ the set of first $p$ integers. For a subset $\bI\subseteq[p]$, we let $|\bI|$ denote its cardinality. For an $p\times p$ matrix $\bSigma$ and set of indices $\bI\subseteq[p]$, $\bJ\subseteq[p]$, we use $\bSigma_{\bI\bJ}$ to denote the $|\bI|\times|\bJ|$ sub-matrix formed by rows in $\bI$ and columns in $\bJ$. Likewise, for a vector $\bbeta\in\bR^p$, $\bbeta_{\bI}$ is the restriction of $\bbeta$ to indices in $\bI$. The following Theorem shows that, under general covariance $\bSigma$, the phase transition curve exists and depends on the asymmetry parameter $\Delta$.
	\begin{theorem}\label{thm2}
		Let $\{\bbeta_0(p),\bw(p),\bSigma(p),\bX(p)\}_{p\in\bN}$ be a converging sequence of instances and $\bw(p)=0$. Assume $p_{\bbeta_0}\in{\cal F}_{\epsilon,\Delta}$. Then the phase space $0\le\delta,\epsilon\le 1$ can be divided into two components separated by a curve $\delta_c=\delta(\epsilon)$. Above this curve, LASSO algorithm (\ref{lasso}) perfectly recovers the sparse signal $\bbeta_0$ with high probability, i.e. $\frac{1}{p}\|\hat{\bbeta}(\lambda)-\bbeta_0\|\rightarrow 0$ after appropriately choosing the tuning parameter $\lambda$. Below this curve, we have $\hat{\bbeta}\ne\bbeta_0$ with high probability. For fixed $\epsilon$, the $\delta_c$ is determined by
		\begin{eqnarray}\label{phasecurve}
		\delta_c=\inf_{\alpha}M(\epsilon,\Delta,\alpha),
		\end{eqnarray}	 
		where 
		\begin{eqnarray}\nn
		&&M(\epsilon,\Delta,\alpha)\\\label{mfun}
		&=&\lim_{p\rightarrow\infty}\frac{1}{p}E\{((\bSigma^{1/2}\bz)_{\cal A}-\alpha\text{sign}(\hat{\bbeta}_{\cal A}))^T\bSigma_{{\cal A}{\cal A}}^{-1}((\bSigma^{1/2}\bz)_{\cal A}-\alpha\text{sign}(\hat{\bbeta}_{\cal A}))\},
		\end{eqnarray}	 
		where the active set ${\cal A}={\cal B}\cup{\bar{\cal B}}$ with ${\cal B}=\{j:\beta_{0,j}\ne 0\}$ and ${\bar{\cal B}}$ the active set of LASSO problem 
		\begin{eqnarray}\label{blasso}
		\bar{\bbeta}=\text{argmin}_{\bbeta\in\bR^{\bar{p}}}\left\{\frac{1}{2}\|\bar{\by}-\bar{\bX}\bbeta\|_2^2+\alpha\|\bbeta\|_1\right\}
		\end{eqnarray} 
		with 
		\begin{eqnarray}\nn
		\bar{\bX}&=&(\bSigma_{{\cal B}^c{\cal B}^c}-\bSigma_{{\cal B}^c{\cal B}}\bSigma_{{\cal B}{\cal B}}^{-1}\bSigma_{{\cal B}{\cal B}^c})^{1/2},\\\nn
		\bar{\by}&=&\bar{\bX}^{-1}\left[(\bSigma^{1/2}\bz)_{{\cal B}^c}-\bSigma_{{\cal B}^c{\cal B}}\bSigma_{{\cal B}{\cal B}}^{-1}\{(\bSigma^{1/2}\bz)_{\cal B}-\alpha\text{sign}(\bbeta_{0,{\cal B}})\}\right],
		\end{eqnarray} 
		where $\bar{p}=|{\cal B}^c|$ and $\bz\sim N(0,\bI_{p\times p})$ is independent of $p_{\beta_0}$.
	\end{theorem}
	Note that all the nonzero components of $\bbeta_0$ contribute to function (\ref{mfun}). Some zero components also have contribution if they are selected by the LASSO problem (\ref{blasso}).
	Theorem \ref{thm2} shows that the LASSO phase transition is independent of the actual distribution of $p_{\beta_0}$ but depends on the positive-negative asymmetry of the nonzero components of $\bbeta_0$, i.e. depends on $\epsilon_+=\#\{\bbeta_0\textgreater 0\}/p$ and $\epsilon_-=\#\{\bbeta_0\textless 0\}/p$ with $\epsilon=\epsilon_++\epsilon_-$ and $\Delta=(\epsilon_+-\epsilon_-)/(\epsilon_++\epsilon_-)$. 
	
	The following two Corollaries provide the explicit phase transition curves for two special covariance matrices.
	\begin{corollary}\label{corollary1}
	For $\bSigma=\bI_{p\times p}$, the LASSO phase transition curve is determined by
	\begin{eqnarray}\nn
	\delta&=&\frac{2\phi(\alpha)}{\alpha+2\phi(\alpha)-2\alpha\Phi(-\alpha)},\\\label{phaseiid}
	\epsilon&=&\frac{2\phi(\alpha)-2\alpha\Phi(-\alpha)}{\alpha+2\phi(\alpha)-2\alpha\Phi(-\alpha)},
	\end{eqnarray}	 
	\end{corollary}
	This is equivalent to the result provided in \cite{dmm2009} based on techniques of combinatorial geometry. 
	
	\begin{corollary}\label{corollary2}
	For block-diagonal matrix $\bSigma$ with block  $\left(\begin{array}{cc}1&\rho\\\rho&1\end{array}\right)$, the LASSO phase transition curve is determined by
	\begin{eqnarray}\nn
	\delta&=&\epsilon^2A(\alpha,\Delta)+\epsilon(1-\epsilon)B(\alpha)+(1-\epsilon)^2C(\alpha),\\\label{phasedep}
	\epsilon&=&\frac{2C^\prime(\alpha)-B^\prime(\alpha)+\sqrt{B^\prime(\alpha)^2-4\frac{\partial A(\alpha,\Delta)}{\partial\alpha}C^\prime(\alpha)}}{2\{\frac{\partial A(\alpha,\Delta)}{\partial\alpha}-B^\prime(\alpha)+C^\prime(\alpha)\}},
	\end{eqnarray}	 
	where 
	\begin{eqnarray}\label{afun}
	A(\alpha,\Delta)&=&1+\frac{\alpha^2}{2}\left(\frac{1+\Delta^2}{1-\rho}+\frac{1-\Delta^2}{1+\rho}\right),\\\label{bfun}
	B(\alpha)&=&B_1(\alpha)+B_2(\alpha)+B_3(\alpha),\\\label{cfun}
	C(\alpha)&=&C_1(\alpha)+C_2(\alpha)+C_3(\alpha)+C_4(\alpha),
	\end{eqnarray}	 
	where 
	\begin{eqnarray}\nn
	B_1(\alpha)&=&E(\xi_1-\alpha)^2I(|\xi_2-\rho\xi_1+\rho\alpha|\le\alpha)+E(\xi_2-\alpha)^2I(|\xi_1-\rho\xi_2+\rho\alpha|\le\alpha),\\\nn
	B_2(\alpha)&=&E\frac{(\xi_1-\alpha)^2+(\xi_2-\alpha)^2-2\rho(\xi_1-\alpha)(\xi_2-\alpha)}{1-\rho^2}\{I(\xi_2-\rho\xi_1+\rho\alpha\ge\alpha)\\\nn
	&&+I(\xi_1-\rho\xi_2+\rho\alpha\ge\alpha)\},\\\nn
	B_3(\alpha)&=&E\frac{(\xi_1-\alpha)^2+(\xi_2+\alpha)^2-2\rho(\xi_1-\alpha)(\xi_2+\alpha)}{1-\rho^2}I(\xi_2-\rho\xi_1+\rho\alpha\le -\alpha)\\\nn
	&&+E\frac{(\xi_1+\alpha)^2+(\xi_2-\alpha)^2-2\rho(\xi_1+\alpha)(\xi_2-\alpha)}{1-\rho^2}I(\xi_1-\rho\xi_2+\rho\alpha\le -\alpha),\\\nn
	C_1(\alpha)&=&E(\xi_1-\alpha)^2I(|\xi_2-\rho\xi_1+\rho\alpha|\le\alpha)I(\xi_1\ge\alpha)\\\nn
	&&+E(\xi_1+\alpha)^2I(|\xi_2-\rho\xi_1+\rho\alpha|\le\alpha)I(\xi_1\le -\alpha),\\\nn
	C_2(\alpha)&=&E(\xi_2-\alpha)^2I(|\xi_1-\rho\xi_2+\rho\alpha|\le\alpha)I(\xi_2\ge\alpha)\\\nn
	&&+E(\xi_2+\alpha)^2I(|\xi_1-\rho\xi_2+\rho\alpha|\le\alpha)I(\xi_2\le -\alpha),\\\nn
	C_3(\alpha)&=&E\frac{(\xi_1-\alpha)^2+(\xi_2-\alpha)^2-2\rho(\xi_1-\alpha)(\xi_2-\alpha)}{1-\rho^2}\\\nn
	&&I(\xi_1-\rho\xi_2+\rho\alpha\ge\alpha)I(\xi_2-\rho\xi_1+\rho\alpha\ge\alpha)\\\nn
	&&+E\frac{(\xi_1-\alpha)^2+(\xi_2+\alpha)^2-2\rho(\xi_1-\alpha)(\xi_2+\alpha)}{1-\rho^2}\\\nn
	&&I(\xi_1-\rho\xi_2+\rho\alpha\ge\alpha)I(\xi_2-\rho\xi_1+\rho\alpha\le -\alpha),\\\nn
	C_4(\alpha)&=&E\frac{(\xi_1+\alpha)^2+(\xi_2-\alpha)^2-2\rho(\xi_1+\alpha)(\xi_2-\alpha)}{1-\rho^2}\\\nn
	&&I(\xi_1-\rho\xi_2+\rho\alpha\le -\alpha)I(\xi_2-\rho\xi_1+\rho\alpha\ge\alpha)\\\nn
	&&+E\frac{(\xi_1+\alpha)^2+(\xi_2+\alpha)^2-2\rho(\xi_1+\alpha)(\xi_2+\alpha)}{1-\rho^2}\\\nn
	&&I(\xi_1-\rho\xi_2+\rho\alpha\le -\alpha)I(\xi_2-\rho\xi_1+\rho\alpha\le -\alpha),
	\end{eqnarray}	 
	where 
	\begin{eqnarray}\nn
	\xi_1&=&\frac{\sqrt{1+\rho}+\sqrt{1-\rho}}{2}z_1+\frac{\sqrt{1+\rho}-\sqrt{1-\rho}}{2}z_2,\\\nn
	\xi_2&=&\frac{\sqrt{1+\rho}-\sqrt{1-\rho}}{2}z_1+\frac{\sqrt{1+\rho}+\sqrt{1-\rho}}{2}z_2,
	\end{eqnarray}	 
	and $(z_1,z_2)\sim N(0,\bI_{2\times 2})$.
		\end{corollary}

	For general $\bSigma$, it is difficult to derive closed form analytic result for $\delta_c$ due to the complicated expression of (\ref{mfun}) and LASSO problem (\ref{blasso}). We provide Monte Carlo based numerical solutions in following section.
	
	\section{Numerical illustration}\label{preliminary} 
	
	In this section, we present some numerical studies to support our theoretical results in Section \ref{method} and Section \ref{sec3}. Our studies are based on simulations on finite size systems of moderate dimensions. We compute asymptotic LASSO risks and compare them with simulation results in Section \ref{sec41}. In Section \ref{sec42}, we verify our theoretical prediction on LASSO phase transition through Monte Carlo simulations. In Section \ref{sec43}, we study the dependence of LASSO phase transition on the covariance structure $\bSigma$ and positive negative asymmetrical parameter $\Delta$ under various settings.
	
We consider block-diagonal covariance matrix with AR(1) block structure. For this choice, we can easily verify that Condition 5 is satisfied with limit
\begin{eqnarray}\label{finited}
\lim_{p\rightarrow\infty}{\cal E}^{(p)}(a,b)=\frac{1}{s}E\left\{\frac{1}{2}\|\hat{\bbeta}-\bbeta_0-\sqrt{a}\bSigma_s^{-\frac{1}{2}}\bz\|^2_{\bSigma_s}+b\|\hat{\bbeta}\|_1\right\},
\end{eqnarray}
where $\bSigma_s$ is the block matrix with length $s$ and $\hat{\bbeta}=\veta_b(\bbeta_0+\sqrt{a}\bSigma_s^{-\frac{1}{2}}\bz)$ with covariance matrix $\bSigma_s$ and $\bz\sim N(0,\bI_{s\times s})$. Similarly, we can verify that Condition 6 is also satisfied. We use $s=2, 10, 20, 50$ in our numeric studies.

%	Our theoretical results for asymptotic risk and phase transition are characterized by limiting formulas (\ref{lrisk}) and (\ref{phasecurve}) with $p\rightarrow\infty$. For general $\bSigma$, their closed form expressions cannot be obtained and we have to use numerical methods. To compute a quantity $v$ in the limit of $p\rightarrow\infty$, we first choose several values of $p$ denoted as $p_i$, $i=1,\cdots,m$. Then, for each finite dimension $p_i$, we calculate the corresponding estimation $v_i$ numerically. According to central limit theorem, we have $v_i\sim N(v,\sigma^2/p_i)$. Then we estimate $v$ by minimizing the log-likelihood
%	$\sum_{i=1}^mp_i(v_i-v)^2$ which ends up with the weighted least square estimator $\hat{v}=\frac{\sum_{i=1}^mw_iv_i}{\sum_{i=1}^mp_i}$ as our theoretical prediction for the limiting quantity of $v$. In our numerical experiments, we choose $m=5$ and the 5 dimensions are 300, 600, 900, 1200, and 1500.   
	
		\subsection{LASSO risk}\label{sec41}
		We compute LASSO risk using (\ref{lrisk}) with $\tau_\star^2$ determined by solving the fixed point equation of $\tau^2=\psi(\tau^2,\alpha(\lambda)\tau)$, where $\alpha(\lambda)$ is defined in (\ref{alphalambda}). We use the bisection method to numerically solve the non-linear equation $f(\tau^2)=\tau^2-\psi(\tau^2,\alpha\tau)=0$.
%		since $f(\tau^2)$ is monotone increasing and moreover for $\alpha\textgreater\alpha_{min}$, we have $f(0)\textless 0$ and $f(\tau^2)\textgreater 0$ as $\tau^2$ large enough. 
		
		For each setting, we generate 100 data sets with $p=400$ consisting of design matrix $\bX\sim N(0,\bSigma)$ and measurement vector $\by=\bX\bbeta_0+\bw$ obtained from independent signal vector $\bbeta_0$ and independent noise vector $\bw$. For each data set, we obtain the LASSO optimum estimator $\hat{\bbeta}(\lambda)$ using $glmnet$, an efficient package for fitting lasso or elastic-net regularization path for linear and generalized linear regression models. For each case, the dependence of MSE as a function of tuning parameter $\lambda$ is plotted as shown in Figure \ref{figure3}. Here the random error $w_i\overset{i.i.d.}{\sim}N(0,1)$ and the magnitude for nonzero components of $\bbeta_0$ are sampled from uniform $[1,2]$. The agreement is remarkably good already for $p,n$ of a few hundred.
%		, and the deviations are consistent with statistical fluctuations.  
		
		\begin{figure}[hbtp] \hspace{-1cm}
			\begin{minipage}[t]{0.49\linewidth}
				\begin{center}
		\includegraphics[angle=-90,width=0.95\textwidth]{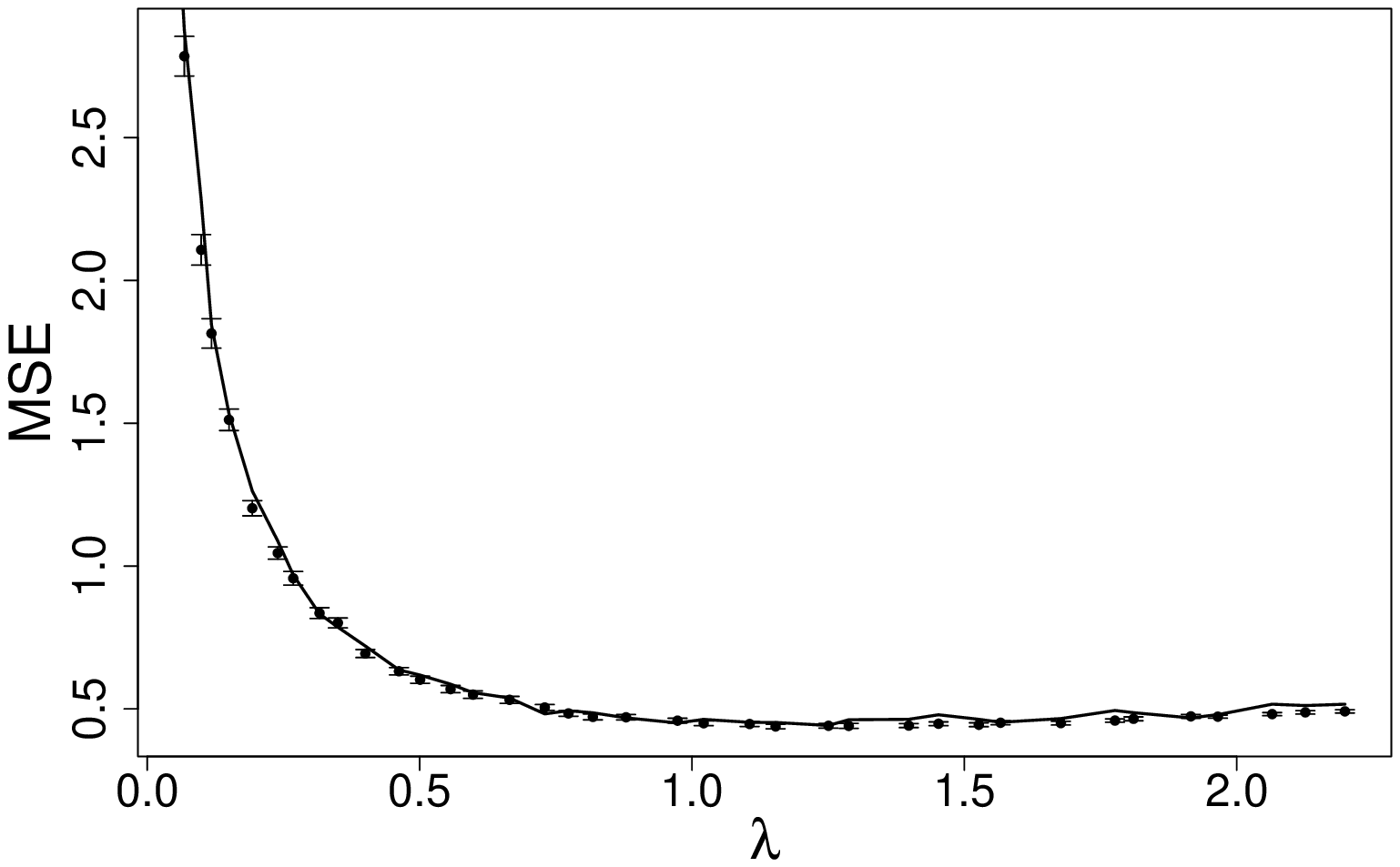}
		
				\end{center}
			\end{minipage}
			\hspace{0.3cm}
			\begin{minipage}[t]{0.49\linewidth}
				\begin{center}
		\includegraphics[angle=-90,width=0.95\textwidth]{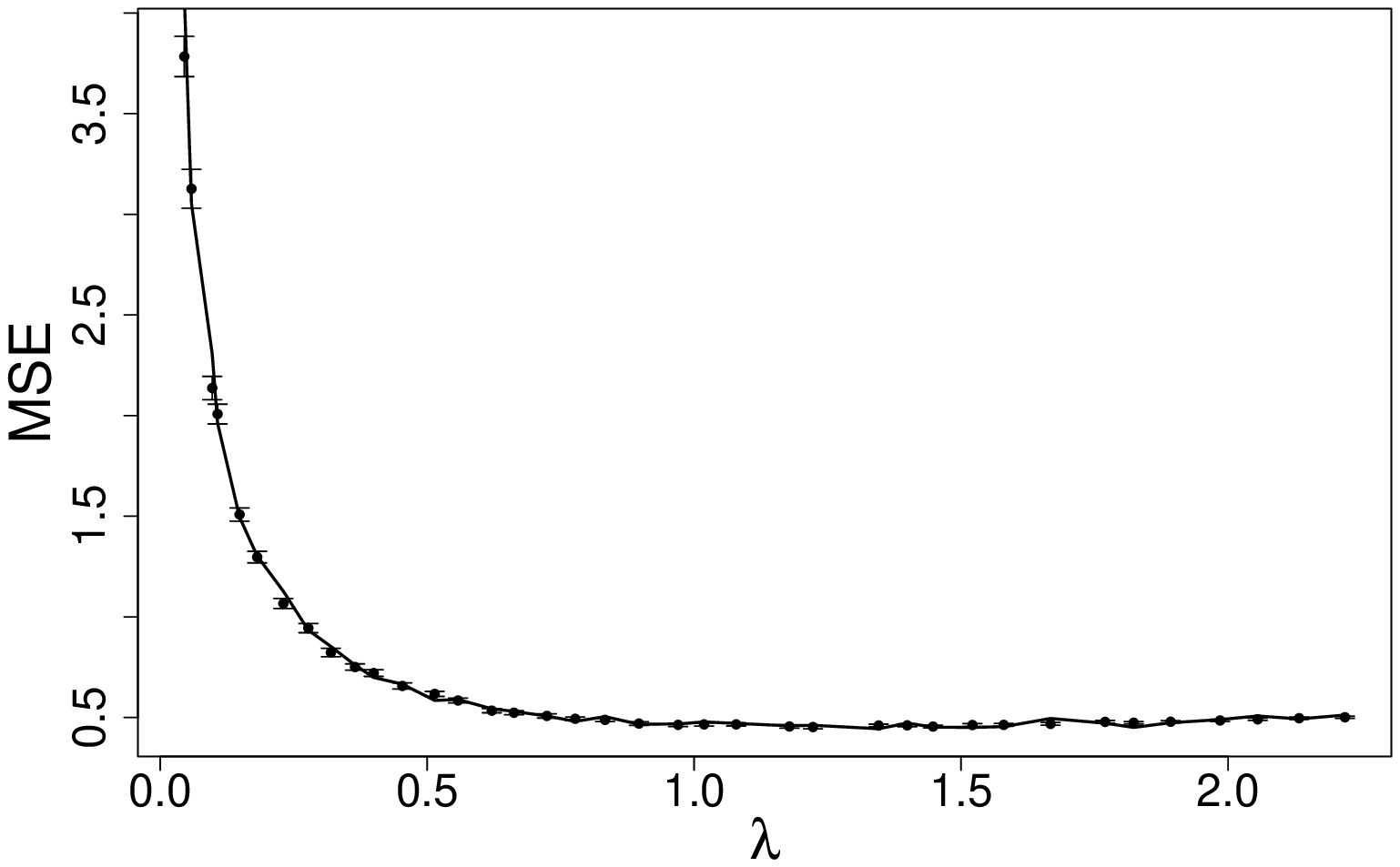}

				\end{center}
			\end{minipage}
			\vspace{-0cm}
			\caption{LASSO MSE as a function of the regularization parameter $\lambda$ compared to the asymptotic prediction. The solid curves represent theoretical prediction using (\ref{lrisk}) and the error bars are summaries over 100 simulated data with $p=400$. Here the covariance matrix is block-diagonal with AR(1) block structure $\Sigma_{s,ij}=\rho^{|i-j|}$ with $s=2$ and $\rho=0.5$. The under-sampling $\delta=1$ and the sparsity $\epsilon=0.15$. Left panel is for $\Delta=0$ and right panel is for $\Delta=1$.}
			\label{figure2}
		\end{figure}
		
		\subsection{Phase transition verification}\label{sec42}
		
		For noiseless case, we compare the theoretical phase transition with the empirical one estimated by applying the following optimization algorithm to simulated data. 
		\begin{eqnarray}\nn
		&&\text{minimize}\|\bbeta\|_1,\\\nn
		&&\text{subject to }\by=\bX\bbeta.
		\end{eqnarray}	 
		Using the similar procedure as in \cite{dmm2009}, we first fix a grid of 31 $\epsilon$ values between 0.05 and 0.95. For each $\epsilon$, we consider a series of $\delta$ values between $\max(0,\delta_c(\epsilon)-0.2)$ and $\min(1,\delta_c(\epsilon)+0.2)$, where $\delta_c(\epsilon)$ is the theoretically expected phase transition based on Theorem \ref{thm2}. We then have a grid of $\delta,\epsilon$ values in parameter space $[0,1]^2$. At each $\delta,\epsilon$, we generate $m=100$ problem instances $(\bX,\bbeta_0)$ with size $p=500$. Then $\by=\bX\bbeta_0$. For the $i$th problem instance, we obtain an output $\hat{\bbeta}_i$ by using the rq.lasso.fit function in package rqPen to the $i$th simulated data. We set the success indicator variable $S_i=1$ if $\frac{\|\hat{\bbeta}_i-\bbeta_0\|}{\|\bbeta_0\|}\le 10^{-4}$ and $S_i=0$ otherwise. Then at each $(\delta,\epsilon)$ combination, we have $S=\sum_{i=1}^{m}S_i$.   
		
\begin{figure}[hbtp] \hspace{-1cm}
		\begin{minipage}[t]{0.49\linewidth}
			\begin{center}
	\includegraphics[angle=-90,width=0.95\textwidth]{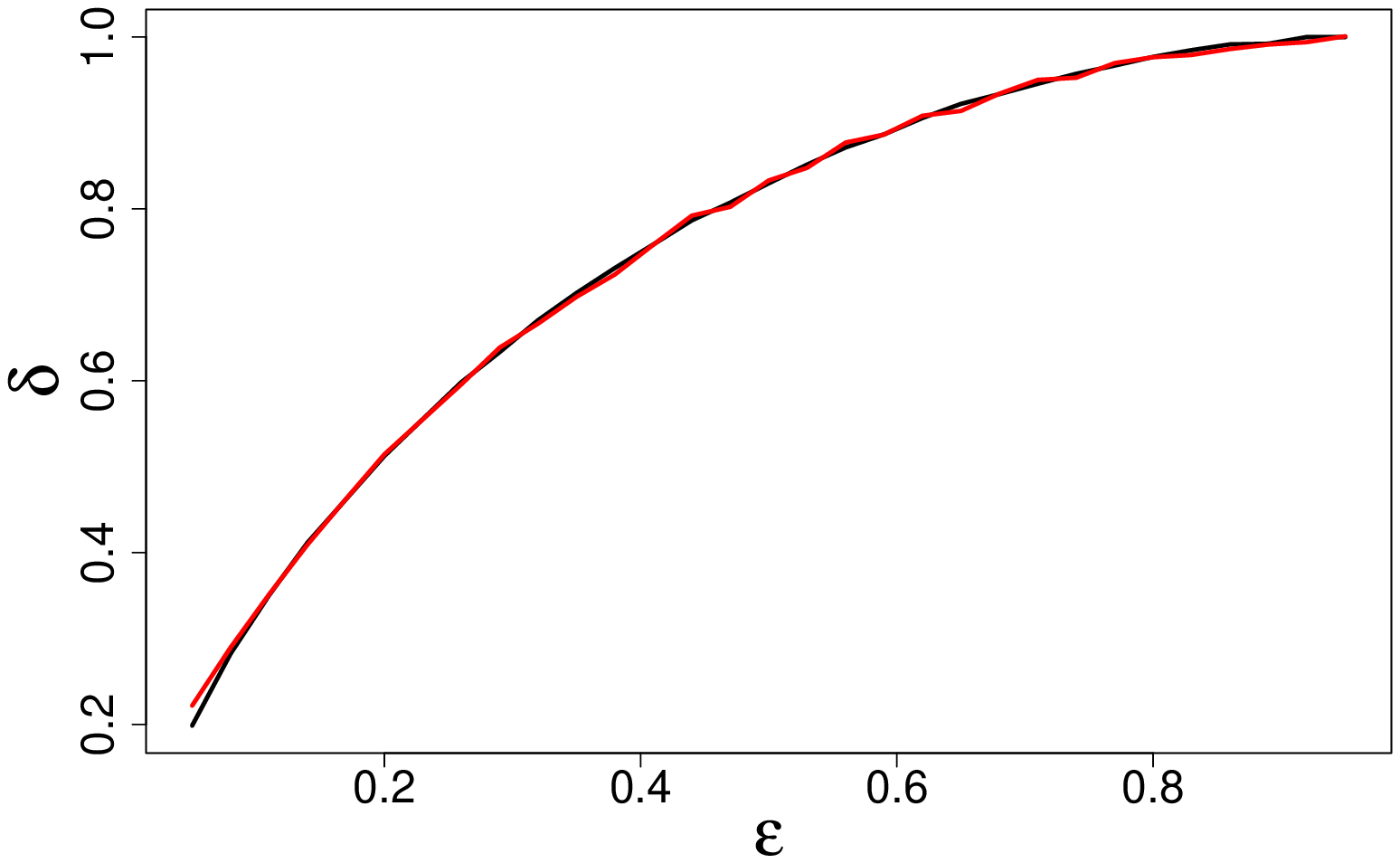}
			\end{center}
		\end{minipage}
		\hspace{0.3cm}
		\begin{minipage}[t]{0.49\linewidth}
			\begin{center}
	\includegraphics[angle=-90,width=0.95\textwidth]{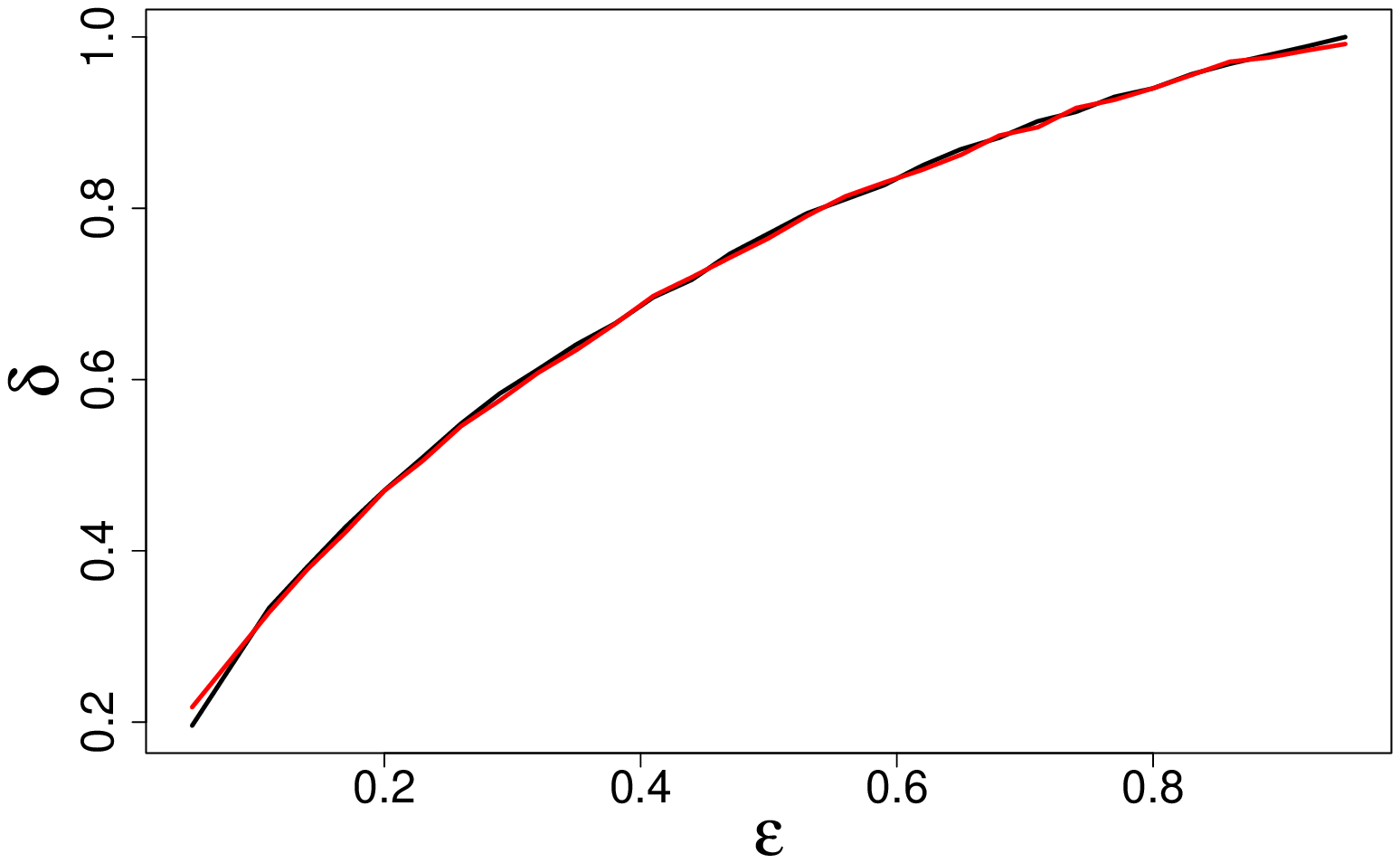}
			\end{center}
		\end{minipage}
		\vspace{-0cm}
		\caption{Compare the theoretical phase transition curve with the one determined by simulation studies with $p=500$. Here the covariance matrix is based on AR(1) model with $\rho=0.5$. The black curves represent the theoretical estimations while the red curves represent the simulation results. Left panel is for $\Delta=0$ and right panel is for $\Delta=1$.}
		\label{figure3}
	\end{figure}

	We analyze the simulated data-set to estimate the phase transition. At each fixed value of $\epsilon$ in our grid, we model the dependence of $S$ on $\delta$ using logistic regression. We assume that $S$ follows a binomial $B(\pi,100)$ distribution with $\text{logit}(\pi)=a+b\delta$. We define the phase transition as the value of $\delta$ at which the success probability $\pi=0.5$. In terms of the fitted parameters $\hat{a},\hat{b}$, we have the estimated phase transition $\hat{\delta}(\epsilon)=-\hat{a}/\hat{b}$. Figure \ref{figure3} shows that the agreement between the estimated phase transition curve based on the simulated finite-size systems and the analytical curve based on asymptotic theorem is remarkably good. We have tried different distributions for the random error $\bw$ and nonzero components of $\bbeta_0$ and found that our phase transition results are dependent of those choices as illustrated by Theorem \ref{thm2}. 
	
	\subsection{Phase transition under different dependent settings}\label{sec43}
	
\begin{figure}[hbtp] \hspace{-1cm}
		\begin{minipage}[t]{0.49\linewidth}
			\begin{center}
	\includegraphics[angle=-90,width=0.95\textwidth]{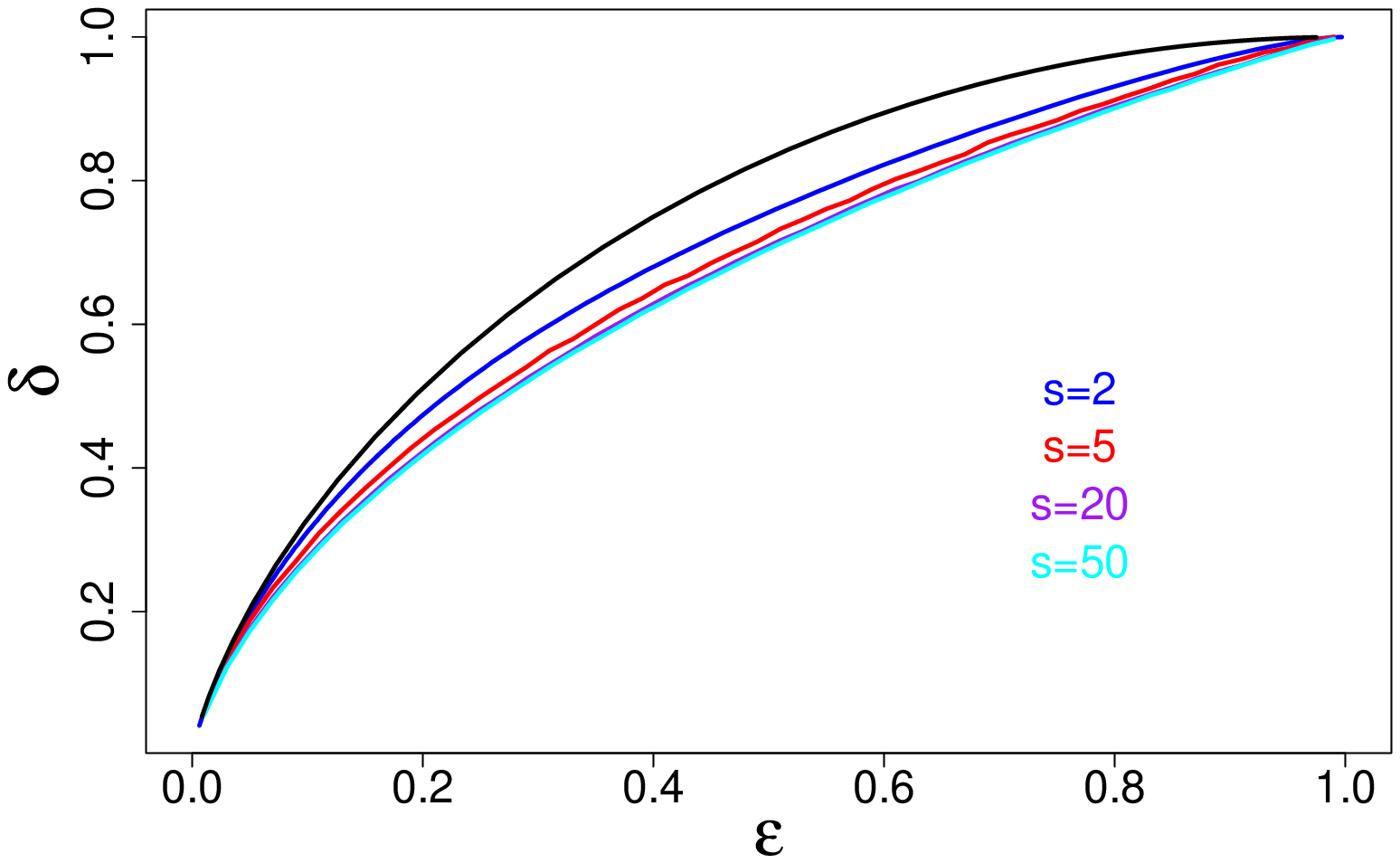}
			\end{center}
		\end{minipage}
		\hspace{0.3cm}
		\begin{minipage}[t]{0.49\linewidth}
			\begin{center}
	\includegraphics[angle=-90,width=0.95\textwidth]{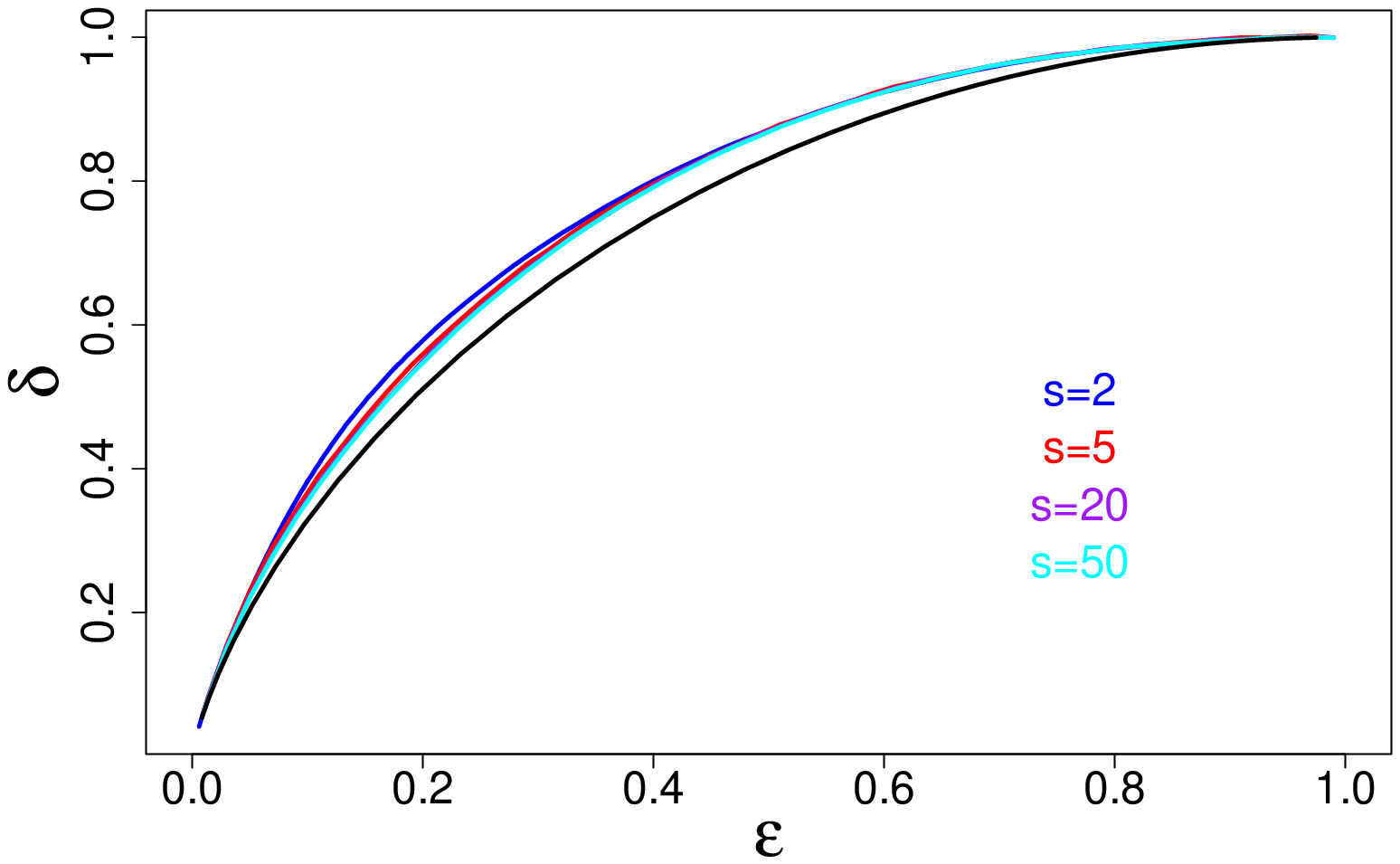}
			\end{center}
		\end{minipage}
		\vspace{-0cm}
		\caption{The values of $\delta_c(\epsilon)$ as a function of $\epsilon$ for several different values of block length $s$ with fixed $\Delta=1$. Here the block covariance matrix is based on AR(1) model with $\Sigma_{s,ij}=\rho^{|i-j|}$. Left panel is for $\rho=0.9$ and right panel is for $\rho=-0.9$.}
		\label{figure4}
	\end{figure}

	In this section, we study the dependence of phase transition on the block length $s$, correlation coefficient $\rho$, and asymmetric coefficient $\Delta$.
	Figure \ref{figure4} shows the change of phase transition boundaries with the block length $s$ for fixed $\Delta=1$ and $\rho$. As $s$ increases, the boundary moves further away from the i.i.d. boundary. For large $s$, in order to make a perfect recovery, less samples are needed under positive correlation $\rho=0.9$ and more samples are needed under negative correlation $\rho=-0.9$. When $s$ is big enough, e.g. $s=20$, the boundaries only change slightly for further increasing of $s$. 
	
\begin{figure}[hbtp] \hspace{-1cm}
		\begin{minipage}[t]{0.49\linewidth}
			\begin{center}
	\includegraphics[angle=-90,width=0.95\textwidth]{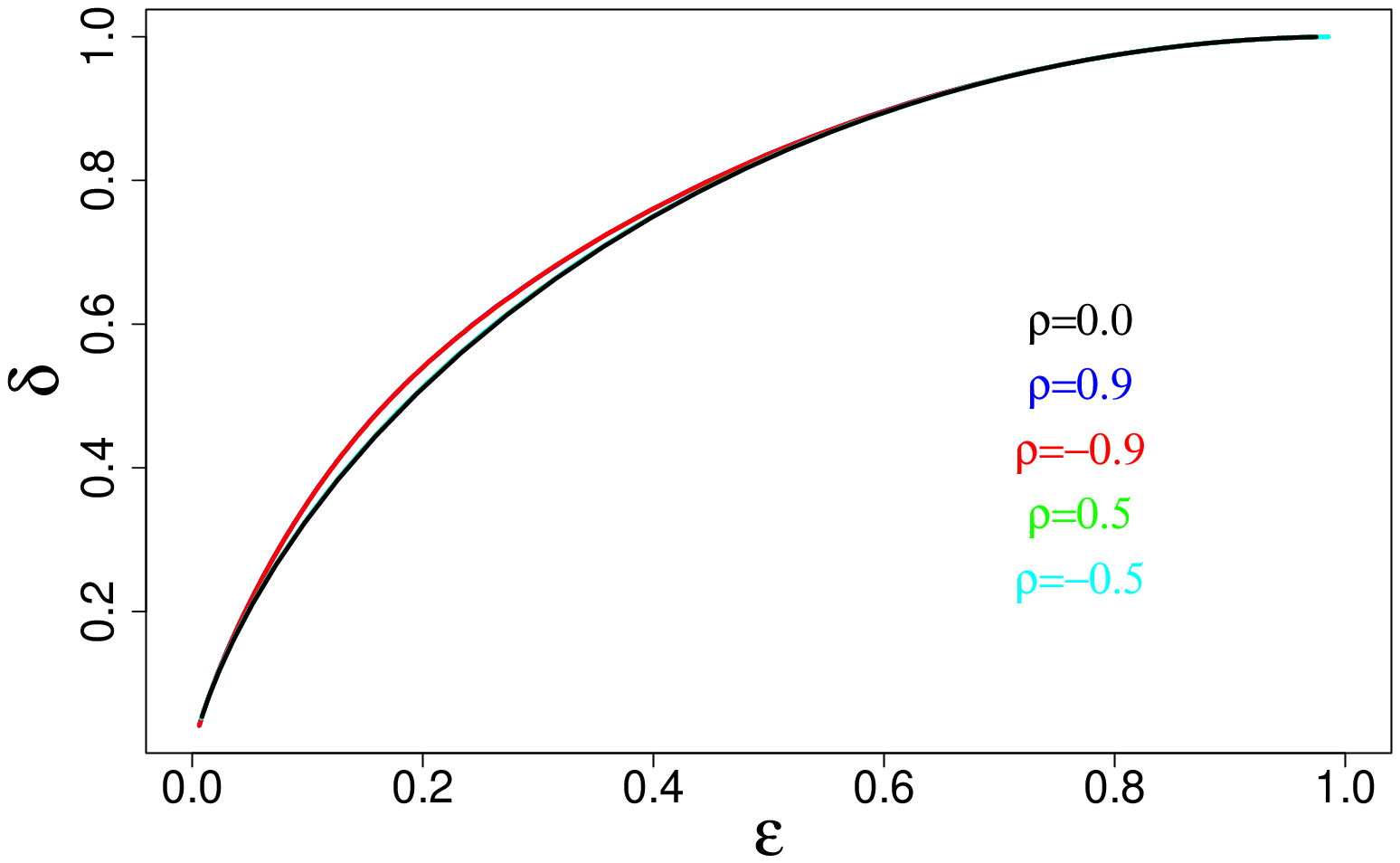}
			\end{center}
		\end{minipage}
		\hspace{0.3cm}
		\begin{minipage}[t]{0.49\linewidth}
			\begin{center}
	\includegraphics[angle=-90,width=0.95\textwidth]{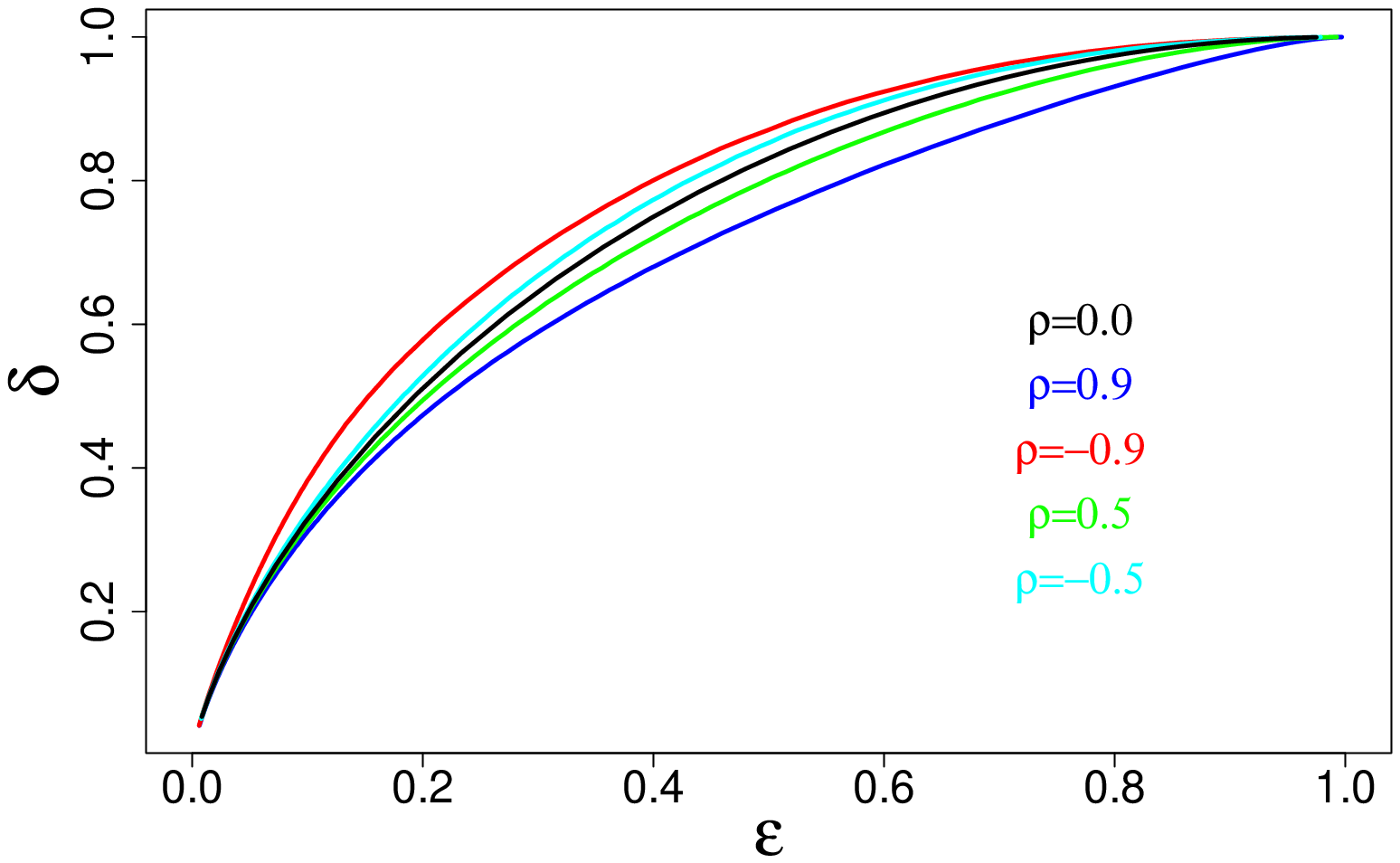}
			\end{center}
		\end{minipage}
		\vspace{-0cm}
		\caption{The values of $\delta_c(\epsilon)$ as a function of $\rho$ with fixed $s=2$ and $\Delta$. Left panel is for $\Delta=0$ and right panel is for $\Delta=1$.}
		\label{figure5}
	\end{figure}

	Figure \ref{figure5} shows the dependence of phase transition on $\rho$ for fixed $s=2$ and $\Delta$. If the distribution of $p_{\beta_0}$ is positive-negative symmetric, i.e. $\Delta=0$, the boundaries are almost independent of $\rho$ and very close to the Donoho-Tanner phase transition observed in \cite{Donoho4273} as illustrated by the left panel of Figure \ref{figure5}. If the distribution of the nonzero components of $p_{\beta_0}$ is highly skewed, e.g. $\Delta=1$, the phase transition curves fall below the Donoho-Tanner phase transition curve for $\rho\textgreater 0$ and above it for $\rho\textless 0$. As is clear from the right panel of Figure \ref{figure5}, for asymmetrically distributed signal $\bbeta_0$, the performance can be improved by increasing the correlation of covariance matrix $\bSigma$.  
	
\begin{figure}[hbtp] \hspace{-1cm}
		\begin{minipage}[t]{0.49\linewidth}
			\begin{center}
				\includegraphics[angle=-90,width=0.95\textwidth]{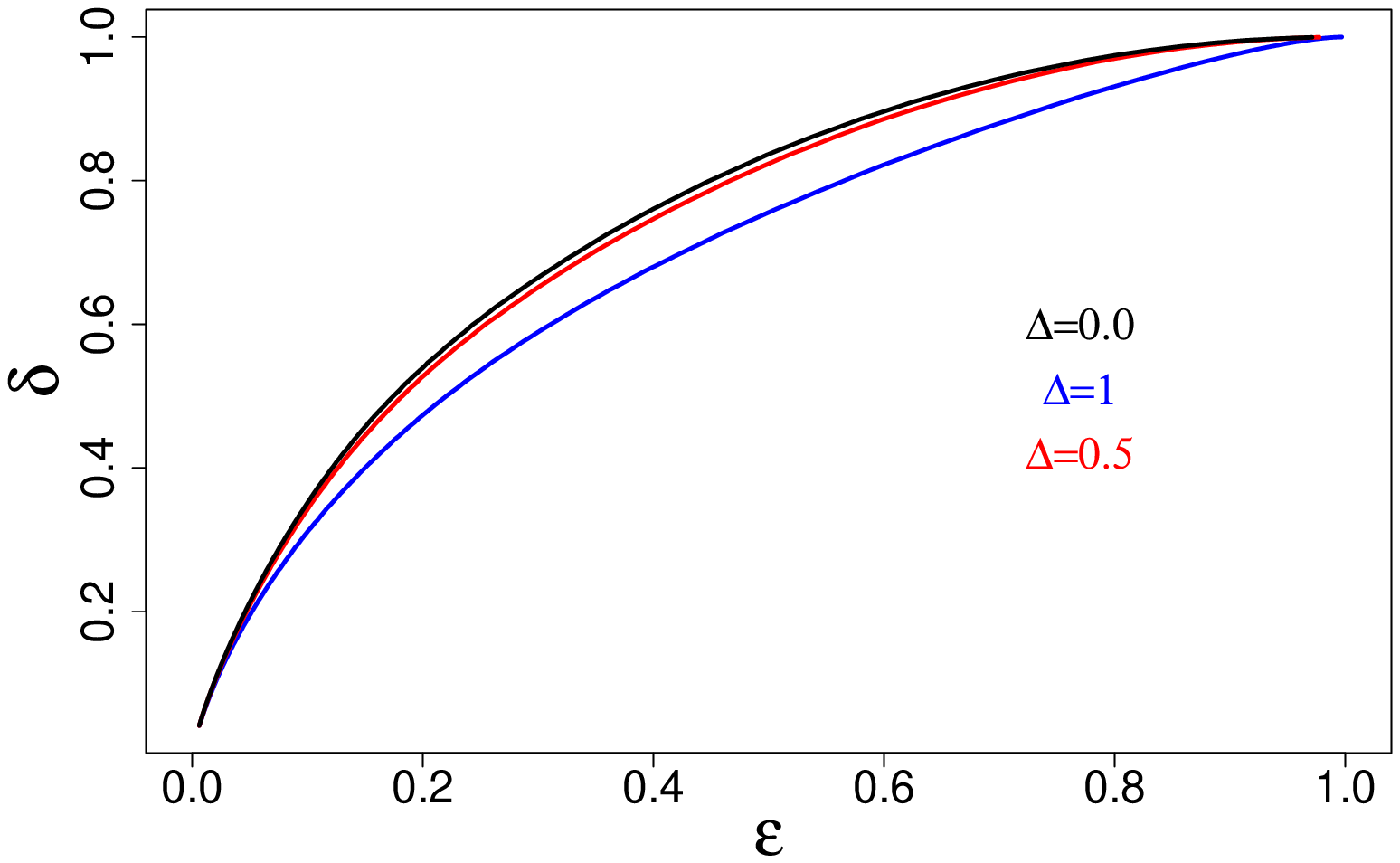}
			\end{center}
		\end{minipage}
		\hspace{0.3cm}
		\begin{minipage}[t]{0.49\linewidth}
			\begin{center}
				\includegraphics[angle=-90,width=0.95\textwidth]{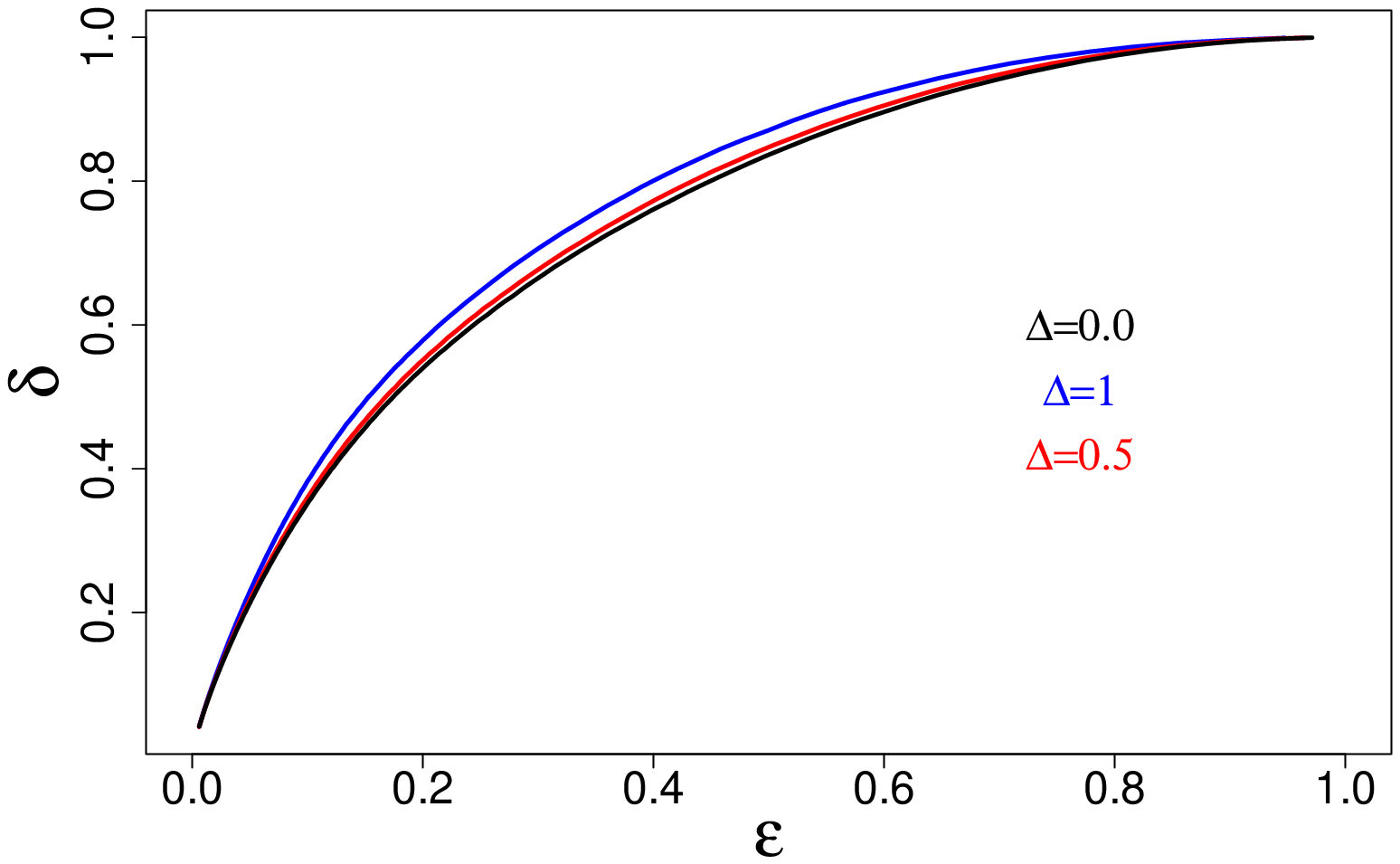}
			\end{center}
		\end{minipage}
		\vspace{-0cm}
		\caption{The dependence of the phase transition curve on $\Delta$ for fixed $s=2$ and $\rho$. Left panel is for $\rho=0.9$ and right panel is for $\rho=-0.9$.}
		\label{figure6}
	\end{figure}

	The phase transition curves for different $\Delta$ with fixed $\rho$ are exhibited in Figure \ref{figure6}. For positive correlation, at the same sparsity level $\epsilon$, the number of measurements $\delta$ that is required for successful recovery decreases as we increase $\Delta$ as shown by the left panel of Figure \ref{figure6}. For negative correlation, the conclusion is opposite as shown by the right panel of Figure \ref{figure6}.
%	The right panel of Figure \ref{figure5} shows the phase transition curves for three different spiked population models. In the first model, we choose $\sigma=1$, $r=10$ and $\lambda_1=\cdots=\lambda_r=60$. In the second model, we choose $\sigma=1$, $r=100$ and $\lambda_1=\cdots=\lambda_r=60$. In the third model, we choose $\sigma=1$, $r=10$ and $\lambda_1=2^{10},\lambda_2=2^{9}$, $\cdots$, $\lambda_r=2^1$. As is clear from the plots, all the phase transition curves are almost the same and equivalent to Donoho-Tanner phase transition curve under spiked population setting. Also our simulation shows that this conclusion holds for different choices of asymmetric index $\Delta$.
	
	\section{Proof of the main results}
	We prove Theorem \ref{thm1} using the limiting distribution of the approximate message passing (AMP) estimator. The AMP algorithm is a recently developed efficient iterative algorithm for solving the optimization problem (\ref{lasso}). In order to define AMP algorithm, we need to use the soft-thresholding operation  $\veta_\theta:\bR^p\rightarrow\bR^p$ defined in (\ref{veta}). For an arbitrary sequence of thresholds $\{\theta_t\}_{t\ge 0}$, the AMP constructs a sequence of estimates $\bbeta^{t}\in\bR^p$, and residuals $\bz^t\in\bR^n$, according to the iteration
	\begin{eqnarray}\nn
	\bbeta^{t+1}&=&\veta_{\theta_t}(\bSigma^{-1}\bX^T\bz^t+\bbeta^t),\\\label{vamp0}
	\bz^t&=&\by-\bX\bbeta^t+\frac{1}{p\delta}\bz^{t-1}\text{div}\veta_{\theta_{t-1}}(\bSigma^{-1}\bX^T\bz^{t-1}+\bbeta^{t-1}),
	\end{eqnarray}
	where $\text{div}\veta_\theta(\bv)$ is the divergence of the soft thresholding function. The algorithm (\ref{vamp0}) is mainly designed for theoretical analysis rather than practical use due to the fact that $\bSigma$ is usually unknown. The following proposition shows the relation between the fixed-point solution of AMP algorithm (\ref{vamp0}) and the optimization solution of LASSO problem (\ref{lasso}).
	\begin{proposition}\label{prop0}
		Any fixed point $\bbeta^t=\bbeta_\star,\bz^t=\bz_\star$ of the AMP iteration (\ref{vamp0}) with $\theta_t=\theta_\star$ is a minimizer of the LASSO cost function (\ref{lasso}) with 
		\begin{eqnarray}\label{calibr}
		\lambda=\theta_\star\left\{1-\frac{1}{p\delta}\text{div}\veta_{\theta_\star}(\bSigma^{-1}\bX^T\bz_\star+\bbeta_\star)\right\}.
		\end{eqnarray}
	\end{proposition}	
	
	For a converging sequence of instances $\{\bbeta_0(p),\bw(p),\bSigma(p),\bX(p)\}$, the asymptotic behavior of the recursion (\ref{vamp0}) can be characterized as follows. Define the sequence $\{\tau_t^2\}_{t\ge 0}$ by setting $\tau_0^2=\sigma^2_w+\lim_{p\rightarrow\infty}E\{\|\bbeta_0\|^2_{\bSigma}\}/(p\delta)$ (for $\bbeta_0\sim p_{\beta_0})$ and letting, for all $t\ge 0$:
	\begin{eqnarray}\label{psi1}
	\tau_{t+1}^2&=&\psi(\tau^2_t,\theta_t),
	\end{eqnarray}
	where the function $\psi(\cdot,\cdot)$ is defined in (\ref{psi}) which depends implicitly on the law $p_{\beta_0}$. 
	%We say a vector function $\bf:\bR^p\times\bR^p\rightarrow \bR^p$ is pseudo-Lipschitz if there exist a constant $L\textgreater 0$ such that for all $\bx,\by\in\bR^p: \|\bf(\bx,\by)\|^2\le L(1+\|\bx\|^2+\|\by\|^2)\|\bx-\by\|^2$. 
	The next proposition shows that the behavior of AMP can be tracked by the above one dimensional recursion which was often referred to as state evolution.
	\begin{proposition}\label{prop11}
		Let $\{\bbeta_0(p),\bw(p),\bSigma(p),\bX(p)\}_{p\in\bN}$ be a converging sequence of instances and let sequence $\varphi_p:\bR^p\times\bR^p\rightarrow\bR,~p\ge 1$ be uniformly pseudo-Lipschitz functions. Then 
		\begin{eqnarray}\nn
		\varphi_p(\bbeta^{t+1},\bbeta_0)\overset{P}{\approx}E\varphi_p(\veta_{\theta_t}(\bbeta_0+\tau_t\bSigma^{-1/2}\bz),\bbeta_0),
		\end{eqnarray}
		where $\bz\sim N(0,\bI_{p\times p})$ is independent of $\bbeta_0\sim p_{\beta_0}$ and the sequence $\{\tau_t\}_{t\ge 0}$ is given by the recursion (\ref{psi1}).
	\end{proposition}	
	
	In order to establish the connection with LASSO, a specific policy has to be chosen for the thresholds $\{\theta_t\}_{t\ge 0}$. Throughout this paper we will take $\theta_t=\alpha\tau_t$ with $\alpha$ is fixed. The sequence $\{\tau_t\}_{t\ge 0}$ is given by the recursion
	\begin{eqnarray}\label{taut}
	\tau^2_{t+1}=\psi(\tau^2_t,\alpha\tau_t).
	\end{eqnarray}
	We prove Theorem \ref{thm1} by proving the following result.
	\begin{theorem}\label{thm3}
		Assume the hypothesis of Theorem \ref{thm1}. Let $\hat{\bbeta}(\lambda;p)$ be the LASSO estimator for instance $\{\bbeta_0(p),\bw(p)$, $\bSigma(p),\bX(p)\}$ and denote by $\{\bbeta^t(\alpha;p)\}_{t\ge 0}$ the sequence of estimators produced by AMP algorithm (\ref{vamp0}) with $\theta_t=\alpha(\lambda)\tau_t$, where $\alpha(\lambda)$ is the calibration mapping between $\alpha$ and $\lambda$ defined in (\ref{alphalambda}) and $\tau_t$ is updated by the recursion (\ref{taut}). Then
		\begin{eqnarray}\nn
		\lim_{t\rightarrow\infty}\lim_{p\rightarrow\infty}\frac{1}{p}\|\bbeta^t(\alpha;p)-\hat{\bbeta}(\lambda;p)\|^2=0.
		\end{eqnarray}
	\end{theorem}
	As mentioned by \cite{BayatiM12}, Theorem \ref{thm3} requires taking the limit of infinite dimensions $p\rightarrow\infty$ before the limit of an infinite number of $t\rightarrow\infty$.
	
	\section{Discussion}\label{discussion}
	
	This paper focuses on the behavior of LASSO for learning the sparse coefficient vector in high-dimensional setting. We rigorously analyze the asymptotic behavior of LASSO for nonstandard Gaussian design models where the row of design matrix $\bX$ are drawn independently from distribution $N(0,\bSigma)$. We first obtain the formula for the asymptotic mean square error (AMSE) characterized through a series of non-linear equations. Then we present an accurate characterization of the phase transition curve $\delta_c=\delta(\epsilon)$ for separating successful from unsuccessful reconstruction of $\bbeta_0$ by LASSO in the noiseless case $\by=\bX\bbeta_0$. Our results show that the values of the non-zero elements of $\bbeta_0$ do not have any effect on the phase transition curve. However, for general $\bSigma$, the phase boundary $\delta_c$ not only depends on the sparsity coefficient $\epsilon$ but also depends on the signed sparsity pattern of the nonzero components of $\bbeta_0$. This is in sharp contrast to the result for i.i.d. case where $\delta_c$ is completely determined by $\epsilon$ regardless of the distribution of $\bbeta_0$.
	
	\cite{7987040} shows that, in the noiseless setting, the $l_q$-regularized least squares exhibits the same phase transition for every $0\le q\textless 1$ and this phase transition is much better than that of LASSO. However, in the noisy setting, there is a major difference between the performance of $l_p$-regularized least squares with different values of $q$. For instance, $q=0$ and $q=1$ outperform the other values of $q$ for very small and very large measurement noises. \cite{weng2018} further reveals some of the limitations and misleading features of the phase transition analysis. To overcome these limitations, they propose the small error analysis for $l_q$-regularized least squares to describe when phase transition analysis is reliable. \cite{AccuratePrediction} applied the AMP framework to a wider range of shrinkers including firm shrinkage and minimax shrinkage. Particularly, they show that the phase transition curve for AMP firm shrinkage and AMP minimax shrinkage are slightly better than that for LASSO.
	
	An interesting future research direction is to generalize the results derived in \cite{7987040,weng2018,AccuratePrediction} from the case of $\bSigma=\bI_{p\times p}$ to the case of $\bSigma\ne\bI_{p\times p}$. Our goal is to provide more accurate comparison for different regularizers in general setting for $\bSigma$. One of the major challenges in this direction is to establish the correspondence between regularized least square methods and specific AMP algorithms.
	
	\cite{Rangan} introduces a class of generalized approximate message passing (GAMP) algorithms that cope with the case where the noisy measurement vector $\by$ can be non-linear function of the noiseless measurement $\bX\bbeta_0$. \cite{glm1} evaluate the asymptotic behavior of GLAM in standard Gaussian setting and locate the associated sharp phase transitions separating learnable and nonlearnable regions in phase space. Another interesting future direction is to generalize these GLM results from the case of i.i.d. design matrix to the case of general design matrix.
	
	This work deals with the phase transition in noiseless case. For i.i.d. design matrix, \cite{dmm} studied the phase transition behavior in the noisy case by introducing a quantity called noise sensitivity which is proportional to the mean-squared error of LASSO estimator. They found a boundary curve in the phase space $0\le\epsilon,\delta\le 1$ such that the noise sensitivity is bounded above the curve and unbounded below the curve. This phase boundary is identical to the phase transition curve in the noiseless case for i.i.d. design. We plan to investigate if there is a similar phenomenon for LASSO phase transition with non-zero noise under non-i.i.d. design.
	
	\section*{Acknowledgments} 
	The author thanks the editor, associate editor, and two referees for many helpful comments and suggestions which led to a much improved presentation. This research is supported in part by Division of Mathematical Sciences (National Science Foundation) Grant DMS-1916411.
	
	\appendix
	
	\section{Proofs}
	
	\subsection{Proof of Proposition \ref{prop0}}
	
	\begin{proof}
		
		First we need to prove that the fixed-point of iteration (\ref{vamp0}) is a solution of (\ref{lasso}). Toward this end, the first equation of (\ref{vamp0}) implies that
		\begin{eqnarray}\nn
		\bSigma\{\bbeta_\star-(\bSigma^{-1}\bX^T\bz_\star+\bbeta_\star)\}+\theta_\star\partial\|\bbeta_\star\|_1=0.
		\end{eqnarray} 
		Therefore
		\begin{eqnarray}\nn
		\bX^T\bz_\star=\theta_\star\partial\|\bbeta_\star\|_1.
		\end{eqnarray} 
		The second equation of (\ref{vamp0}) implies that
		\begin{eqnarray}\nn
		\left(1-\omega_\star\right)\bz_\star=\by-\bX\bbeta_\star,
		\end{eqnarray} 
		where 
		\begin{eqnarray}\label{omegastar}
\omega_\star=\frac{1}{p\delta}\text{div}\veta_{\theta_\star}(\bSigma^{-1}\bX^T\bz_\star+\bbeta_\star). 
		\end{eqnarray} 
Therefore
		\begin{eqnarray}\nn
		\bX^T(\by-\bX\bbeta_\star)=\theta_\star\left(1-\omega_\star\right)\partial\|\bbeta_\star\|_1,
		\end{eqnarray} 
		which is the solution of (\ref{lasso}) for appropriately choosing tuning parameter $\lambda=\theta_\star\left(1-\omega_\star\right)$. 
		
	\end{proof}
	
	%\subsection{Proof of Proposition \ref{prop0}}
	%First we need to prove that the fixed-point of iteration (\ref{originali}) is a solution of (\ref{newprob}). Toward this end, the first equation of (\ref{originali}) implies that
	%\begin{eqnarray}\nn
	%\tilde{\bx}_\star-(\tilde{\bA}^T\bz_\star+\tilde{\bx}_\star)+\theta_\star\bSigma^{-1/2}\partial\|\bSigma^{-1/2}\bx_\star\|_1=0.
	%\end{eqnarray} 
	%Therefore
	%\begin{eqnarray}\nn
	%\tilde{\bA}^T\bz_\star=\theta_\star\bSigma^{-1/2}\partial\|\bSigma^{-1/2}\bx_\star\|_1.
	%\end{eqnarray} 
	%The second equation of (\ref{originali}) implies that
	%\begin{eqnarray}\nn
	%\left(1-\frac{\omega_\star}{\delta}\right)\bz_\star=\by-\tilde{\bA}\tilde{\bx}_\star.
	%\end{eqnarray} 
	%Therefore
	%\begin{eqnarray}\nn
	%\tilde{\bA}^T(\by-\tilde{\bA}\tilde{\bx}_\star)=\theta_\star\left(1-\frac{\omega_\star}{\delta}\right)\bSigma^{-1/2}\partial\|\bSigma^{-1/2}\bx_\star\|_1,
	%\end{eqnarray} 
	%which is the solution of (\ref{newprob}) for appropriately choosing tuning parameter $\lambda=\theta_\star\left(1-\frac{\omega_\star}{\delta}\right)$. 
	
	\subsection{Proof of Proposition \ref{prop11}}
	\begin{proof}
		
		Since the entries of $\bX$ are not i.i.d. normal, we do transformation $\tilde{\bX}=\bX\bSigma^{-1/2}$ and consider a different problem from (\ref{lasso})
		\begin{eqnarray}\label{newprob}
		\hat{\tilde{\bbeta}}&=&\text{argmin}_{\tilde{\bbeta}}{\tilde{\cal C}}(\tilde{\bbeta}),
		\end{eqnarray}
		where
		\begin{eqnarray}\nn
		\tilde{{\cal C}}(\tilde{\bbeta})&=&\frac{1}{2}\|\by-\tilde{\bX}\tilde{\bbeta}\|^2+\lambda\|\bSigma^{-1/2}\tilde{\bbeta}\|_1.
		\end{eqnarray}
		Here the design matrix $\tilde{\bX}$ has i.i.d. normal entries but the penalty term is not component-wise. The AMP algorithm for solving $\tilde{\bbeta}$ in (\ref{newprob}) constructs a sequence of estimates $\tilde{\bbeta}^t\in\bR^p$, and residuals $\bz^t\in\bR^n$, according to the iteration
		\begin{eqnarray}\nn
		\tilde{\bbeta}^{t+1}&=&\tilde{\veta}_{\theta_t}(\tilde{\bX}^T\bz^t+\tilde{\bbeta}^t),\\\label{originali}
		\bz^t&=&\by-\tilde{\bX}\tilde{\bbeta}^t+\frac{1}{p\delta}\bz^{t-1}\text{div}\tilde{\veta}_{\theta_{t-1}}(\tilde{\bX}^T\bz^{t-1}+\tilde{\bbeta}^{t-1}),
		\end{eqnarray}
		initialized with $\tilde{\bbeta}^0=0\in\bR^p$, where
		\begin{eqnarray}\label{etatilde}
		\tilde{\veta}_{\theta}(\bv)=\text{argmin}_{\bbeta\in\bR^p}\left\{\frac{1}{2}\|\bbeta-\bv\|^2+\theta\|\bSigma^{-1/2}\bbeta\|_1\right\}.
		\end{eqnarray} 
		Comparing (\ref{etatilde}) and (\ref{veta}), we have 
		\begin{eqnarray}\nn
		\tilde{\veta}_{\theta}(\bv)&=&\text{argmin}_{\bbeta\in\bR^p}\left\{\frac{1}{2}\|\bSigma^{-1/2}\bbeta-\bSigma^{-1/2}\bv\|_{\bSigma}^2+\theta\|\bSigma^{-1/2}\bbeta\|_1\right\}\\\nn
		&=&\bSigma^{1/2}\veta_\theta(\bSigma^{-1/2}\bv). 
		\end{eqnarray}
		Substituting $\bbeta=\bSigma^{-1/2}\tilde{\bbeta}$ into (\ref{originali}), the AMP update for $\bbeta^{t+1}$ is
		\begin{eqnarray}\nn
		\bbeta^{t+1}&=&\bSigma^{-1/2}\tilde{\bbeta}^{t+1}=\bSigma^{-1/2}\tilde{\veta}_{\theta_t}(\tilde{\bX}^T\bz^t+\tilde{\bbeta}^t)\\\nn
		&=&\veta_{\theta_t}(\bSigma^{-1/2}(\tilde{\bX}^T\bz^t+\tilde{\bbeta}^t))=\veta_{\theta_t}(\bSigma^{-1}\bX^T\bz^t+\bbeta^t)\\\nn
		\bz^t&=&\by-\bX\bbeta^t+\frac{1}{p\delta}\bz^{t-1}\text{div}\veta_{\theta_{t-1}}\left(\bSigma^{-1/2}(\tilde{\bX}^T\bz^{t-1}+\tilde{\bbeta}^{t-1})\right)\\\nn
		&=&\by-\bX\bbeta^t+\frac{1}{p\delta}\bz^{t-1}\text{div}\veta_{\theta_{t-1}}\left(\bSigma^{-1}\bX^T\bz^{t-1}+\bbeta^{t-1}\right)
		\end{eqnarray}
		which is equal to the AMP (\ref{vamp0}) constructed for solving the original problem (\ref{lasso}).
		
		The asymptotic property of AMP algorithm (\ref{originali}) has been established in \cite{nonseparable}. It can be verified that the assumptions (C1)-(C6) of Theorem 14 in \cite{nonseparable} are satisfied for the AMP iteration problem (\ref{originali}) by using the Conditions 1-6 introduced in the definition of converging sequences. More specifically, assumption (C1) is trivial. Assumptions (C3) and (C4) can be implied by Conditions 1 and 2 respectively. Assumption (C2) is satisfied due to the fact that both $1/\lambda_{min}(\bSigma)$ and $\lambda_{max}(\bSigma)$ are bounded. Assumptions (C5) and (C6) can be implied by Condition 6. 
%		because the quadratic forms introduced in (C5) and (C6) are just two special cases of the general quadratic form defined in Condition 5. 
		 Applying Theorem 14 in \cite{nonseparable}, for any sequence $\tilde{\varphi}_p:(\bR^p)^2\rightarrow\bR,~p\ge 1$, of uniformly pseudo-Lipschitz functions, we obtain
		\begin{eqnarray}\label{separablese}
		\tilde{\varphi}_p\left(\tilde{\bbeta}^{t+1},\tilde{\bbeta}_0\right)\overset{P}{\approx}E\tilde{\varphi}_p\left(\tilde{\veta}_{\theta_t}(\tilde{\bbeta}_0+\tau_t\bz),\tilde{\bbeta}_0\right),
		\end{eqnarray}
		where $\bz\sim N(0,\bI_{p\times p})$ is independent of $\tilde{\bbeta}_0$ and $\tau_t$ is determined by the following state evolution recursion
		\begin{eqnarray}\nn
		\tau_0^2&=&\sigma_w^2+\frac{1}{p\delta}E\|\tilde{\bbeta}_0\|^2,\\\nn
		\tau_{t+1}^2&=&\sigma_w^2+\frac{1}{p\delta}E\left(\|\tilde{\veta}_{\theta_t}(\tilde{\bbeta}_0+\tau_t\bz)-\tilde{\bbeta}_0\|^2\right),
		\end{eqnarray}
		where $\tilde{\bbeta}_0=\bSigma^{1/2}\bbeta_0$. 
		
		Define sequence of functions: $\tilde{\varphi}_p\left(\bx,\by\right)=\varphi_p\left(\bSigma^{-1/2}\bx,\bSigma^{-1/2}\by\right)$ which is also uniformly pseudo-Lipschitz due to the fact that $\bSigma^{-1/2}$ is well-conditioned. Therefore, the distributional limit of $\bbeta^{t+1}=\bSigma^{-1/2}\tilde{\bbeta}^{t+1}$ can be described by  
		\begin{eqnarray}\nn
		\varphi_p\left(\bbeta^{t+1},\bbeta_0\right)&=&\varphi_p\left(\bSigma^{-1/2}\tilde{\bbeta}^{t+1},\bSigma^{-1/2}\tilde{\bbeta}_0\right)=\tilde{\varphi}_p\left(\tilde{\bbeta}^{t+1},\tilde{\bbeta}_0\right)\\\nn
		&\overset{P}{\approx}&E\tilde{\varphi}_p\left(\tilde{\veta}_{\theta_t}(\tilde{\bbeta}_0+\tau_t\bz),\tilde{\bbeta}_0\right)\\\nn
		&=&E\varphi_p\left(\bSigma^{-1/2}\tilde{\veta}_{\theta_t}(\tilde{\bbeta}_0+\tau_t\bz),\bSigma^{-1/2}\tilde{\bbeta}_0\right)\\\nn
		&=&E\varphi_p\left(\veta_{\theta_t}(\bSigma^{-1/2}(\tilde{\bbeta}_0+\tau_t\bz)),\bbeta_0\right)\\\nn
		&=&E\varphi_p\left(\veta_{\theta_t}(\bbeta_0+\tau_t\bSigma^{-1/2}\bz),\bbeta_0\right).
		\end{eqnarray} 
	\end{proof}
	
	\subsection{Proof of Proposition \ref{prop1}}
	\begin{proof}
		In order to prove Proposition \ref{prop1}, we need the following Lemma.
\begin{lemma}\label{lm2}
	For any fixed $\alpha\textgreater 0$, the function $\psi(\tau^2,\alpha\tau)$ is strictly increasing and concave with respect to $\tau^2$.  
\end{lemma}	
	We first prove that $f(\alpha)=1$ has a unique solution when $\delta\textless 1$. From the definition (\ref{veta}), we get
	\begin{eqnarray}\label{etalargetau}
	\veta_\alpha(\bSigma^{-1/2}\bz)&=&\hat{\bbeta}\\\nn
	&=&\text{argmin}_{\bbeta\in\bR^p}\left\{\frac{1}{2}\|\bbeta-\bSigma^{-1/2}\bz\|^2_{\bSigma}+\alpha\|\bbeta\|_1\right\},
	\end{eqnarray} 
	which is equivalent to the solution of LASSO problem with $\bX=\bSigma^{1/2}$, $\by=\bz$, and $\lambda=\alpha$. It can be easily verified that $f(\alpha)=\frac{1}{p\delta}E\|\hat{\by}\|^2=\frac{1}{p\delta}E\|\bX\hat{\bbeta}\|^2$ with $f(0)=1/\delta$ and $f(\infty)=0$. Thus in order to find unique solution of $f(\alpha)=1$, it is enough if we can prove that $f(\alpha)$ is a strictly decreasing function. Denote ${\cal A}=\{j:\hat{\beta}_j\ne 0\}$ the active set of LASSO solution $\hat{\bbeta}$. From (\ref{etalargetau}), we obtain
	\begin{eqnarray}\nn
	\bSigma(\hat{\bbeta}-\bSigma^{-1/2}\bz)+\alpha\partial\|\hat{\bbeta}\|_1=0,
	\end{eqnarray} 
	which implies
	\begin{eqnarray}\nn
	\bSigma_{{\cal A}{\cal A}}(\hat{\bbeta}_{\cal A}-(\bSigma^{-1/2}\bz)_{\cal A})-\bSigma_{{\cal A}{\cal A}^c}(\bSigma^{-1/2}\bz)_{{\cal A}^c}+\alpha\text{sign}(\hat{\bbeta}_{\cal A})=0.
	\end{eqnarray}
	Taking derivative over $\alpha$ on both side, we obtain 
	\begin{eqnarray}\nn
	\bSigma_{{\cal A}{\cal A}}\frac{\partial\hat{\bbeta}_{\cal A}}{\partial\alpha}+\text{sign}(\hat{\bbeta}_{\cal A})+h(\alpha,\bz)=0,
	\end{eqnarray} 
	where $h(\alpha,\bz)$ is the contribution from the changing of active set ${\cal A}$ with $\alpha$. Since $\|\veta_\alpha(\bSigma^{-1/2}\bz)\|_{\bSigma}^2$ is continuous across the entire space $\bz\in\bR^p$, according to the discussion before (\ref{dpsi}) in Section \ref{a13}, the term $h(\alpha,\bz)$ disappears after taking expectation over $\bz$. Therefore
	\begin{eqnarray}\nn
	\frac{df(\alpha)}{d\alpha}=\frac{2}{p\delta}E\left(\hat{\bbeta}_{\cal A}^T\bSigma_{{\cal A}{\cal A}}\frac{\partial\hat{\bbeta}_{\cal A}}{\partial\alpha}\right)=-\frac{2}{p\delta}E\left(\hat{\bbeta}_{\cal A}^T\text{sign}(\hat{\bbeta}_{\cal A})\right)\textless 0,
	\end{eqnarray} 
	and we prove that $f(\alpha)$ is a decreasing function from $1/\delta$ to 0 as $\alpha$ increasing from 0 to $\infty$. Hence $f(\alpha)=1$ has a unique solution denoted by $\alpha_{min}$. 
	
	Next we prove that for fixed $\alpha\textgreater\alpha_{min}$, the solution of equation (\ref{fixedpoint}) exists. According to the definition (\ref{psi}), we have
	\begin{eqnarray}\nn
	\lim_{\tau^2\rightarrow\infty}E\left(\|\veta_{\alpha\tau}(\bbeta_0+\tau\bSigma^{-1/2}\bz)-\bbeta_0\|_{\bSigma}^2\right)&\rightarrow&E\left(\|\veta_\alpha(\bSigma^{-1/2}\bz)\|_{\bSigma}^2\right)\tau^2,
	\end{eqnarray} 
which implies
	\begin{eqnarray}\nn
	\lim_{\tau^2\rightarrow\infty}\frac{\psi(\tau^2,\alpha\tau)}{\tau^2}&=&f(\alpha)
	\end{eqnarray} 
	based on the definition (\ref{falpha}). From Lemma 1, we have that $\psi(\tau^2,\alpha\tau)$ is strictly increasing and concave function.
	Further, we have $\psi(\tau^2,\alpha\tau)|_{\tau^2=0}=\sigma_w^2\textgreater 0$. Therefore, in order for the fixed point equation $\tau^2=\psi(\tau^2,\alpha\tau)$ to have solutions, it is enough to show that $f(\alpha)\textless 1$ for $\alpha\textgreater\alpha_{min}(\delta)$. This can be obtained from the fact that $f(\alpha)$ is decreasing and $f(\alpha_{min})=1$. Thus we conclude that $\psi(\tau^2,\alpha\tau)<\tau^2$ as $\tau^2\rightarrow\infty$ and prove that the solution of (\ref{fixedpoint}) exists and is unique.

	\end{proof}	
	
	\subsection{Proof of Proposition \ref{prop3}}
	\begin{proof}	
		Consider a system of equations
		\begin{eqnarray}\label{u1}
		\tau^2&=&\psi(\tau^2,\theta),\\\label{u2}
		\theta&=&1-\frac{1}{\delta}E\left\langle\veta_{\theta}(\bbeta_0+\tau\bSigma^{-1/2}\bz),\bz\right\rangle.
\end{eqnarray}
According to Theorem 1 in \cite{celentano2020lasso}, for $\sigma_w^2\textgreater 0$, equations (\ref{u1}) and (\ref{u2}) have a unique solution denoted by $\tau^\star$, $\theta^\star$. Therefore, for any give $\lambda$, let $\alpha=\lambda/(\theta^\star\tau^\star)$, then $\alpha$ satisfies equation (\ref{lambdaalpha}) and is also unique.

	\end{proof}	
	
	\subsection{Proof of Theorem \ref{thm3}}
	\begin{proof}	
	The proof of Theorem \ref{thm3} is based on a series of Lemmas. The first Lemma implies that, asymptotically for large $p$, the AMP estimates converge. 
	\begin{lemma}\label{lm1}
		The estimates $\{\bbeta^t\}_{t\ge 0}$ and residuals $\{\bz^t\}_{t\ge 0}$ of AMP (\ref{vamp0}) almost surely satisfy
		\begin{eqnarray}\nn
		\lim_{t\rightarrow\infty}\lim_{p\rightarrow\infty}\frac{1}{p}\|\bbeta^t-\bbeta^{t-1}\|^2=0,&&\lim_{t\rightarrow\infty}\lim_{p\rightarrow\infty}\frac{1}{p}\|\bz^t-\bz^{t-1}\|^2=0.
		\end{eqnarray}
	\end{lemma}

Denote $\sigma_{\min}(\bX)$ and $\sigma_{\max}(\bX)$ the maximum and minimum non-zero singular value of $\bX$. Then the second Lemma implies that with high probability, $\sigma_{\min}(\bX)$ is lower bounded and $\sigma_{\max}(\bX)$ is upper bounded.  
	\begin{lemma}\label{lm1.1} 
For every $t\ge 0$, there exists $c_5\textgreater 0$ such that 
	\begin{eqnarray}\nn
\bP\left(c_5^{-1}\le\sigma_{\min}(\bX)\le\sigma_{\max}(\bX)\le c_5\right)\textgreater 1-2\exp(-t^2/2).
	\end{eqnarray}
\end{lemma}

	According to the first equation of (\ref{vamp0}), denote the subgradient $\bv^t\in\partial\|\bbeta^t\|_1$ such that
	\begin{eqnarray}\label{bvt}
	\bSigma\{\bbeta^t-(\bSigma^{-1}\bX^T\bz^{t-1}+\bbeta^{t-1})\}+\theta_{t-1}\bv^t=0.
	\end{eqnarray}
Then the next Lemma implies that with high probability, the subgradient $\bv^t$ cannot have too many coordinates with magnitude close to 1.
	\begin{lemma}\label{csparsity}
	For large enough $t$, there exists $c,C,c_2\textgreater 0$ such that
	\begin{eqnarray}\nn
	\bP\left(\frac{\left|j\in[p]:|v^t_j|\ge 1-c_2\right|}{n}\ge 1-\omega^\star/2\right)\le C\exp(-cn),
	\end{eqnarray}	
where $\omega^\star$ is defined in (\ref{omegastar}) and 
	\begin{eqnarray}\nn
\omega^\star=\frac{1}{n}E\left(\left\|\veta_{\theta_\star}\left(\bbeta_0+\tau_\star\bSigma^{-1/2}\bz\right)\right\|_0\right).
	\end{eqnarray}	
\end{lemma}

Define the minimum singular value of $\bX$ over a set $S\subset[p]$ by 
	\begin{eqnarray}\nn
\kappa_-(\bX,S)=\inf\left\{\|\bX\bw\|_2:~\text{supp}(\bw)\subset S,\|\bw\|_2=1\right\},
\end{eqnarray}
and the $s$ sparse singular value by
	\begin{eqnarray}\nn
\kappa_-(\bX,s)=\min_{|S|\le s}\kappa_-(\bX,S).
\end{eqnarray}
Then the next Lemma implies that $\kappa_-(\bX,s)$ is lower bounded with high probability.
	\begin{lemma}\label{lm1.2}
	For every $c_4\ge 0$, there exists $C,c\textgreater 0$ such that
		\begin{eqnarray}\nn
\bP\left(\kappa_-(\bX,n(1-\omega^\star/4))\le c_4\right)\le Ce^{-cn}.
\end{eqnarray}	  
\end{lemma}

	We are now ready to prove Theorem \ref{thm3}. The remainder of the argument takes place on the high-probability event determined by Lemmas \ref{lm1.1}, \ref{csparsity}, and \ref{lm1.2}.
	
Let $\br=\hat{\bbeta}-\bbeta^t$ denote the distance between the LASSO optimum and the AMP estimate at $t$-th iteration, then
	\begin{eqnarray}\nn
	0&\ge&\frac{{\cal C}(\bbeta^t+\br)-{\cal C}(\bbeta^t)}{p}\\\nn
	&=&\frac{1}{2p}\|\by-\bX(\bbeta^t+\br)\|^2+\frac{\lambda}{p}\|\bbeta^t+\br\|_1-\frac{1}{2p}\|\by-\bX\bbeta^t\|^2-\frac{\lambda}{p}\|\bbeta^t\|_1\\\nn
	&=&\frac{1}{2p}\|\bX\br\|^2-\frac{\br^T\bX^T(\by-\bX\bbeta^t)}{p}+\frac{\lambda}{p}(\|\bbeta^t+\br\|_1-\|\bbeta^t\|_1).
	\end{eqnarray}
	Then by using equation (\ref{vamp0}) we have
	\begin{eqnarray}\label{zero}
	0&\ge&\underbrace{\frac{1}{2p}\|\bX\br\|^2}_{\text{I}}+\underbrace{\frac{1}{p}\langle\br,\text{sg}{\cal C}(\bbeta^t)\rangle}_{\text{II}}+\underbrace{\frac{\lambda}{p}(\|\bbeta^t+\br\|_1-\|\bbeta^t\|_1-\br^T\bv^{t})}_{\text{III}}.
	\end{eqnarray}
	where the sub-gradient $\text{sg}{\cal C}(\bbeta^t)=-\bX^T(\by-\bX\bbeta^t)+\lambda\bv^{t}$ and $\bv^{t}$ is defined in (\ref{bvt}). 
	
	Let's first take a look at the second term of (\ref{zero}). Substituting (\ref{vamp0}) and $\bv^t$ from (\ref{bvt}), we obtain
	\begin{eqnarray}\nn
	\text{sg}{\cal C}(\bbeta^t)&=&\bX^T(\omega_t\bz^{t-1}-\bz^t)-\frac{\lambda}{\theta_{t-1}}\{\bSigma(\bbeta^t-\bbeta^{t-1})-\bX^T\bz^{t-1}\}\\\nn
	&=&\frac{\lambda-\theta_{t-1}(1-\omega_t)}{\theta_{t-1}}\bX^T\bz^{t-1}-\bX^T(\bz^{t}-\bz^{t-1})-\frac{\lambda}{\theta_{t-1}}\bSigma(\bbeta^t-\bbeta^{t-1}),
	\end{eqnarray}
	where $\omega_t=\text{div}\veta_{\theta_{t-1}}(\bSigma^{-1}\bX^T\bz^{t-1}+\bbeta^{t-1})/p/\delta$. Hence
	\begin{eqnarray}\nn
	\frac{1}{\sqrt{p}}\|\text{sg}{\cal C}(\bbeta^t)\|&\le&\frac{|\lambda-\theta_{t-1}(1-\omega_t)|}{\theta_{t-1}}\sigma_{max}(\bX)\frac{\|\bz^{t-1}\|}{\sqrt{p}}+\sigma_{max}(\bX)\frac{\|\bz^{t}-\bz^{t-1}\|}{\sqrt{p}}\\\nn
	&&+\frac{\lambda}{\theta_{t-1}}\sigma_{max}(\bSigma)\frac{\|\bbeta^t-\bbeta^{t-1}\|}{\sqrt{p}}.
	\end{eqnarray}
	By Lemmas \ref{lm1}, \ref{lm1.1} and the fact that $\lambda_{max}(\bSigma)$ is bounded as $p\rightarrow\infty$, we deduce that the last two terms converge to 0 as $p\rightarrow\infty$ and then $t\rightarrow\infty$. For the first term, using state evolution, we obtain $\frac{\|\bz^{t-1}\|}{\sqrt{p}}=O(1)$. Finally, using the calibration relation (\ref{calibr}), we get
	\begin{eqnarray}\nn
	\lim_{t\rightarrow\infty}\lim_{p\rightarrow\infty}\frac{|\lambda-\theta_{t-1}(1-\omega_t)|}{\theta_{t-1}}\overset{a.s.}{=}\frac{1}{\theta_\star}|\lambda-\theta_\star(1-\omega_\star)|=0.
	\end{eqnarray}
	Therefore $\frac{1}{\sqrt{p}}\|\text{sg}{\cal C}(\bbeta^t)\|\rightarrow 0$ almost surely. Since $\frac{\|\hat{\bbeta}\|}{\sqrt{p}}=O(1)$ and $\frac{\|\bbeta^t\|}{\sqrt{p}}=O(1)$, we get that $\frac{\|\br\|}{\sqrt{p}}=O(1)$ and hence the second term of (\ref{zero}) $\langle\br,\text{sg}{\cal C}(\bbeta^t)\rangle\rightarrow 0$ almost surely. From (\ref{zero}), we have
	\begin{eqnarray}\nn
	\frac{1}{2p}\|\bX\br\|^2+\frac{\lambda}{p}(\|\bbeta^t+\br\|_1-\|\bbeta^t\|_1-\br^T\bv^{t})\le c_1\varepsilon.
	\end{eqnarray}	
	Both the first and third terms on the right-hand side of (\ref{zero}) are non-negative. The first one is trivial. Denote $S\equiv\{j\in\bN:\bbeta^t_j\ne 0\}$ the support of $\bbeta^t$. The third one is non-negative since 
	\begin{eqnarray}\nn
	&&\|\bbeta^t+\br\|_1-\|\bbeta^t\|_1-\br^T\bv^{t}\\\nn
	&=&\|\bbeta_S^t+\br_S\|_1-\|\bbeta_S^t\|_1-\br_S^T\text{sign}(\bbeta_S^{t})+\|\br_{\bar{S}}\|_1-\br_{\bar{S}}^T\bv_{\bar{S}}^{t}\\\nn
	&=&(\bbeta_S^t+\br_S)\{\text{sign}(\bbeta_S^t+\br_S)-\text{sign}(\bbeta_S^t)\}+\|\br_{\bar{S}}\|_1-\br_{\bar{S}}^T\bv_{\bar{S}}^{t}\ge 0.
	\end{eqnarray}
	Since $(\bbeta_S^t+\br_S)\{\text{sign}(\bbeta_S^t+\br_S)-\text{sign}(\bbeta_S^t)\}\ge 0$ and $\|\bv_{\bar{S}}^t\|_1\le 1$, we have 
	\begin{eqnarray}\label{xr}
	\frac{\|\bX\br\|^2}{p}&\le&\xi_1(\varepsilon),\\\label{sbar}
	\|\br_{\bar{S}}\|_1-\br_{\bar{S}}^T\bv_{\bar{S}}^{t}&\le&p\xi_1(\varepsilon),
	\end{eqnarray}
where $\xi_1(\varepsilon)\rightarrow 0$ as $\varepsilon\rightarrow 0$.

	Consider $\br=\br^\perp+\br^\parallel$ with $\br^\parallel\in\text{ker}(\bX)$ and $\br^\perp\perp\text{ker}(\bX)$. It follows from (\ref{xr}) and Lemma \ref{lm1.1} that 
	\begin{eqnarray}\label{rperp}
	\|\br^\perp\|^2\le pc_5\xi_1(\varepsilon).
	\end{eqnarray}
	We need to prove an analogous bound for $\br^\parallel$. Note that $\|\br^\perp_{\bar{S}}\|_1\le\sqrt{p}\|\br^\perp_{\bar{S}}\|_2\le\sqrt{p}\|\br^\perp\|_2\le p\sqrt{\xi_1(\varepsilon)}$, from (\ref{sbar}), we get
	\begin{eqnarray}\label{rpara}
	\|\br^\parallel_{\bar{S}}\|_1-(\br^\parallel_{\bar{S}})^T\bv^{t\parallel}_{\bar{S}}\le p\xi_2(\varepsilon).
	\end{eqnarray}
	Define $S(c_2)\equiv\{j\in\bN:|v^t_j|\ge 1-c_2\}$, then $\bar{S}(c_2)\subseteq\bar{S}$. We have
	\begin{eqnarray}\label{rpara}
	\|\br^\parallel_{\bar{S}}\|_1-(\br^\parallel_{\bar{S}})^T\bv^{t\parallel}_{\bar{S}}\ge\|\br^\parallel_{\bar{S}(c_2)}\|_1-|\br^\parallel_{\bar{S}(c_2)}|^T|\bv^{t\parallel}_{\bar{S}(c_2)}|\ge c_2\|\br^\parallel_{\bar{S}(c_2)}\|_1.
	\end{eqnarray}
	Therefore using (\ref{rpara}), we have
	\begin{eqnarray}\label{rparabd}
	\|\br^\parallel_{\bar{S}(c_2)}\|_1\le c_2^{-1}p\xi_2(\varepsilon).
	\end{eqnarray}
	Denote $c_3=\delta\omega^\star/4$. Then from Lemma \ref{csparsity}, we have $|S(c_2)|\le n-2pc_3$. Thus if $|\bar{S}(c_2)|\le pc_3/2$, one obtains $p\le n-3pc_3/2$. In this case, $\text{ker}(\bX)=\{0\}$ and the proof is concluded. Let us now consider the case $|\bar{S}(c_2)|\ge pc_3/2$. Then partition $\bar{S}(c_2)=\cup_{l=1}^KS_l$, where $pc_3/2\le|S_l|\le pc_3$, and for each $i\in S_l$, $j\in S_{l+1}$, $|r_i^\parallel|\ge|r_j^\parallel|$. Also define $\bar{S}_+\equiv\cup_{l=2}^KS_l\subseteq\bar{S}(c_2)$. Since, for any $i\in S_l$, $|r^\parallel_i|\le\|\br^\parallel_{S_{l-1}}\|_1/|S_{l-1}|$, we have
	\begin{eqnarray}\nn
	\|\br^\parallel_{\bar{S}_+}\|_2^2&=&\sum_{l=2}^K\|\br^\parallel_{S_l}\|_2^2\le\sum_{l=2}^K|S_l|\left(\frac{\|\br^\parallel_{S_{l-1}}\|_1}{|S_{l-1}|}\right)^2\\\nn
	&\le&\frac{4}{pc_3}\sum_{l=2}^K\|\br^\parallel_{S_{l-1}}\|_1^2\le\frac{4}{pc_3}\left(\sum_{l=2}^K\|\br^\parallel_{S_{l-1}}\|_1\right)^2\\\label{rparasp}
	&\le&\frac{4}{pc_3}\|\br^\parallel_{\bar{S}(c_2)}\|_1^2\le\frac{4\xi_2(\varepsilon)^2}{c_2^2c_3}p\equiv p\xi_3(\varepsilon).
	\end{eqnarray}
	To conclude the proof, it is sufficient to prove an analogous bound for $\|\br^\parallel_{S_+}\|_2^2$ with $S_+=S(c_2)\cup S_1$. Since $|S_1|\le pc_3$ and $|S(c_2)|\le n-2pc_3$, we have $|S_+|\le n-pc_3$ and by  Lemma \ref{lm1.2} that $\sigma_{min}(\bX_{S_+})\ge c_4$. Since $0=\bX\br^\parallel=\bX_{S_+}\br^\parallel_{S_+}+\bX_{\bar{S}_+}\br^\parallel_{\bar{S}_+}$, we have
	\begin{eqnarray}\label{rparas}
	c_4^2\|\br^\parallel_{S_+}\|_2^2\le\|\bX_{\bar{S}_+}\br^\parallel_{\bar{S}_+}\|_2^2=\|\bX_{S_+}\br^\parallel_{S_+}\|^2\le c_5\|\br^\parallel_{\bar{S}_+}\|_2^2\le c_5p\xi_3(\varepsilon).
	\end{eqnarray}
Combining (\ref{rperp}), (\ref{rparasp}), and (\ref{rparas}), we conclude the proof.
	\end{proof}	
	
	\subsection{Proof of Theorem \ref{thm2}}\label{a13}
	\begin{proof}
		Since there is no measurement noise, i.e. $\sigma_w^2=0$, we have $\psi(\tau^2,\alpha\tau)|_{\tau^2=0}=0$. Thus in order for the fixed point equation $\tau^2=\psi(\tau^2,\alpha\tau)$ to have unique solution $\tau_\star^2=0$, we need to have $\inf_\alpha\frac{d\psi(\tau^2,\alpha\tau)}{d\tau^2}|_{\tau^2=0}\le 1$ due to the fact that $\psi(\tau^2,\alpha\tau)$ is a increasing and concave function of $\tau^2$ for fixed $\alpha$. Since $\psi(\tau^2,\alpha\tau)$ decreases with $\delta$, the critical value $\delta_c$ is defined as 
		\begin{eqnarray}\label{deltac}
		\delta_c=\inf\left\{\delta: \inf_\alpha\frac{d\psi(\tau^2,\alpha\tau)}{d\tau^2}|_{\tau^2=0}\le 1\right\}.
		\end{eqnarray} 
		Then for any $\delta\textgreater\delta_c$, we have unique solution $\tau_\star^2=0$; for any $\delta\textless\delta_c$, we also have solution $\tau_\star^2\textgreater 0$. According to Theorem \ref{thm1}, we can consider the following solution
		\begin{eqnarray}\nn
		\hat{\bbeta}=\text{argmin}_{\bbeta\in\bR^p}\left\{\frac{1}{2}\|\bbeta-\bbeta_0-\tau\bSigma^{-1/2}\bz\|^2_{\bSigma}+\alpha\tau\|\bbeta\|_1\right\},
		\end{eqnarray} 
		where $\bz\sim N(0,\bI_{p\times p})$ is independent of $p_{\beta_0}$. Define ${\cal A}=\{j:\hat{\beta}_j\ne 0\}$, then we have $\hat{\bbeta}_{{\cal A}^c}=0$ and
		\begin{eqnarray}\nn
		\{\bSigma(\hat{\bbeta}-\bbeta_0)-\tau\bSigma^{1/2}\bz\}_{\cal A}+\alpha\tau\text{sign}(\hat{\bbeta}_{\cal A})=0,
		\end{eqnarray} 
		which implies
		\begin{eqnarray}\nn
		\bSigma_{{\cal A}{\cal A}}(\hat{\bbeta}_{\cal A}-\bbeta_{0,{\cal A}})=\tau(\bSigma^{1/2}\bz)_{\cal A}-\alpha\tau\text{sign}(\hat{\bbeta}_{\cal A})+\bSigma_{{\cal A}{\cal A}^c}\bbeta_{0,{\cal A}^c},
		\end{eqnarray} 
		and
		\begin{eqnarray}\label{lasso1}
		&&\hat{\bbeta}_{\cal A}-\bbeta_{0,{\cal A}}\\\nn
		&=&\bSigma_{{\cal A}{\cal A}}^{-1}\{\tau(\bSigma^{1/2}\bz)_{\cal A}-\alpha\tau\text{sign}(\hat{\bbeta}_{\cal A})+\bSigma_{{\cal A}{\cal A}^c}\bbeta_{0,{\cal A}^c}\}.
		\end{eqnarray} 
		Substituting into the definition, we get
		\begin{eqnarray}\nn
		&&E\{\|\hat{\bbeta}-\bbeta_0\|^2_{\bSigma}\}\\\nn
		&=&E\{(\hat{\bbeta}_{\cal A}-\bbeta_{0,{\cal A}})^T\bSigma_{{\cal A}{\cal A}}(\hat{\bbeta}_{\cal A}-\bbeta_{0,{\cal A}})\\\nn
		&&~~~-2(\hat{\bbeta}_{\cal A}-\bbeta_{0,{\cal A}})^T\bSigma_{{\cal A}{\cal A}^c}\bbeta_{0,{\cal A}^c}+\bbeta_{0,{\cal A}^c}^T\bSigma_{{\cal A}^c{\cal A}^c}\bbeta_{0,{\cal A}^c}\}\\\nn
		&=&E\{\tau^2((\bSigma^{1/2}\bz)_{\cal A}-\alpha\text{sign}(\hat{\bbeta}_{\cal A}))^T\bSigma_{{\cal A}{\cal A}}^{-1}((\bSigma^{1/2}\bz)_{\cal A}-\alpha\text{sign}(\hat{\bbeta}_{\cal A}))\}\\\label{psidtau}
		&&+E\{\bbeta_{0,{\cal A}^c}^T(\bSigma_{{\cal A}^c{\cal A}^c}-\bSigma_{{\cal A}^c{\cal A}}\bSigma_{{\cal A}{\cal A}}^{-1}\bSigma_{{\cal A}{\cal A}^c})\bbeta_{0,{\cal A}^c}\}.
		\end{eqnarray} 
		To perform the integrals over $\bz\in\bR^p$, we divide the $p$-dimensional space into regions such that the active set of $\hat{\bbeta}(\bz)$ keeps the same in each region and changes by one variable between two neighboring regions that share a common boundary hyperplane. In each region, the sign of $\bbeta(\bz)$ also keeps the same. A illustration of this space separation is shown in Figure \ref{figure2d} for a simple two dimensional example. Let $S_i$ and $S_j$ denote two neighboring regions that share a common hyperplane $F_{ij}$ determined by equation $g_{ij}(\bz,\tau)=0$ with $g_{ij}(\bz,\tau)\textgreater 0$ in $S_i$ and $g_{ij}(\bz,\tau)\textless 0$ in $S_j$. Denote $f_i(\bz,\tau)$ the function form of $\|\hat{\bbeta}(\bz,\tau)-\bbeta_0\|^2_{\bSigma}$ in region $S_i$. Then $f_i(\bz,\tau)$ is differentiable over $\tau^2$ inside $S_i$ and the derivative of $Ef_i(\bz,\tau)I(\bz\in S_i)$ over $\tau^2$ involves integrals over face $F_{ij}$ with respect to $d-1$
 dimensional measure $\sigma_{F_{ij}}(\cdot)$. An application of Stokes's theorem, as in Theorem 1 of \cite{baddeley_1977}, establishes differentiability of this integral which is given by $\sigma_{F_{ij}}(f_i(\bz,\tau)\frac{\partial g_{ij}(\bz,\tau)}{\partial\tau^2})$. Similarly, we can obtain the boundary contribution of $F_{ij}$ to the derivative of $Ef_j(\bz,\tau)I(\bz\in S_j)$ over $\tau^2$ which is given by $-\sigma_{F_{ij}}(f_j(\bz,\tau)\frac{\partial g_{ij}(\bz,\tau)}{\partial\tau^2})$. Since $\hat{\bbeta}(\bz,\tau)$ is continuous across $F_{ij}$, we have $f_{i}(\bz,\tau)=f_{j}(\bz,\tau)$ on $F_{ij}$ and thus the contributions of the boundary effects due to $F_{ij}$ cancel each other between the derivative of $Ef_i(\bz,\tau)I(\bz\in S_i)$ over $\tau^2$ and the derivative of $Ef_j(\bz,\tau)I(\bz\in S_j)$ over $\tau^2$. Therefore, in taking derivative over $\tau^2$ for (\ref{psidtau}), the boundary effects are canceled and one gets	  
		\begin{eqnarray}\nn
		&&\frac{d\psi(\tau^2,\alpha\tau)}{d\tau^2}\\\label{dpsi}
		&=&\lim_{p\rightarrow\infty}\frac{1}{p\delta}E\{((\bSigma^{1/2}\bz)_{\cal A}-\alpha\text{sign}(\hat{\bbeta}_{\cal A}))^T\bSigma_{{\cal A}{\cal A}}^{-1}((\bSigma^{1/2}\bz)_{\cal A}-\alpha\text{sign}(\hat{\bbeta}_{\cal A}))\},
		\end{eqnarray} 
		which only depends on the sign of the non-zero components of $\hat{\bbeta}$. 

We need to consider situations as $\tau^2\rightarrow 0$. Since $\hat{\bbeta}\rightarrow \bbeta_0$, we have $\hat{\bbeta}=\bbeta_0+o_P(1)$ as $\tau^2\rightarrow 0$. Let ${\cal B}=\{j:\beta_{0,j}\ne 0\}$, clearly ${\cal B}\subseteq{\cal A}$ and ${\cal A}^c\subseteq{\cal B}^c$ as $\tau^2\rightarrow 0$. For ${\cal B}$ part, from (\ref{lasso1}), we obtain
		\begin{eqnarray}\nn
		\{\bSigma(\hat{\bbeta}-\bbeta_0)-\tau\bSigma^{1/2}\bz\}_{\cal B}+\alpha\tau\text{sign}(\hat{\bbeta}_{{\cal B}})=0,
		\end{eqnarray} 
		which implies
		\begin{eqnarray}\nn
		&&\bSigma_{{\cal B}{\cal B}}(\hat{\bbeta}_{\cal B}-\bbeta_{0,{\cal B}})\\\nn
		&=&\tau(\bSigma^{1/2}\bz)_{\cal B}-\alpha\tau\text{sign}(\hat{\bbeta}_{{\cal B}})-\bSigma_{{\cal B}{\cal B}^c}\hat{\bbeta}_{{\cal B}^c},
		\end{eqnarray} 
		and thus
		\begin{eqnarray}\label{bset}
		&&\hat{\bbeta}_{\cal B}-\bbeta_{0,{\cal B}}\\\nn
		&=&\bSigma_{{\cal B}{\cal B}}^{-1}\{\tau(\bSigma^{1/2}\bz)_{\cal B}-\alpha\tau\text{sign}(\hat{\bbeta}_{{\cal B}})-\bSigma_{{\cal B}{\cal B}^c}\hat{\bbeta}_{{\cal B}^c}\}.
		\end{eqnarray} 
		For ${\cal B}^c$ part, we have
		\begin{eqnarray}\nn
		\{\bSigma(\hat{\bbeta}-\bbeta_0)-\tau\bSigma^{1/2}\bz\}_{{\cal B}^c}+\alpha\tau\partial\|\hat{\bbeta}_{{\cal B}^c}\|_1=0
		\end{eqnarray} 
		which implies
		\begin{eqnarray}\nn
		&&\bSigma_{{\cal B}^c{\cal B}}(\hat{\bbeta}_{\cal B}-\bbeta_{0,{\cal B}})\\\nn
		&=&\tau(\bSigma^{1/2}\bz)_{{\cal B}^c}-\alpha\tau\partial\|\hat{\bbeta}_{{\cal B}^c}\|_1-\bSigma_{{\cal B}^c{\cal B}^c}\hat{\bbeta}_{{\cal B}^c}.
		\end{eqnarray} 
		Using  (\ref{bset}), we have
		\begin{eqnarray}\nn
		&&\bSigma_{{\cal B}^c{\cal B}}\bSigma_{{\cal B}{\cal B}}^{-1}\{\tau(\bSigma^{1/2}\bz)_{\cal B}-\alpha\tau\text{sign}(\hat{\bbeta}_{{\cal B}})-\bSigma_{{\cal B}{\cal B}^c}\hat{\bbeta}_{{\cal B}^c}\}\\\nn
		&=&\tau(\bSigma^{1/2}\bz)_{{\cal B}^c}-\alpha\tau\partial\|\hat{\bbeta}_{{\cal B}^c}\|_1-\bSigma_{{\cal B}^c{\cal B}^c}\hat{\bbeta}_{{\cal B}^c}.
		\end{eqnarray} 
		The final equation for $\hat{\bbeta}_{{\cal B}^c}$ is
		\begin{eqnarray}\nn
		(\bSigma_{{\cal B}^c{\cal B}^c}-\bSigma_{{\cal B}^c{\cal B}}\bSigma_{{\cal B}{\cal B}}^{-1}\bSigma_{{\cal B}{\cal B}^c})\hat{\bbeta}_{{\cal B}^c}-\tau(\bSigma^{1/2}\bz)_{{\cal B}^c}\\\nn
		+\tau\bSigma_{{\cal B}^c{\cal B}}\bSigma_{{\cal B}{\cal B}}^{-1}\{(\bSigma^{1/2}\bz)_{\cal B}-\alpha\text{sign}(\hat{\bbeta}_{{\cal B}})\}+\alpha\tau\partial\|\hat{\bbeta}_{{\cal B}^c}\|_1=0.
		\end{eqnarray} 
		Therefore $\hat{\bbeta}_{{\cal B}^c}$ is equivalent to the solution of the following LASSO problem 
		\begin{eqnarray}\nn
		\hat{\bbeta}_{{\cal B}^c}=\text{argmin}_{\bbeta\in\bR^p}\left\{\frac{1}{2}\|\by-\bX\bbeta\|_2^2+\lambda\|\bbeta\|_1\right\}
		\end{eqnarray} 
		with 
		\begin{eqnarray}\nn
		\bX&=&(\bSigma_{{\cal B}^c{\cal B}^c}-\bSigma_{{\cal B}^c{\cal B}}\bSigma_{{\cal B}{\cal B}}^{-1}\bSigma_{{\cal B}{\cal B}^c})^{1/2},\\\nn
		\by&=&\tau\bX^{-1}\left[(\bSigma^{1/2}\bz)_{{\cal B}^c}-\bSigma_{{\cal B}^c{\cal B}}\bSigma_{{\cal B}{\cal B}}^{-1}\{(\bSigma^{1/2}\bz)_{\cal B}-\alpha\text{sign}(\bbeta_{0,{\cal B}})\}\right],
		\end{eqnarray} 
		and $\lambda=\alpha\tau$. Since (\ref{dpsi}) only involves the sign of $\hat{\bbeta}$, without loss of generality, we can take $\tau=1$. Therefore $\hat{\bbeta}_{{\cal B}^c}$ is independent of the actual distribution of $\bbeta_0$ but depends on $\epsilon$ and $\Delta$. Denote $\bar{\cal B}=\{j:j\in{\cal B}^c\text{ and }\hat{\beta}_j\ne 0\}$, then we have ${\cal A}={\cal B}\cup\bar{\cal B}$. Define function
		\begin{eqnarray}\nn
		M(\epsilon,\Delta,\alpha)=\lim_{p\rightarrow\infty}\frac{1}{p}E\{((\bSigma^{1/2}\bz)_{\cal A}-\alpha\text{sign}(\hat{\bbeta}_{\cal A}))^T\bSigma_{{\cal A}{\cal A}}^{-1}((\bSigma^{1/2}\bz)_{\cal A}-\alpha\text{sign}(\hat{\bbeta}_{\cal A}))\},
		\end{eqnarray}	
	which exists according to Condition 5. Substituting (\ref{dpsi}) into (\ref{deltac}), we obtain 
		\begin{eqnarray}\nn
		\delta_c=\inf_{\alpha}M(\epsilon,\Delta,\alpha).
		\end{eqnarray}	 
	\end{proof}
	
	\subsection{Proof of Lemma \ref{lm2}}
	\begin{proof}
		For fixed $\alpha$, in order to prove that $\psi(\tau^2,\alpha\tau)$ is an increasing and concave function of $\tau^2$, we need to show that $\frac{d\psi(\tau^2,\alpha\tau)}{d\tau^2}\textgreater 0$ and $\frac{d^2\psi(\tau^2,\alpha\tau)}{(d\tau^2)^2}\textless 0$. Since $\bSigma_{{{\cal A}}{{\cal A}}}^{-1}$ is positive definite, from (\ref{dpsi}), we get $\frac{d\psi(\tau^2,\alpha\tau)}{d\tau^2}\textgreater 0$ and prove that $\psi(\tau^2,\alpha\tau)$ is an increasing function of $\tau^2$. 
		
		We need to take further derivative over $\tau^2$ to obtain $\frac{d^2\psi(\tau^2,\alpha\tau)}{(d\tau^2)^2}$. Toward this end, consider the LASSO problem with
		\begin{eqnarray}\nn
		\bX=\bSigma^{1/2},&&
		\by=\tau\bz+\bSigma^{1/2}\bbeta_0,
		\end{eqnarray} 
		and $\lambda=\alpha\tau$. Following the discussion in deriving (\ref{dpsi}), we can divide the $p$-dimensional space $\bz\in\bR^p$ into regions such that the active set and the sign of each variable are fixed in each region. Denote by ${\cal A}_i$ and ${\cal A}_j$ the active sets in two neighboring regions $S_i$ and $S_j$ respectively. Further denote by $F_{ij}$ the boundary hyperplane between $S_i$ and $S_j$. Assume that $|{\cal A}_i|=k$, $|{\cal A}_j|=k-1$, and denote $\bx_k$ the active variable that drops when moving from $S_i$ to $S_j$. Therefore, ${\cal A}_j\subset{\cal A}_j$ and ${\cal A}_i\setminus{\cal A}_j=\bx_k$. Then, from (\ref{lasso1}), we obtain that the solution of $\hat{\bbeta}$ inside $S_i$ is differentiable over $\tau^2$ and can be written as $\hat{\bbeta}_{{\cal S}_i}=\bSigma_{{{\cal A}_i}{{\cal A}_i}}^{-1}\left\{(\bSigma^{1/2}\by)_{{\cal A}_i}-\alpha\tau\text{sign}(\hat{\bbeta}_{{\cal S}_i})\right\}$. Assume that the $k$-th component of $\hat{\bbeta}_{{\cal S}_i}$, i.e. $\hat{\bbeta}_{{\cal S}_i}[k]\textgreater 0$ in $S_i$ and $\hat{\bbeta}_{{\cal S}_i}[k]=0$ in $S_j$, then the boundary hyperplane $F_{ij}$ is determined by equation
		\begin{eqnarray}\label{gx}
			g_{ij}(\bz,\tau)=\be_{(k)}^T\hat{\bbeta}_{{\cal S}_i}=\be_{(k)}^T\bSigma_{{{\cal A}_i}{{\cal A}_i}}^{-1}\left\{(\bSigma^{1/2}\by)_{{\cal A}_i}-\alpha\tau\text{sign}(\hat{\bbeta}_{{\cal S}_i})\right\}=0,	
		\end{eqnarray} 
		where $\be_{(k)}$ represents the $k$-th coordinate vector for $\hat{\bbeta}_{{\cal A}_i}$. Denote by $\bar{S}_i$ and $\bar{S}_j$ the other two neighboring regions that have the same active sets but opposite sign of variables comparing to $S_i$ to $S_j$, i.e. $\text{sign}(\hat{\bbeta}_{\bar{\cal S}_i})=-\text{sign}(\hat{\bbeta}_{{\cal S}_i})$ and $\text{sign}(\hat{\bbeta}_{\bar{\cal S}_j})=-\text{sign}(\hat{\bbeta}_{{\cal S}_j})$. Then their boundary hyperplane $\bar{F}_{ij}$ is determined by equation
		\begin{eqnarray}\nn
\bar{g}_{ij}(\bz,\tau)=\be_{(k)}^T\hat{\bbeta}_{\bar{\cal S}_i}=\be_{(k)}^T\bSigma_{{{\cal A}_i}{{\cal A}_i}}^{-1}\left\{(\bSigma^{1/2}\by)_{{\cal A}_i}-\alpha\tau\text{sign}(\hat{\bbeta}_{\bar{\cal S}_i})\right\}=0,	
\end{eqnarray} 
		
		Denote $f_i(\bz,\tau)$ the integrand inside the expectation on the right hand side of (\ref{dpsi}) in region $S_i$, i.e.
		\begin{eqnarray}\nn
		f_i(\bz,\tau)=((\bSigma^{1/2}\bz)_{{\cal A}_i}-\alpha\text{sign}(\hat{\bbeta}_{{\cal A}_i}))^T\bSigma_{{{\cal A}_i}{{\cal A}_i}}^{-1}((\bSigma^{1/2}\bz)_{{\cal A}_i}-\alpha\text{sign}(\hat{\bbeta}_{{\cal A}_i})), 
		\end{eqnarray}
		which does not depend on $\tau^2$ explicitly, thus the dependence of the expectation on $\tau^2$ only comes from the boundary effects. From (\ref{psidtau}), the continuity of $\|\hat{\bbeta}(\bz,\tau)-\bbeta_0\|^2_{\bSigma}$ leads to
		\begin{eqnarray}\nn
		&&\tau^2[(\bSigma^{1/2}\bz)_{{\cal A}_{i}}-\alpha\text{sign}(\hat{\bbeta}_{{\cal A}_{i}})]^T\bSigma_{{{\cal A}_{i}}{{\cal A}_{i}}}^{-1}[(\bSigma^{1/2}\bz)_{{\cal A}_{i}}-\alpha\text{sign}(\hat{\bbeta}_{{\cal A}_{i}})]\\\nn
		&&+\bbeta_{0,{{\cal A}^c_{i}}}^T(\bSigma_{{{\cal A}^c_{i}}{{\cal A}^c_{i}}}-\bSigma_{{{\cal A}^c_{i}}{{\cal A}_{i}}}\bSigma_{{{\cal A}_{i}}{{\cal A}_{i}}}^{-1}\bSigma_{{{\cal A}_{i}}{{\cal A}^c_{i}}})\bbeta_{0,{{\cal A}^c_{i}}}\\\nn
		&=&\tau^2[(\bSigma^{1/2}\bz)_{{\cal A}_{j}}-\alpha\text{sign}(\hat{\bbeta}_{{\cal A}_{j}})]^T\bSigma_{{{\cal A}_{j}}{{\cal A}_{j}}}^{-1}[(\bSigma^{1/2}\bz)_{{\cal A}_{j}}-\alpha\text{sign}(\hat{\bbeta}_{{\cal A}_{j}})]\\\label{identity}
		&&+\bbeta_{0,{{\cal A}^c_{j}}}^T(\bSigma_{{{\cal A}^c_{j}}{{\cal A}^c_{j}}}-\bSigma_{{{\cal A}^c_{j}}{{\cal A}_{j}}}\bSigma_{{{\cal A}_{j}}{{\cal A}_{j}}}^{-1}\bSigma_{{{\cal A}_{j}}{{\cal A}^c_{j}}})\bbeta_{0,{{\cal A}^c_{j}}}.
		\end{eqnarray}
		Therefore, the difference of the integrand function on (\ref{dpsi}) caused by the change of active set from region $S_i$ to region $S_j$ can be written as
\begin{eqnarray}\nn
\Delta_{ij}&=&f_i(\bz,\tau)-f_j(\bz,\tau)\\\nn
&=&[(\bSigma^{1/2}\bz)_{{\cal A}_{i}}-\alpha\text{sign}(\hat{\bbeta}_{{\cal A}_{i}})]^T\bSigma_{{{\cal A}_{i}}{{\cal A}_{i}}}^{-1}[(\bSigma^{1/2}\bz)_{{\cal A}_{i}}-\alpha\text{sign}(\hat{\bbeta}_{{\cal A}_{i}})]\\\nn
&&-[(\bSigma^{1/2}\bz)_{{\cal A}_{j}}-\alpha\text{sign}(\hat{\bbeta}_{{\cal A}_{j}})]^T\bSigma_{{{\cal A}_{j}}{{\cal A}_{j}}}^{-1}[(\bSigma^{1/2}\bz)_{{\cal A}_{j}}-\alpha\text{sign}(\hat{\bbeta}_{{\cal A}_{j}})]\\\nn
&=&\{\bbeta_{0,{{\cal A}^c_{j}}}^T(\bSigma_{{{\cal A}^c_{j}}{{\cal A}^c_{j}}}-\bSigma_{{{\cal A}^c_{j}}{{\cal A}_{j}}}\bSigma_{{{\cal A}_{j}}{{\cal A}_{j}}}^{-1}\bSigma_{{{\cal A}_{j}}{{\cal A}^c_{j}}})\bbeta_{0,{{\cal A}^c_{j}}}\\\label{identity}
&&-\bbeta_{0,{{\cal A}^c_i}}^T(\bSigma_{{{\cal A}^c_i}{{\cal A}^c_i}}-\bSigma_{{{\cal A}^c_i}{{\cal A}_{i}}}\bSigma_{{{\cal A}_{i}}{{\cal A}_{i}}}^{-1}\bSigma_{{{\cal A}_{i}}{{\cal A}^c_i}})\bbeta_{0,{{\cal A}^c_i}}\}/\tau^2,
\end{eqnarray}
which only depends on the active sets ${\cal A}_i$ and ${\cal A}_j$. Therefore, we also have  $\bar{\Delta}_{ij}=\Delta_{ij}$, where $\bar{\Delta}_{ij}$ represents the difference of the integrand function caused by the change of active set from region $\bar{S}_i$ to region $\bar{S}_j$. 

		According to Stokes's theorem shown in Theorem 1 of \cite{baddeley_1977}, the contribution of boundary $F_{ij}$ to the derivative of integral $Ef_i(\bz,\tau)I(\bz\in S_i)+Ef_j(\bz,\tau)I(\bz\in S_j)$ over $\tau^2$ is given by $\sigma_{F_{ij}}(\Delta_{ij}\frac{\partial g_{ij}(\bz,\tau)}{\partial\tau^2})$. Similarly, we derive that the contribution of boundary $\bar{F}_{ij}$ to the derivative of integral $E\bar{f}_i(\bz,\tau)I(\bz\in \bar{S}_i)+E\bar{f}_j(\bz,\tau)I(\bz\in \bar{S}_j)$ over $\tau^2$ is given by $-\sigma_{\bar{F}_{ij}}(\Delta_{ij}\frac{\partial \bar{g}_{ij}(\bz,\tau)}{\partial\tau^2})$. Define
\begin{eqnarray}\nn
		\ba_k&=&\bSigma^{1/2}_{,{\cal A}_i}\bSigma_{{{\cal A}_i}{{\cal A}_i}}^{-1}\be_{(k)},\\\nn b_k&=&\be_{(k)}^T\bSigma_{{{\cal A}_i}{{\cal A}_i}}^{-1}\text{sign}(\hat{\bbeta}_{{\cal S}_i}),\\\nn c_k&=&\be_{(k)}^T\bSigma_{{{\cal A}_i}{{\cal A}_i}}^{-1}(\bSigma\bbeta_0)_{{\cal A}_i}.
 \end{eqnarray}
 Then from (\ref{gx}), we have $g_{ij}(\bz,\tau)=\tau\ba_k^T\bz-\alpha\tau b_k+c_k$. Therefore, $\frac{\partial g_{ij}(\bz,\tau)}{\partial\tau^2}=\frac{1}{2\tau}(\ba_k^T\bz-\alpha b_k)$. We obtain the boundary contributions of $F_{ij}$ and $\bar{F}_{ij}$ as 
\begin{eqnarray}\nn
\sigma_{F_{ij}}\left\{\frac{\Delta_{ij}c_k}{2\tau^3\|\ba_k\|}\left[\phi\left(\frac{c_k}{\tau\|\ba_k\|}+\frac{\alpha b_k}{\|\ba_k\|}\right)-\phi\left(\frac{c_k}{\tau\|\ba_k\|}-\frac{\alpha b_k}{\|\ba_k\|}\right)\right]\right\}.
\end{eqnarray}
From (\ref{identity}), since ${\cal A}_j\subset{\cal A}_i$, we get ${\cal A}^c_j\supset{\cal A}^c_i$ and hence $\Delta_{ij}\ge 0$. Then we conclude that the boundary contribution is less than or equal to zero since $x(\phi(x+c)-\phi(x-c))\le 0$ for any $x$ and $c\ge 0$. This complete the proof of the concavity of function $\psi(\tau^2,\alpha\tau)$.
		\end{proof}
	
	\subsection{Proof of Lemma \ref{lm1}}
	\begin{proof}		
		We begin with the convergence of the state evolution (\ref{taut}) iteration described by the following lemma which can be immediately proved using the concavity of $\psi(\tau^2,\alpha\tau)$ over $\tau^2$. 
		\begin{lemma}\label{lm110}
			For any $\alpha\ge\alpha_{min}$. The iteration (\ref{taut}) converges to the unique solution of the fixed-point equation $\tau_\star^2=\psi(\tau_\star^2,\alpha\tau_\star)$, i.e. $\tau_t^2\rightarrow\tau_\star^2$ as $t\rightarrow\infty$. 
		\end{lemma}	
		Next we need to generalize state evolution to compute large system limits for functions of $\bbeta^t$, $\bbeta^s$, with $t\ne s$. To this purpose, we define the covariances $\{\tau_{s,t}\}_{s,t\ge 0}$ recursively by
		\begin{eqnarray}\nn
		\tau_{s+1,t+1}&=&\sigma_w^2+\lim_{p\rightarrow\infty}\frac{1}{p\delta}E\left\{[\veta_{\theta_s}(\bbeta_0+\bSigma^{-1/2}\bz_s)-\bbeta_0]^T\bSigma\right.\\\label{psig}
		&&~~~~~~~~~~~~~\left.[\veta_{\theta_t}(\bbeta_0+\bSigma^{-1/2}\bz_t)-\bbeta_0]\right\},
		\end{eqnarray}
		where $(\bz_s,\bz_t)$ jointly Gaussian, independent from $\bbeta_0\sim p_{\beta_0}$ with mean 0 and covariance given by $E(\bz_s\bz_s^T)=\tau_{s,s}\bI_{p\times p}=\tau_s^2\bI_{p\times p}$, $E(\bz_t\bz_t^T)=\tau_{t,t}\bI_{p\times p}=\tau_t^2\bI_{p\times p}$, and $E(\bz_s\bz_t^T)=\tau_{s,t}\bI_{p\times p}$. The boundary condition is fixed by letting $\tau_{0,0}=\sigma^2_w+E\{\|\bbeta\|^2_{\bSigma}\}/\delta$ and $\tau_{0,1}=\sigma^2_w+\lim_{p\rightarrow\infty}E\{[\bbeta_0-\veta_{\theta_0}(\bbeta_0+\bSigma^{-1/2}\bz_0)]^T\bSigma\bbeta_0\}/p/\delta$. With this definition, we have the following generalization of Proposition \ref{prop1}.
		\begin{lemma}\label{lm111}
			Let $\{\bbeta_0(p),\bw(p),\bSigma(p),\bX(p)\}_{p\in\bN}$ be a converging sequence of instances and let sequence $\varphi_p:(\bR^p)^3\rightarrow\bR,~p\ge 1$ be uniformly pseudo-Lipschitz functions. Then for all $s\ge 0$ and $t\ge 0$, we get
			\begin{eqnarray}\nn
			\varphi_p(\bbeta^{s+1},\bbeta^{t+1},\bbeta_0)\overset{P}{\sim}\varphi_p(\veta_{\theta_s}(\bbeta_0+\bSigma^{-1/2}\bz_s),\veta_{\theta_t}(\bbeta_0+\bSigma^{-1/2}\bz_t),\bbeta_0),
			\end{eqnarray}
			where $(\bz_s,\bz_t)$ jointly Gaussian, independent from $\bbeta_0\sim p_{\beta_0}$ with mean 0 and covariance given by $E(\bz_s\bz_s^T)=\tau_s^2\bI_{p\times p}$, $E(\bz_t\bz_t^T)=\tau_t^2\bI_{p\times p}$, and $E(\bz_s\bz_t^T)=\tau_{s,t}\bI_{p\times p}$. The recursion $\tau_{s,,t}$ for all $s,t\ge 0$ is determined by (\ref{psi}) and (\ref{psig}).
		\end{lemma}	
		\parindent=0pt
		Proof of Lemma \ref{lm1}. Define sequence of $\{y_t\}_{t\ge 0}$ as
		\begin{eqnarray}\nn
		y_t&=&\lim_{p\rightarrow\infty}\frac{1}{p\delta}E\left\|\veta_{\theta_t}(\bbeta_0+\bSigma^{-1/2}\bz_t)-\veta_{\theta_{t-1}}(\bbeta_0+\bSigma^{-1/2}\bz_{t-1})\right\|^2_{\bSigma}.
		\end{eqnarray}
		From (\ref{psig}), we have
		\begin{eqnarray}\label{yt}
		y_t&=&\tau_{t}^2+\tau_{t-1}^2-2\tau_{t,t-1}.
		\end{eqnarray}
		Take $\theta_t=\alpha\tau_t$ with $\alpha$ is fixed, then according to Lemma \ref{lm110}, we have $\tau^2_{t}\rightarrow\tau^2_\star$ and $\theta_{t}\rightarrow\theta_\star=\alpha\tau_\star$ as $t\rightarrow\infty$. We will show that $y_t\rightarrow 0$ which in turn yields $\tau_{t,t-1}\rightarrow\tau^2_\star$ based on (\ref{yt}). For large enough $t$, we have the representation as follows in terms of the two independent random vectors $\bz,\bw\sim N(0,\bI_{p\times p})$:
		\begin{eqnarray}\nn
		y_t&=&\lim_{p\rightarrow\infty}\frac{1}{p\delta}E\left\|\veta_{\theta_\star}\left(\bbeta_0+\sqrt{\tau_\star^2-\frac{y_{t-1}}{4}}\bSigma^{-1/2}\bz+\sqrt{\frac{y_{t-1}}{4}}\bSigma^{-1/2}\bw\right)\right.\\\nn
		&&~~~~~~~~~~~~~~~\left.-\veta_{\theta_\star}\left(\bbeta_0+\sqrt{\tau_\star^2-\frac{y_{t-1}}{4}}\bSigma^{-1/2}\bz-\sqrt{\frac{y_{t-1}}{4}}\bSigma^{-1/2}\bw\right)\right\|^2_{\bSigma}.
		\end{eqnarray}
		Consider $y_t$ as a function of $y_{t-1}$ denoted by $y_t=R(y_{t-1})$. A straightforward calculation yields
		\begin{eqnarray}\nn
		R^\prime(y_{t-1})&=&\lim_{p\rightarrow\infty}\frac{1}{p\delta}E\left(Tr\left[\bSigma^{-1}\left\{\nabla\veta_{\theta_\star}\left(\bbeta_0+\bSigma^{-1/2}\bz_t\right)\right\}^T\right.\right.\\\nn
		&&~~~~~~~~~~~\left.\left.\bSigma\nabla\veta_{\theta_\star}\left(\bbeta_0+\bSigma^{-1/2}\bz_{t-1}\right)\right]\right),
		\end{eqnarray}
		where
		\begin{eqnarray}\nn
		\bz_t=\sqrt{\tau_\star^2-\frac{y_{t-1}}{4}}\bz+\sqrt{\frac{y_{t-1}}{4}}\bw,&&
		\bz_{t-1}=\sqrt{\tau_\star^2-\frac{y_{t-1}}{4}}\bz-\sqrt{\frac{y_{t-1}}{4}}\bw,
		\end{eqnarray}
		and $\nabla$ denotes the vector differential operator. For $y_{t-1}=0$, we have $\bz_t=\bz_{t-1}$ and
		\begin{eqnarray}\label{yt0}
		R^\prime(0)&=&\lim_{p\rightarrow\infty}\frac{1}{p\delta}E\left(Tr\left[\bSigma^{-1}\left\{\nabla\hat{\veta}\right\}^T\bSigma\nabla\hat{\veta}\right]\right),
		\end{eqnarray}
		where $\hat{\veta}=\veta_{\alpha\tau_\star}\left(\bbeta_0+\tau_\star\bSigma^{-1/2}\bz\right)$. Denote ${\cal A}=\{j:\hat{\eta}_j\ne 0\}$. From the definition (\ref{veta}), we get
		\begin{eqnarray}\nn
		\bSigma(\hat{\veta}-(\bbeta_0+\bSigma^{-1/2}\bz))+\theta_\star\partial\|\hat{\veta}\|_1=0,
		\end{eqnarray}
		which implies that
		\begin{eqnarray}\nn
		\bSigma_{{\cal A}{\cal A}}(\hat{\veta}_{\cal A}-(\bbeta_0+\bSigma^{-1/2}\bz)_{\cal A})-\bSigma_{{\cal A}{\cal A}^c}(\bbeta_0+\bSigma^{-1/2}\bz)_{{\cal A}^c}+\theta_\star\text{sign}(\hat{\veta}_{\cal A}))=0.
		\end{eqnarray}
		Taking derivatives, we obtain
		\begin{eqnarray}\nn
		\bSigma_{{\cal A}{\cal A}}(\nabla\hat{\veta})_{{\cal A}{\cal A}}=\bSigma_{{\cal A}{\cal A}}&\text{and}&
		\bSigma_{{\cal A}{\cal A}}(\nabla\hat{\veta})_{{\cal A}{\cal A}^c}=\bSigma_{{\cal A}{\cal A}^c}.
		\end{eqnarray}
		Substituting into (\ref{yt0}), we obtain
		\begin{eqnarray}\nn
		&&R^\prime(0)\\\nn
		&=&\lim_{p\rightarrow\infty}\frac{1}{p\delta}E\left(Tr\left[\left\{(\bSigma^{-1})_{{\cal A}{\cal A}}\left[(\nabla\hat{\veta})_{{\cal A}{\cal A}}\right]^T+(\bSigma^{-1})_{{\cal A}{\cal A}^c}\left[(\nabla\hat{\veta})_{{\cal A}{\cal A}^c}\right]^T\right\}\bSigma_{{\cal A}{\cal A}}\right.\right.\\\nn
		&&~~~~~~~~~~+\left.\left.\left\{(\bSigma^{-1})_{{\cal A}^c{\cal A}}\left[(\nabla\hat{\veta})_{{\cal A}{\cal A}}\right]^T+(\bSigma^{-1})_{{\cal A}^c{\cal A}^c}\left[(\nabla\hat{\veta})_{{\cal A}{\cal A}^c}\right]^T\right\}\bSigma_{{\cal A}{\cal A}^c}\right]\right)\\\nn
		&=&\lim_{p\rightarrow\infty}\frac{1}{p\delta}E\left(Tr\left[\left[(\nabla\hat{\veta})_{{\cal A}{\cal A}}\right]^T\left\{\bSigma_{{\cal A}{\cal A}}(\bSigma^{-1})_{{\cal A}{\cal A}}+\bSigma_{{\cal A}{\cal A}^c}(\bSigma^{-1})_{{\cal A}^c{\cal A}}\right\}\right.\right.\\\nn
		&&~~~~~~~~~~+\left.\left.\left[(\nabla\hat{\veta})_{{\cal A}{\cal A}^c}\right]^T\left\{\bSigma_{{\cal A}{\cal A}}(\bSigma^{-1})_{{\cal A}{\cal A}^c}+\bSigma_{{\cal A}{\cal A}^c}(\bSigma^{-1})_{{\cal A}^c{\cal A}^c}\right\}\right]\right)\\\nn
		&=&\lim_{p\rightarrow\infty}\frac{1}{p\delta}E\{\text{div}(\hat{\veta})\}=\lim_{p\rightarrow\infty}\frac{1}{p\delta}E\left\{\sum_{j=1}^pI(\hat{\eta}_j\ne 0)\right\}\le 1,
		\end{eqnarray}
		for any $\alpha\ge\alpha_{min}$ according to Propositions (\ref{prop1}) and  (\ref{prop3}). By the argument in \cite{BayatiM12}, the covariance of $z_t$ and $z_{t-1}$ is $\tau_\star^2-y_{t-1}/2$ decreasing with $y_{t-1}$ which implies that $R^\prime(y_{t-1})$ is a decreasing function. Moreover $R(0)=0$. Therefore $R(y)$ is concave with $R^\prime(0)\le 1$ and $R(0)=0$. For any $y_0\textgreater 0$, the iteration procedure $y_t=R(y_{t-1})$ leads to a convergent result with $y_t\xrightarrow{t\rightarrow\infty}0$. Therefore,
		\begin{eqnarray}\nn
	&&\lim_{p\rightarrow\infty}\frac{1}{p}\left\|\bbeta^{t+1}-\bbeta^t\right\|^2\\\nn
	&=&\lim_{p\rightarrow\infty}\frac{1}{p}E\left\|\veta_{\theta_t}(\bbeta_0+\bSigma^{-1/2}\bz_t)-\veta_{\theta_{t-1}}(\bbeta_0+\bSigma^{-1/2}\bz_{t-1})\right\|^2
	\end{eqnarray}
	which vanishes as $t\rightarrow\infty$. The statement of $\lim_{t\rightarrow\infty}\lim_{p\rightarrow\infty}\frac{1}{p}\|\bz^t-\bz^{t-1}\|^2=0$ can be proved similarly.
	\end{proof}	
	
	\subsection{Proof of Lemma \ref{lm111}}
	\begin{proof}
		Applying Corollary 2 of \cite{nonseparable} to the AMP iteration (\ref{originali}), for any sequence  $\tilde{\varphi}_p:(\bR^p)^3\rightarrow\bR,~p\ge 1$, of uniformly pseudo-Lipschitz functions, we obtain \begin{eqnarray}\label{pl3}
		\tilde{\varphi}_p\left(\tilde{\bbeta}^{t+1},\tilde{\bbeta}^{s+1},\tilde{\bbeta}_0\right)&\overset{P}{\approx}&\tilde{\varphi}_p\left(\tilde{\veta}_{\theta_t}(\tilde{\bbeta}_0+\bz_t),\tilde{\veta}_{\theta_t}(\tilde{\bbeta}_0+\bz_s),\tilde{\bbeta}_0\right),
		\end{eqnarray}
		where $(\bz_s,\bz_t)$ jointly Gaussian, independent from $\bbeta_0\sim p_{\beta_0}$ with mean 0 and covariance given by $E(\bz_s\bz_s^T)=\tau_s^2\bI_{p\times p}$, $E(\bz_t\bz_t^T)=\tau_t^2\bI_{p\times p}$, and $E(\bz_s\bz_t^T)=\tau_{s,t}\bI_{p\times p}$. The recursion $\tau_{s,,t}$ for all $s,t\ge 0$ is determined by
		\begin{eqnarray}\nn
		\tau_{t+1}^2&=&\sigma_w^2+\lim_{p\rightarrow\infty}\frac{1}{p\delta}E\left(\|\tilde{\veta}_{\theta_t}(\tilde{\bbeta}_0+\tau_t\bz)-\tilde{\bbeta}_0\|^2\right),\\\nn
		\tau_{s+1}^2&=&\sigma_w^2+\lim_{p\rightarrow\infty}\frac{1}{p\delta}E\left(\|\tilde{\veta}_{\theta_s}(\tilde{\bbeta}_0+\tau_s\bz)-\tilde{\bbeta}_0\|^2\right),\\\nn
		\tau_{t+1,s+1}&=&\sigma_w^2+\lim_{p\rightarrow\infty}\frac{1}{p\delta}E\left(\tilde{\veta}_{\theta_t}(\tilde{\bbeta}_0+\tau_t\bz)-\tilde{\bbeta}_0\right)\left(\tilde{\veta}_{\theta_s}(\tilde{\bbeta}_0+\tau_t\bz)-\tilde{\bbeta}_0\right).
		\end{eqnarray}
		Then define sequence of functions: $\tilde{\varphi}_p\left(\bx,\by,\bz\right)=\varphi_p\left(\bSigma^{-1/2}\bx,\bSigma^{-1/2}\by,\bSigma^{-1/2}\bz\right)$ which is also uniformly pseudo-Lipschitz. We then obtain the distributional limit for $\bbeta^{t+1}=\bSigma^{-1/2}\tilde{\bbeta}^{t+1}$ and $\bbeta^{s+1}=\bSigma^{-1/2}\tilde{\bbeta}^{s+1}$ using (\ref{pl3}). 
	\end{proof}
			
	\subsection{Proof of Lemma \ref{lm1.1}}
\begin{proof}
	The matrix $\bX=\tilde{\bX}\bSigma^{1/2}$, where $\tilde{\bX}$ has entries distributed i.i.d. $N(0,1/n)$. Thus, one has (\cite{vershynin2011introduction}, Corollary 5.35)
	\begin{eqnarray}\nn
	\bP\left(\sqrt{\delta}-1-t\le\sigma_{\min}(\tilde{\bX})\le\sigma_{\max}(\tilde{\bX})\le\sqrt{\delta}+1+t\right)\ge 1-2\exp(-t^2/2).
	\end{eqnarray}
	From the fact that
	\begin{eqnarray}\nn
	\sigma_{\min}(\bX)\ge\sigma_{\min}(\tilde{\bX})\sigma_{\min}(\bSigma^{1/2}),&\text{and}&
	\sigma_{\max}(\bX)\le\sigma_{\max}(\tilde{\bX})\sigma_{\max}(\bSigma^{1/2}),
	\end{eqnarray}
	We conclude that, for every $t\ge 0$, there exists $c_5\textgreater 0$ such that 
	\begin{eqnarray}\nn
	\bP\left(c_5^{-1}\le\sigma_{\min}(\bX)\le\sigma_{\max}(\bX)\le c_5\right)\textgreater 1-2\exp(-t^2/2).
	\end{eqnarray}
	
\end{proof}

	\subsection{Proof of Lemma \ref{csparsity}}
	\begin{proof}
		Define $S(c_2)=\{j\in[p]:|v^t_j|\ge 1-c_2\}$, we have almost surely
	\begin{eqnarray}\nn
	\frac{|S(c_2)|}{p}&=&\frac{1}{p}\sum_{i=1}^p\vI\left\{\frac{1}{\theta_{t-1}}|\bX^T\bz^{t-1}+\bSigma(\bbeta^{t-1}-\bbeta^t)|_i\ge 1-c_2\right\}\\\nn
	&\rightarrow&\frac{1}{p}\sum_{i=1}^p\vI\left\{\frac{1}{\theta_{t-1}}|\bSigma\{\bbeta_0+\tau_{t-1}\bSigma^{-1/2}\bz-\eta_{\theta_{t-1}}(\bbeta_0+\tau_{t-1}\bSigma^{-1/2}\bz)\}|_i\ge 1-c_2\right\}.
	\end{eqnarray}
           Let us write $\bar{\bSigma}=\bSigma/\lambda_{\min},\bar{\tau}_{t-1}=\tau_{t-1}/\lambda_{\min}^{1/2},\bar{\theta}_{t-1}=\theta_{t-1}/\lambda_{\min}$, so that
	\begin{eqnarray}\nn
	\hat{\bbeta}&=&\eta_{\theta_{t-1}}(\bbeta_0+\tau_{t-1}\bSigma^{-1/2}\bz)\\\nn
&=&\text{argmin}_{\bbeta\in\bR^p}\left\{\frac{1}{2}\|{\bSigma}^{1/2}(\bbeta-\bbeta_0)-{\tau}_{t-1}\bz\|_2^2+\theta_{t-1}\|\bbeta\|_1\right\}\\\label{fixlasso}
&=&\text{argmin}_{\bbeta\in\bR^p}\left\{\frac{1}{2}\|\bar{\bSigma}^{1/2}(\bbeta-\bbeta_0)-\bar{\tau}_{t-1}\bz\|_2^2+\bar{\theta}_{t-1}\|\bbeta\|_1\right\}.
	\end{eqnarray}
	The KKT conditions of this optimization problem are
	\begin{eqnarray}\nn
	\bar{\bSigma}^{1/2}(\bar{\tau}_{t-1}\bz+\bar{\bSigma}^{1/2}(\bbeta_0-\hat{\bbeta}))\in\bar{\theta}_{t-1}\partial\|\hat{\bbeta}\|_1.
	\end{eqnarray}
Define $\hat{\by}=\hat{\bbeta}+\bar{\bSigma}^{1/2}(\bar{\tau}_{t-1}\bz+\bar{\bSigma}^{1/2}(\bbeta_0-\hat{\bbeta}))$, we have
	\begin{eqnarray}\nn
	\hat{\bbeta}&=&\eta_{soft}(\hat{\by};\bar{\theta}_{t-1}),
	\end{eqnarray}
	where $\eta_{soft}(x;\alpha)=\text{sign}(x)(|x|-\alpha)_+$ and applies coordinates-wise. 
Define $\vf(\bar{\tau}_{t-1}\bz)=(\bI_{p\times p}-\bar{\bSigma}^{-1})\bar{\bSigma}^{1/2}(\bbeta_0-\hat{\bbeta})$, then $\hat{\by}$ can be written as
	\begin{eqnarray}\nn
	\hat{\by}&=&\bbeta_0+\bar{\bSigma}^{1/2}(\bar{\tau}_{t-1}\bz+(\bI_{p\times p}-\bar{\bSigma}^{-1})\bar{\bSigma}^{1/2}(\bbeta_0-\hat{\bbeta}))\\\nn
&=&\bbeta_0+\bar{\bSigma}^{1/2}(\bar{\tau}_{t-1}\bz+\vf(\bar{\tau}_{t-1}\bz)).
	\end{eqnarray}
Denote $\bsigma_j$ the j-th row of $\bar{\bSigma}^{1/2}$ and $\bsigma_j^T\bz=x$, then $x\sim N(0,\|{\bsigma}_j\|_2^2)$. Let $P_j^\perp$ be the projection operator onto the orthogonal complement of the span of $\bsigma_j$. Then
	\begin{eqnarray}\nn
	\hat{y}_j&=&\bbeta_0+\bar{\tau}_{t-1}\bsigma_j^T\bz+\bsigma_j^T\vf(\bar{\tau}_{t-1}(\bsigma_j^T\bz)\bsigma_j/\|\bsigma_j\|_2^2+\bar{\tau}_{t-1}P_j^\perp\bz)\\\label{yhat}
&=&\bbeta_0+\bar{\tau}_{t-1}x+\bsigma_j^T\vf(\bar{\tau}_{t-1}x\bsigma_j/\|\bsigma_j\|_2^2+\bar{\tau}_{t-1}P_j^\perp\bz)\equiv h(x).
	\end{eqnarray}
By (\ref{fixlasso}), $\bar{\bSigma}^{1/2}(\bbeta_0-\hat{\bbeta})$ is 1-Lipschitz in $\bar{\tau}_{t-1}\bz$. Thus, $\vf(\bar{\tau}_{t-1}\bz)$ is $(1-\kappa_{cond}^{-1})$-Lipschitz in $\bar{\tau}_{t-1}\bz$ and $\bar{\tau}_{t-1}(1-\kappa_{cond}^{-1})/\|\bsigma_j\|_2$-Lipschitz in $x$, where $\kappa_{cond}=\lambda_{\max}/\lambda_{\min}$. For any $x_1,x_2\in\bR$, we have
	\begin{eqnarray}\nn
&&|h(x_1)-h(x_2)|\\\nn
&\ge&\bar{\tau}_{t-1}|x_1-x_2|-\left|\bsigma_j^T\left\{\vf\left(\frac{\bar{\tau}_{t-1}x_1\bsigma_j}{\|\bsigma_j\|_2^2}+\bar{\tau}_{t-1}P_j^\perp\bz\right)-\vf\left(\frac{\bar{\tau}_{t-1}x_2\bsigma_j}{\|\bsigma_j\|_2^2}+\bar{\tau}_{t-1}P_j^\perp\bz\right)\right\}\right|\\\label{hx}
&\ge&\bar{\tau}_{t-1}|x_1-x_2|-\bar{\tau}_{t-1}(1-\kappa_{cond}^{-1})|x_1-x_2|=\bar{\tau}_{t-1}\kappa_{cond}^{-1}|x_1-x_2|.
	\end{eqnarray}
According to (\ref{bvt}), we have
	\begin{eqnarray}\label{equiv}
	\bv^t&=&\frac{1}{\bar{\theta}_{t-1}}(\hat{\by}-\eta_{soft}(\hat{\by};\bar{\theta}_{t-1})).
	\end{eqnarray}
By the definition of $S(c_2)$, one obtains
	\begin{eqnarray}\nn
S(c_2)&=&\{j\in[p]:|\hat{y}_j|\ge\bar{\theta}_{t-1}(1-c_2)\}.
	\end{eqnarray}
Therefore
	\begin{eqnarray}\label{sc2}
\frac{|S(c_2)|}{n}&=&\frac{\{j\in[p]:|\hat{y}_j|\textgreater\bar{\theta}_{t-1}\}}{n}+\frac{\{j\in[p]:1-|\hat{y}_j|/\bar{\theta}_{t-1}\in[0,c_2]\}}{n}.
	\end{eqnarray}
	Consider the function
	\begin{eqnarray}\nn
	g(\hat{\by},c_2)&=&\frac{1}{n}\sum_{j=1}^pg_1(\hat{y}_j,c_2),
\end{eqnarray}
where $g_1(\hat{y},c_2)=\text{min}\left\{1,\left(\frac{|\hat{y}|}{\bar{\theta}_{t-1}c_2}-\frac{1}{c_2}+2\right)_+\right\}$. Since 
	\begin{eqnarray}\nn
	|g(\hat{\by}_1,c_2)-g(\hat{\by}_2,c_2)|&\le&\frac{1}{n}\sum_{j=1}^p\{|g_1(\hat{y}_{1,j},c_2)-g_1(\hat{y}_{2,j},c_2)|\}\\\nn
	&\le&\frac{1}{n}\sum_{j=1}^p\frac{1}{\bar{\theta}_{t-1}c_2}|\hat{y}_{1,j}-\hat{y}_{2,j}|\\\nn
	&\le&\frac{\sqrt{p}}{n\theta_{t-1}c_2}\|\hat{\by}_1-\hat{\by}_2\|_2,
\end{eqnarray}
the function $g(\hat{\by},c_2)$ is $\frac{\sqrt{p}}{n\bar{\theta}_{t-1}c_2}$-Lipschitz in $\hat{\by}$. For all $\hat{\by}$, by definition we have $\frac{|S(c_2)|}{n}\le g(\hat{\by},c_2)\le\frac{|S(2c_2)|}{n}$. Moreover, by (\ref{hx}) and (\ref{sc2}), one obtains	
	\begin{eqnarray}\nn
\vE(g(\hat{\by},c_2))&\le&\vE\left(\frac{\|\hat{\bbeta}\|_0}{n}\right)+\vE\left(\frac{\{j\in[p]:1-|\hat{y}_j|/\bar{\theta}_{t-1}\in[0,2c_2]\}}{n}\right)\\\nn
&\le&1-\omega^\star+\sup_a\vE_x\left(\vI\left(a\le \frac{h(x)}{\bar{\theta}_{t-1}}\le a+4c_2\right)\right)\\\nn
&\le&1-\omega^\star+\sup_a\vE_x\left(\vI\left(\frac{a\kappa_{cond}}{\bar{\tau}_{t-1}}\le\frac{x}{\bar{\theta}_{t-1}}\le \frac{(a+4c_2)\kappa_{cond}}{\bar{\tau}_{t-1}}\right)\right)\\\nn
&\le&1-\omega^\star+\frac{4c_2\kappa_{cond}\bar{\theta}_{t-1}}{\sqrt{2\pi}\bar{\tau}_{t-1}}.
	\end{eqnarray}
From (\ref{yhat}), $\hat{\by}$ is $2\kappa_{cond}^{1/2}\bar{\tau}_{t-1}$-Lipschitz in $\bz$. Therefore, $g(\hat{\by},c_2)$ is $\frac{2\sqrt{p}\kappa_{cond}^{1/2}\bar{\tau}_{t-1}}{n\bar{\theta}_{t-1}c_2}$-Lipschitz in $\bz$. By Gaussian concentration of Lipschitz functions
	\begin{eqnarray}\nn
	&&\bP\left(\frac{|S(c_2)|}{n}\ge 1-\omega^\star+\frac{4c_2\kappa_{cond}\bar{\theta}_{t-1}}{\sqrt{2\pi}\bar{\tau}_{t-1}}+\epsilon\right)\\\nn
	&\le&\bP\left(g(\hat{\by},c_2)\ge 1-\omega^\star+\frac{4c_2\kappa_{cond}\bar{\theta}_{t-1}}{\sqrt{2\pi}\bar{\tau}_{t-1}}+\epsilon\right)\\\nn
	&\le&\bP(g(\hat{\by},c_2)\ge\vE(g(\hat{\by},c_2))+\epsilon)\\\nn
	&\le&\exp\left(-\frac{n\delta\bar{\theta}_{t-1}^2c_2^2}{8\kappa_{cond}\bar{\tau}_{t-1}^2}\epsilon^2\right).
\end{eqnarray}
Absorbing constants appropriately, we conclude there exists $C,c_1\textgreater 0$ such that
	\begin{eqnarray}\nn
\bP\left(\frac{|S(c_2)|}{n}\ge 1-\omega^\star/2\right)&\le&C\exp\left(-nc_1\right).
\end{eqnarray}
  \end{proof}

	\subsection{Proof of Lemma \ref{lm1.2}}
  \begin{proof}

%Consider the event
%	\begin{eqnarray}\nn
%{\cal A}=\{\kappa_-(\bX,n(1-\omega^\star/4))\ge\kappa_{min}\}.
%\end{eqnarray}
%Then there exists $\kappa_{min},C,c\textgreater 0$ such that
%	\begin{eqnarray}\nn
%P({\cal A})\ge 1-C\exp(-cn).
%\end{eqnarray}

Let $k=[n(1-\omega^\star/4)]$ and note that $k\textless p$. Because for $k\textgreater p$, we have $\kappa_-(\bX,n(1-\omega^\star/4))=\kappa_-(\bX,p)$ and thus $\bP(\kappa_-(\bX,n(1-\omega^\star/4))\ge c_4)\ge 1-C\exp(-cn)$.

Because $\kappa_-(\bX,S^\prime)\ge\kappa_-(\bX,S)$ when $S^\prime\subset S$, we have that $\kappa_-(\bX,n(1-\omega^\star/4))=\min_{|S|=k}\kappa_-(\bX,S)$. By a union bound, for any $t\textgreater 0$ 
	\begin{eqnarray}\label{union}
	\bP(\kappa_-(\bX,n(1-\omega^\star/4))\le t)\le\sum_{|S|=k}\bP(\kappa(\bX_S)\le t).
\end{eqnarray}	
The matrix $\bX_S=\tilde{\bX}_S\bSigma^{1/2}_{S,S}$ where $\tilde{\bX}_S$ has entries distribution i.i.d. $N(0,1/n)$. Thus, one has
	\begin{eqnarray}\nn
	\kappa_-(\bX_S)\ge\kappa_-(\tilde{\bX}_S)\kappa_-(\bSigma_{S,S}^{1/2})\ge\kappa_-(\tilde{\bX}_S)\kappa_{min}^{1/2}
\end{eqnarray}	
Invoking the fact that $\tilde{\bX}_S$ has the same distribution for all $|S|=k$, expression (\ref{union}) implies
		\begin{eqnarray}\nn
	\bP(\kappa_-(\bX,n(1-\omega^\star/4))\le t)\le\left(\begin{array}{c}p\\k\end{array}\right)\bP(\kappa_-(\tilde{\bX}_S)\le t/\kappa_{min}^{1/2}).
	\end{eqnarray}	
Let $f_{min}(k,n,\lambda)$ denote the probability density function for the smallest eigenvalue $\kappa_-(\tilde{\bX}_S)$. By Prop. 5.2, pp.553 \cite{edelman}, $f_{min}(k,n,\lambda)$ satisfies
		\begin{eqnarray}\nn
f_{min}(k,n;\lambda)&\le&g_{min}(k,n;\lambda)\\\nn
&\equiv &\frac{\Gamma((n+1)/2)}{\Gamma(k/2)\Gamma((n-k+1)/2)\Gamma((n-k+2)/2)}\\\nn
&&\left(\frac{\pi}{2n\lambda}\right)^{1/2}\left(\frac{n\lambda}{2}\right)^{(n-k)/2}\exp(-n\lambda/2).
	\end{eqnarray}	
It can be verified that the quantity $g_{min}(k,n;\lambda)$ is strictly increasing in $\lambda$ on $[0,(n-k-1)/n)$. Lemma 2.9 of \cite{bct10} states that as $n,k\rightarrow\infty$ with $k/n\rightarrow\rho\in(0,1]$,
		\begin{eqnarray}\nn
g_{min}(k,n;\lambda)\rightarrow p_{min}(n,\lambda)\exp(n\psi_{min}(\lambda,\rho)),
	\end{eqnarray}	
where $p_{min}(n,\lambda)$ is a polynomial in $n,\lambda$, and $\psi_{min}(\lambda,\rho)=H(\rho)+\frac{1}{2}[(1-\rho)\log\lambda+1-\rho+\rho\log\rho-\lambda]$, where $H(\rho)=\rho\log(1/\rho)+(1-\rho)\log(1/(1-\rho))$.	Therefore, for $t/\kappa_{min}^{1/2}\le 1-\rho$, we have
		\begin{eqnarray}\nn
\bP(\kappa_-(\tilde{\bX}_S)\le t/\kappa_{min}^{1/2})&=&\int_0^{t/\kappa_{min}^{1/2}}f_{min}(k,n;\lambda)d\lambda\\\nn
&\le&\int_0^{t/\kappa_{min}^{1/2}}g_{min}(k,n;\lambda)d\lambda\\\nn
&\le& t/\kappa_{min}^{1/2}g_{min}(k,n;t/\kappa_{min}^{1/2})\\\nn
&=&C(n,t/\kappa_{min}^{1/2})\exp(n\psi(\rho,t/\kappa_{min}^{1/2})),
	\end{eqnarray}	
where $C(a,b)$ is a polynomial in $a,b$. To simplify $\left(\begin{array}{c}p\\k\end{array}\right)$, we apply the second of Binet's log gamma formulas \citep{whittaker_watson_1996} and obtain
		\begin{eqnarray}\nn
\frac{1}{n}\log\left(\begin{array}{c}p\\k\end{array}\right)&\rightarrow&\rho\log\frac{1}{\rho\delta}+(\frac{1}{\delta}-\rho)\log\frac{1}{1-\rho\delta}=H(\rho\delta)/\delta. 	
	\end{eqnarray}	
We conclude that	
		\begin{eqnarray}\nn
\bP(\kappa_-(\bX,n(1-\zeta^\star/4))\le t)&\le&C(n,t/\kappa_{min}^{1/2})\exp(n(H(\rho\delta)/\delta+\psi(\rho,t/\kappa_{min}^{1/2}))).
\end{eqnarray}	
Note that $H(\rho)\le 1/2$ for $\rho\in(0,1)$. Thus, there exists $c\textgreater 0$ such that 
		\begin{eqnarray}\nn
H(\rho\delta)/\delta+\psi(\rho,t/\kappa_{min}^{1/2})\le -c
\end{eqnarray}	
for all $\log(t/\kappa_{min}^{1/2})\le -1-\frac{8(1/\delta+1)\log 2}{\omega^\star}-\frac{8c}{\omega^\star}$. Because $C(n,t/\kappa_{min}^{1/2})e^{-cn}$ is upper bounded by a constant $C$, we conclude there exists $C,c\textgreater 0$ such that
		\begin{eqnarray}\nn
\bP(\kappa_-(\bX,n(1-\omega^\star/4))\le t)\le Ce^{-cn}.
\end{eqnarray}	  
  \end{proof}

	\subsection{Proof of Corollary \ref{corollary1}}
\begin{proof}
	For $\bSigma=\bI_{p\times p}$, (\ref{mfun}) can be simplified as 
\begin{eqnarray}\nn
M(\epsilon,\Delta,\alpha)&=&\epsilon_+E(z-\alpha)^2+\epsilon_-E(z+\alpha)^2+(1-\epsilon)E[(z-\alpha)^2I(z\ge\alpha)\\\nn
&&~~~+(z+\alpha)^2I(z\le -\alpha)]\\\nn
&=&\epsilon(1+\alpha^2)+2(1-\epsilon)[(1+\alpha^2)(1-\Phi(\alpha))-\alpha\phi(\alpha)],
\end{eqnarray}	 
where the first term comes from the non-zero components of $\bbeta_0$ and the second term comes from the zero components of $\bbeta_0$. To determine $\delta_c=\inf_\alpha M(\epsilon,\Delta,\alpha)$, we can solve $\frac{\partial M(\epsilon,\Delta,\alpha)}{\partial\alpha}=0$ and thus obtain the phase transition curve as shown in (\ref{phaseiid}).
  \end{proof}

	\subsection{Proof of Corollary \ref{corollary2}}
\begin{proof}
	For block-diagonal matrix with block $\bSigma_s=\left(\begin{array}{cc}1&\rho\\\rho&1\end{array}\right)$, (\ref{mfun}) can be simplified as 
		\begin{eqnarray}\nn
		&&M(\epsilon,\Delta,\alpha)\\\label{mfun2d}
		&=&\frac{1}{2}E\{((\bSigma_s^{1/2}\bz)_{\cal A}-\alpha\text{sign}(\hat{\bbeta}_{\cal A}))^T\bSigma_{s{\cal A}{\cal A}}^{-1}((\bSigma_s^{1/2}\bz)_{\cal A}-\alpha\text{sign}(\hat{\bbeta}_{\cal A}))\},
\end{eqnarray}	 
	where $\bz\sim N(0,\bI_{2\times 2})$. Note that $\bSigma_s^{1/2}=\left(\begin{array}{cc}\rho_1&\rho_2\\\rho_2&\rho_1\end{array}\right)$, where $\rho_1=\frac{\sqrt{1+\rho}}{2}$ and $\rho_2=\frac{\sqrt{1-\rho}}{2}$. 
	
	There are three scenarios. In the first scenario, both components of $\bbeta_0$ are non-zero, which means that ${\cal B}=\{1,2\}$ and ${\cal B}^c=\varnothing$. Its contribution to (\ref{mfun}) can be written as
		\begin{eqnarray}\nn
		M_1(\epsilon,\Delta,\alpha)&=&\epsilon_+^2(1+\frac{\alpha^2}{1+\rho})+\epsilon_-^2(1+\frac{\alpha^2}{1+\rho})+2\epsilon_+\epsilon_-(1+\frac{\alpha^2}{1-\rho})\\\nn
		&=&\epsilon^2A(\alpha,\Delta),
		\end{eqnarray}	 
	where $A(\alpha,\Delta)$ is defined in (\ref{afun}). In the second scenario, only one component of $\bbeta_0$ are non-zero, i.e. ${\cal B}=\{1\}$ or ${\cal B}=\{2\}$. In the situation where ${\cal B}=\{1\}$ and $\beta_{0,1}\textgreater 0$, we need to consider the one-dimensional LASSO problem specified by (\ref{blasso}) with $\bar{x}=\sqrt{1-\rho^2}$ and $\bar{y}=\bar{x}^{-1}(\xi_2-\rho\xi_1+\rho\alpha)$ whose solution is 
		\begin{eqnarray}\nn
		\left\{\begin{array}{ccc}\text{positive} & if & \xi_2-\rho\xi_1+\rho\alpha\ge\alpha\\
		0&if&|\xi_2-\rho\xi_1+\rho\alpha|\textless\alpha\\
		\text{negative} & if & \xi_2-\rho\xi_1+\rho\alpha\le -\alpha\end{array}\right..	
		\end{eqnarray}	 	
		Plugging this result into (\ref{blasso}), we obtain its contribution to (\ref{mfun2d}) to be
		\begin{eqnarray}\nn
		 &&\epsilon_+(1-\epsilon)\{E(\xi_1-\alpha)^2I(|\xi_2-\rho\xi_1+\rho\alpha|\textless\alpha)\\\nn
&+&E\frac{(\xi_1-\alpha)^2+(\xi_2-\alpha)^2-2\rho(\xi_1-\alpha)(\xi_2-\alpha)}{1-\rho^2}I(\xi_2-\rho\xi_1+\rho\alpha\ge\alpha)\\\nn
&+&E\frac{(\xi_1-\alpha)^2+(\xi_2+\alpha)^2-2\rho(\xi_1-\alpha)^2(\xi_2+\alpha)^2}{1-\rho^2}I(\xi_2-\rho\xi_1+\rho\alpha\le -\alpha)\}.
		\end{eqnarray}	 
The other situations in this scenario can be considered in a similar way. The total contribution of the second scenario to (\ref{mfun2d}) is
		\begin{eqnarray}\nn
		M_2(\epsilon,\Delta,\alpha)&=&\epsilon(1-\epsilon)B(\alpha),
		\end{eqnarray}	
		where $B(\alpha)$ is defined in (\ref{bfun}).

In the third scenario, both components of $\bbeta_0$ are zero, i.e. ${\cal B}=\varnothing$ and ${\cal B}^c=\{1,2\}$. According to (\ref{blasso}), we need to consider the following two dimensional LASSO problem  
		\begin{eqnarray}\nn
		\bar{\bbeta}=\text{argmin}_{\bbeta\in\bR^{2}}\left\{\frac{1}{2}\|\bz-\bSigma_s^{1/2}\bbeta\|_2^2+\alpha\|\bbeta\|_1\right\}.
		\end{eqnarray} 
There exists subgradients $\partial\|\beta_1\|_1$ and $\partial\|\beta_2\|_1$ such that
		\begin{eqnarray}\nn
		\bar{\beta}_1+\rho\bar{\beta}_2&=&\xi_1-\alpha\partial\|\bar{\beta}_1\|_1,\\\label{twodequ}
		\rho\bar{\beta}_1+\bar{\beta}_2&=&\xi_2-\alpha\partial\|\bar{\beta}_2\|_1.
		\end{eqnarray} 
By dividing the two dimensional space into nine regions (as illustrated by Figure \ref{figure2d}), we obtain the following solution for $\bar{\bbeta}$
 		\begin{eqnarray}\label{ninearea}
 \left\{\begin{array}{ccc}\bar{\bbeta}_1=\bar{\bbeta}_2=0 & if & \|\xi_1\|\textless\alpha~\&~ \|\xi_2\|\textless\alpha\\
 \bar{\bbeta}_1\textgreater 0,~\bar{\bbeta}_2=0&if&\|\xi_1\|\ge\alpha~\&~|\xi_2-\rho\xi_1+\rho\alpha|\textless\alpha\\
 \bar{\bbeta}_1\textless 0,~\bar{\bbeta}_2=0&if&\|\xi_1\|\le -\alpha~\&~|\xi_2-\rho\xi_1-\rho\alpha|\textless\alpha\\
  \bar{\bbeta}_1 =0,~\bar{\bbeta}_2\textgreater 0&if&\|\xi_2\|\ge\alpha~\&~|\xi_1-\rho\xi_2+\rho\alpha|\textless\alpha\\
 \bar{\bbeta}_1=0,~\bar{\bbeta}_2\textless 0&if&\|\xi_2\|\le -\alpha~\&~|\xi_1-\rho\xi_2-\rho\alpha|\textless\alpha\\
 \bar{\bbeta}_1\textgreater 0,~\bar{\bbeta}_2\textgreater 0&if&\xi_1-\rho\xi_2+\rho\alpha\ge\alpha~\&~\xi_2-\rho\xi_1+\rho\alpha\ge\alpha\\
 \bar{\bbeta}_1\textgreater 0,~\bar{\bbeta}_2\textless 0&if&\xi_1-\rho\xi_2-\rho\alpha\ge\alpha~\&~\xi_2-\rho\xi_1+\rho\alpha\le -\alpha\\
 \bar{\bbeta}_1\textless 0,~\bar{\bbeta}_2\textgreater 0&if&\xi_1-\rho\xi_2+\rho\alpha\le -\alpha~\&~\xi_2-\rho\xi_1-\rho\alpha\ge\alpha\\
 \bar{\bbeta}_1\textless 0,~\bar{\bbeta}_2\textless 0&if&\xi_1-\rho\xi_2-\rho\alpha\le -\alpha~\&~\xi_2-\rho\xi_1-\rho\alpha\le -\alpha
 \end{array}\right..	
 \end{eqnarray}	 	
Substituting into (\ref{blasso}), the total contribution of the third scenario to (\ref{mfun2d}) can be written as
		\begin{eqnarray}\nn
M_3(\epsilon,\Delta,\alpha)&=&(1-\epsilon)^2C(\alpha),
\end{eqnarray}	
where $C(\alpha)$ is defined in (\ref{cfun}). Therefore 
		\begin{eqnarray}\nn
M(\epsilon,\Delta,\alpha)&=&M_1(\epsilon,\Delta,\alpha)+M_2(\epsilon,\Delta,\alpha)+M_3(\epsilon,\Delta,\alpha)\\\label{mfund}
&=&\epsilon^2A(\alpha,\Delta)+\epsilon(1-\epsilon)B(\alpha)+(1-\epsilon)^2C(\alpha).
\end{eqnarray}	 
To get $\delta_c$, we need to solve the equation $\frac{\partial M(\epsilon,\Delta,\alpha)}{\partial\alpha}=0$ for $\epsilon$ which is given by
		\begin{eqnarray}\nn
\epsilon=\frac{2C^\prime(\alpha)-B^\prime(\alpha)+\sqrt{B^\prime(\alpha)^2-4\frac{\partial A(\alpha,\Delta)}{\partial\alpha}C^\prime(\alpha)}}{2\{\frac{\partial A(\alpha,\Delta)}{\partial\alpha}-B^\prime(\alpha)+C^\prime(\alpha)\}}.
\end{eqnarray}	
 Substituting into (\ref{mfund}), we conclude that the transition curve is determined by (\ref{phasedep}).
 	\begin{figure}[hbtp]
 	\vspace{0cm}
 	\begin{center}
 		\includegraphics[angle=-90,width=\textwidth]{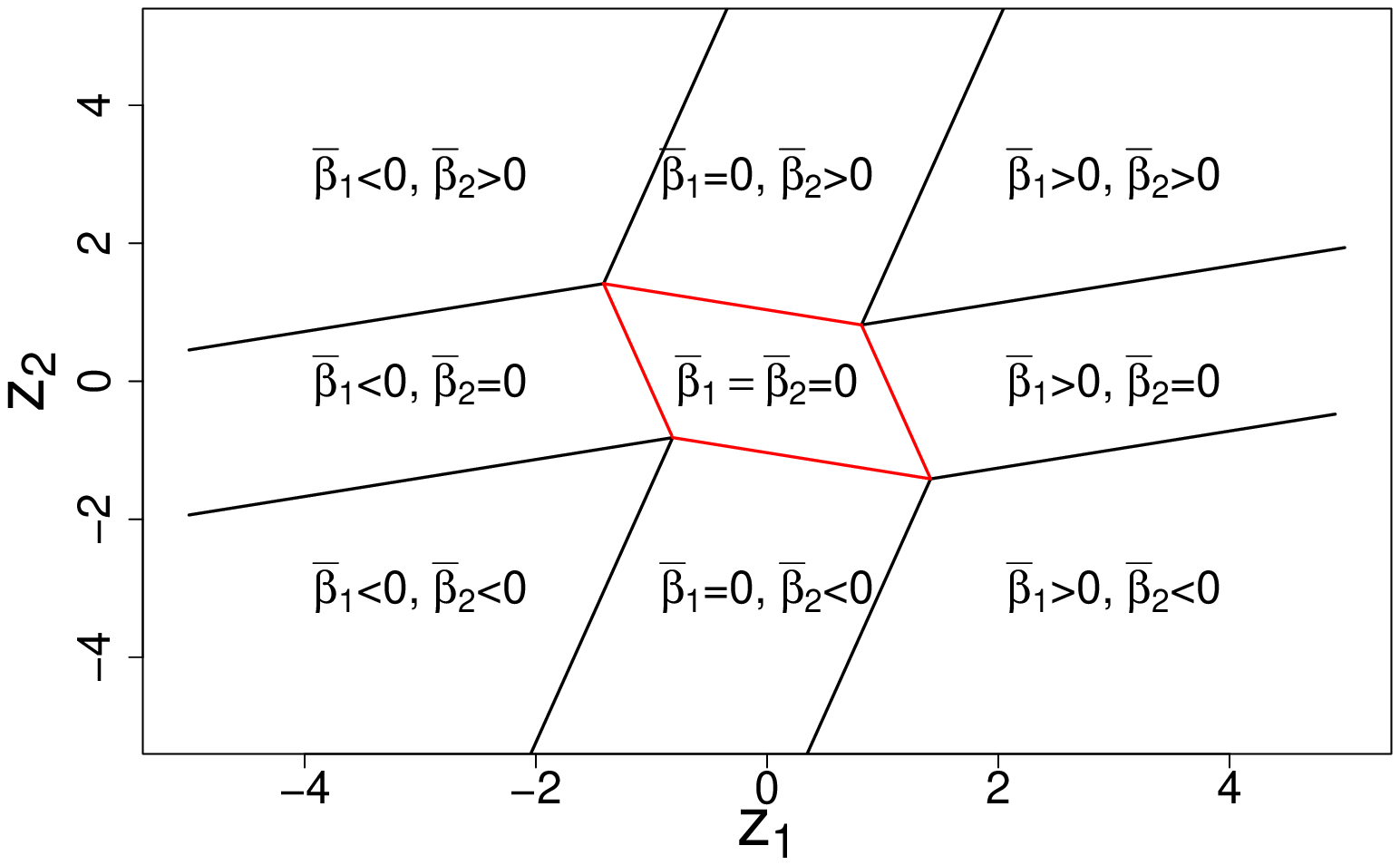}
 	\end{center} 
 	\vspace{-0.5cm}
 	\caption{Illustration of the solution (\ref{ninearea}) for equation (\ref{twodequ}) in two dimensional space. Here $\rho=0.5$ and $\alpha=1$.}  
 	\label{figure2d}
 \end{figure}
 
	\end{proof}

%\begin{proof}
%	\begin{lemma}\label{lm2a}
%	For any fixed $\alpha\textgreater 0$, the equation 
%		\begin{eqnarray}\nn
%	\tau^2=\psi(\tau^2,\alpha\tau) 
%\end{eqnarray}	  
%	has a unique solution.  
%\end{lemma}
%
%  \end{proof}

	%\section*{Acknowledgments}
	%And this is an acknowledgments section with a heading that was produced by the
	%$\backslash$section* command. Thank you all for helping me writing this
	%\LaTeX\ sample file. See \ref{suppA} for the supplementary material example.
	%
	%\begin{supplement}
	%\sname{Supplement A}\label{suppA}
	%\stitle{Title of the Supplement A}
	%\slink[url]{http://www.e-publications.org/ims/support/dowload/imsart-ims.zip}
	%\sdescription{}
	%\end{supplement}

	\bibliographystyle{chicago} \bibliography{biblist}
	
	\end{document}